\documentclass[useAMS,usenatbib]{mn2e}
\usepackage{graphicx}
\usepackage{lscape}
\usepackage{perpage}
\usepackage{color}
\MakePerPage{footnote}

\newcommand{\beq}	{\begin{equation}}
\newcommand{\eeq}	{\end{equation}}
\newcommand{\beqa}	{\begin{eqnarray}}
\newcommand{\eeqa}	{\end{eqnarray}}
\newcommand{\e}	        {$^{-1}$}
\newcommand{\ee}	{$^{-2}$}
\newcommand{\eee}	{$^{-3}$}
\newcommand{\dis}		{\displaystyle}

\newcommand{\calm}	{{\cal M}}

\newcommand{\calr}	{{\cal R}}

\newcommand{\avg}[1]    {{\langle #1 \rangle}} 

\newcommand{\vecB}	{{\bf B}}

\newcommand{\vecF}      {{\bf F}}
\newcommand{\vecg}	{{\bf g}}

\newcommand\fnm		{\footnotemark}
\newcommand\fnt		{\footnotetext}

\newcommand{\alfven}    {{Alfv$\acute{\rm e}$n }}

\newcommand{\alfvenic}  {{Alfv$\acute{\rm e}$nic }}

\newcommand{\Lor}		{{\rm L}}
\newcommand{\los}		{{\rm los}}

\newcommand\med		{{\rm med}}
\newcommand{\muphi}	{\mu_{\Phi}}
\newcommand{\muphio}	{\mu_{\Phi,0}}
\newcommand{\muplos}	{\mu_{\Phi,{\rm los}}}
\newcommand{\nh}	{n_{\rm H}}

\newcommand{\nH}	{n_{\rm H}}
\newcommand{\nlos}  {\bar n_{\rm H,\,los}}

\newcommand{\rbeam}  {r_{\rm beam}}

\newcommand{\va}	{v_{\rm A}}

\newcommand{\vrms}	{v_{\rm rms}}

\newcommand{\avir}      {\alpha_{\rm vir}}
\newcommand{\bav}		{|\avg{\vecB}_V|}
\newcommand\brms        {B_{\rm rms}}
\newcommand\brmsmu        {B_{{\rm rms},\,\mu}}
\newcommand\brmsv       {B_{{\rm rms,}V}}
\newcommand\brmsr       {B_{{\rm rms,}M}}
\newcommand\blos        {B_{\rm los}}
\newcommand\blosmu      {B_{{\rm los},\,\mu}}
\newcommand\blosr       {B_{\rm los}}

\newcommand\absbmidz  {|B|_{{\rm mid},\,z'}}
\newcommand\bmidz		{B_{{\rm mid},\,z'}}
\newcommand\btot        {B_{\rm tot}}
\newcommand\btotmu        {B_{{\rm tot},\,\mu}}
\newcommand\btotr       {B_{\rm tot}}
\newcommand\btotv       {\bav}

\newcommand{\ma}	{{\calm_{\rm A}}}
\newcommand{\mao}	{{\calm_{\rm A,0}}}

\newcommand{\mug}	{$\mu$G}
\newcommand{\mmug}	{\mu{\rm G}}
\newcommand{\nbh}	{\bar n_{\rm H}}
\newcommand{\nbhf}	{\bar n_{\rm H,\, 4}}
\newcommand{\nbht}	{\bar n_{\rm H,\, 3}}

\newcommand{\sid}	{\sigma_{\rm 1D}}
\newcommand{\snt}       {\sigma_{\rm nt}}
\newcommand{\spc}	{\sigma_{\rm pc}}
\newcommand{\spcs}      {{\sigma_{\rm pc}^*}}

\newcommand{\Nh}	{N_{\rm H}}
\newcommand{\Nhbeam}	{N_{\rm H,\,beam}}
\newcommand{\NH}	{N_{\rm H}}
\newcommand{\tff}	{t_{\rm ff}}

\newcommand{\at}     {\langle \alpha \rangle_t}
\newcommand{\blosna} {\langle \blosr/\nbh^{0.65} \rangle}
\newcommand{\btotna} {\langle \btotr/\nbh^{0.65} \rangle}

\newcommand{\blosnfa} {\langle \blosmu/\nbhf^{0.65} \rangle}
\newcommand{\blosnfat} {\langle \blosmu/\nbhf^{0.65} \rangle_t}

\newcommand{\btotnfa} {\langle \btotmu/\nbhf^{0.65} \rangle}
\newcommand{\btotnfat} {\langle \btotmu/\nbhf^{0.65} \rangle_t}
\newcommand{\blosnam} {(\blosr/\nbh^{0.65})_{\rm med}}
\newcommand{\btotnam} {(\btotr/\nbh^{0.65})_{\rm med}}

\newcommand{\blosnfam} {(\blosmu/\nbhf^{0.65})_{\rm med}}
\newcommand{\blosnfamt} {(\blosmu/\nbhf^{0.65})_{\med,\, t}}

\newcommand{\btotnfam} {(\btotmu/\nbhf^{0.65})_{\rm med}}
\newcommand{\btotnfamt} {(\btotmu/\nbhf^{0.65})_{\med,\, t}}

\newcommand{\btotrm} {\langle \btotr \rangle_t}
\newcommand{\blosrmed} {(\blosr)_{\rm med}}
\newcommand{\btotrmed} {(\btotr)_{\rm med}}

\newcommand\act		{{\rm actual}}
\newcommand{\beam}	{{\rm beam}}
\newcommand\btmin		{B_{\rm tot,\, min}}

\newcommand{\btm}		{B_{\rm tot,\,max}}
\newcommand{\bbtm}		{b_{\rm tot,\,max}}

\newcommand{\bblos}	{b_\los}
\newcommand{\bblosm}	{b_{\los,\,\med}}
\newcommand{\bbtot}	{b_\tot}
\newcommand{\bbtotm}	{b_{\tot,\,\med}}

\newcommand\cs		{c_{\rm s}}
\newcommand\frag		{{\rm frag}}
\newcommand{\ft}		{{\rm ft}}
\newcommand{\kb}		{k_{\rm B}}
\newcommand{\midp}		{{\rm mid}}
\newcommand\pc		{{\rm pc}}
\newcommand{\tot}		{{\rm tot}}
\newcommand{\Hea}	{{\rm H}}
\newcommand{\xlos}	{{x_{\rm los}}}

\title[Magnetized Interstellar Molecular Clouds. I. ]
{Magnetized Interstellar Molecular Clouds. I. Comparison Between Simulations and Zeeman Observations}
\author[Pak Shing Li, Christopher F. McKee, and Richard I. Klein]{Pak Shing Li$^{1}$\thanks{E-mail:psli@astron.berkeley.edu (PSL)}, Christopher F. McKee$^{1,2}$\thanks{E-mail:cmckee@astro.berkeley.edu (CFM)}, and Richard I. Klein$^{1,3}$\thanks{E-mail:klein@astron.berkeley.edu (RIK)}\\
$^{1}$Astronomy Department, University of California, Berkeley, CA 94720\\
$^{2}$Physics Department, University of California, Berkeley, CA 94720\\
$^{3}$Lawrence Livermore National Laboratory,P.O.Box 808, L-30, Livermore, CA 94550}
\begin{document}

\date{}

\pagerange{\pageref{firstpage}--\pageref{lastpage}}  \pubyear{}

\maketitle

\label{firstpage}

\begin{abstract}
The most accurate measurements of magnetic fields in star-forming gas are based on the Zeeman observations analyzed by \citet{cru10}.
We show that their finding that the 3D magnetic field scales approximately as density$^{0.65}$ can also be obtained from analysis of the observed line-of-sight fields. We present two large-scale AMR MHD simulations of several thousand $M_\odot$ of turbulent, isothermal, self-gravitating gas, one with a strong initial magnetic field (\alfven\ Mach number $\mao=1$) and one with a weak initial field ($\mao=10$). We construct samples of the 100 most massive clumps in each simulation and show that they exhibit a power-law relation between field strength and density ($\nbh$) in excellent agreement with the observed one. Our results imply that the average field in molecular clumps in the interstellar medium is $\avg{\btot(\nbh)}\approx 42 \nbhf^{0.65}~\mmug$. Furthermore, the median value of the ratio of the line-of-sight field to density$^{0.65}$ in the simulations is within a factor of about (1.3, 1.7) of the observed value for the strong and weak field cases, respectively. The median value of the mass-to-flux ratio, normalized to the critical value, is 70 per cent of the line-of-sight value.
This is larger than the 50 per cent usually cited for spherical clouds because the actual mass-to-flux ratio depends on the volume-weighted field, whereas the observed one depends on the mass-weighted field. Our results indicate that the typical molecular clump in the ISM is significantly supercritical ($\sim$ factor of 3). The results of our strong-field model are in very good
quantitative agreement with the observations of \citet{li09}, which show a strong correlation in field orientation
between small and large scales. Because there is a negligible correlation in the weak-field model, we conclude that molecular clouds form from strongly magnetized (although magnetically supercritical) gas, in agreement with the conclusion of \citet{li09}.
\end{abstract}


\section{Introduction}

Since the detection of magnetic fields in the interstellar medium (ISM), 
they have played a central role in star formation theory.
The first estimate of the strength of the interstellar field was made by Leverett Davis \citep{dav51}, who inferred
a field strength of $10^{-4}$~G based on a comparison of the observed fluctuations in the orientation of
optical polarization vectors and the observed amplitude of turbulent motions in the ISM. Two years later, 
\citet{cha53} used the same method (with the correction that it is the 1D
velocity dispersion rather than the 3D value that enters) with different values for the density and velocity
dispersion to obtain $B=7.2~\mu$G, remarkably close to the current best estimate for 
the field strength in the diffuse ISM $B=6.0\pm 1.8~\mu$G \citep{hei05b}. (More precisely, Chandrasekhar \& Fermi determined the field in the plane of the sky, so their value for the total field strength
would be $8.3~\mu$G.) This method of determining the
field strength is generally referred to as the Chandrasekhar-Fermi method, but in view of the fact that
it was originally and independently discovered by Davis, as acknowledged by \citet{cha53}, we shall refer to it as the Davis-Chandrasekhar-Fermi (DCF) method.

In a classic paper, \citet{mes56} analyzed the implications of interstellar magnetic fields for star formation. They pointed out that the magnetic energy scales with the gravitational energy, so that the magnetic field sets a lower limit on the mass of a gas cloud that can undergo gravitational collapse.  
In modern terminology, they discovered that there is a magnetic critical mass; masses above this can
undergo gravitational collapse (the magnetically supercritical case), whereas lower masses cannot
(the magnetically subcritical case). They estimated that the magnetic critical mass was at least $500\, M_\odot$.
Subsequent observations have shown that diffuse HI clouds (the only type of interstellar cloud known to
Mestel and Spitzer) are indeed magnetically sub-critical \citep{hei97,hei04}.
Since $500\,M_\odot$ is much larger than the mass of a star, 
\citet{mes56} argued that star formation occurred via ambipolar diffusion, in which the neutrals contracted relative to the ions that were tied to the magnetic field.
They further speculated that the field might be strong enough to prevent the formation of a disc
around the protostar, thereby anticipating what is informally known as the `protostellar accretion-disc crisis' decades before such discs were discovered observationally!

Whereas \citet{mes56} argued that ambipolar diffusion was rapid, subsequent workers have found that it is
about an order of magnitude slower than free-fall collapse \citep{mou87,shu87}. The classical picture
of low-mass star formation was based on the assumption that the molecular cloud cores out of which
low-mass stars form are magnetically subcritical and that gravitational collapse is enabled by quasi-static ambipolar diffusion.  
This model has been challenged by
Zeeman observations of dense molecular cloud clumps inside giant molecular clouds (GMCs), which show that on average the clumps are magnetically supercritical--i.e., that gravity dominates--by a factor of order 2
\citep{cru99,fal08,tro08,cru10}. Zeeman observations measure the net field along the line of sight, and that field is almost never strong enough to render the clumps magnetically subcritical  \citep{cru12}. The only clear counter example that we are aware of comes from DCF observations of a quiescent molecular cloud, which indicated that the clumps in that cloud were uniformly slightly subcritical \citep{mar12}. The significant uncertainties associated with the DCF method are reviewed by \citet{cru12}. 

Although gravitationally bound GMCs should be magnetically supercritical \citep{mck89}, fragmentation
along the field lines should result in magnetically subcritical clumps; indeed, it is for this
reason that \citet{mes56} invoked ambipolar diffusion to enable star formation. The most
likely reason that the clumps inside GMCs are not magnetically subcritical is due to the effects of
the supersonic turbulence observed inside GMCs. Turbulence can accelerate ambipolar diffusion \citep{fat02,zwe02,li12a}.  Furthermore, Lazarian and his collaborators \citep[e.g.][]{laz12,lea12,laz15} have suggested that magnetic reconnection in a turbulent medium (a process they term `reconnection diffusion') leads to rapid diffusion of the magnetic field independent of ambipolar drift; this process is most effective on length scales at which the turbulence is super-\alfvenic.  

Large scale numerical simulations provide a powerful tool for interpreting the observations of magnetic
fields in molecular clouds. Whereas observations of fields are limited to the component along the 
line of sight for Zeeman observations and to the components in the plane of the sky for DCF observations,
the full 3D field is accessible to numerical simulations. The majority of simulations used to study
magnetic fields in molecular clouds are ideal MHD (although see \citealp{ois06,li08,mck10,vaz11,li12a}) and therefore do not include the acceleration of ambipolar diffusion by turbulence. On the other hand, such
simulations can simulate reconnection in a turbulent medium insofar as numerical reconnection mimics
the effects of actual reconnection.

A number of simulations have been carried out to explore super-\alfvenic models of molecular clouds,
in which the energy in turbulence greatly exceeds the energy in the initial, uniform magnetic field
\citep{pad99,pad04,lun08,lun09,col11}.
These simulations did not include self-gravity except that of \citet{col11}, who considered a cloud that was
initially very magnetically supercritical. Comparison of the distribution of
the line-of-sight magnetic field, $\blos$, versus the column density of cloud clumps with observation showed good agreement.  Both \citet{lun09} and \citet{col11} use models with very large initial values of the plasma beta, $\beta =8\pi\rho \cs^2/B^2= 22.2$, corresponding to very weak magnetic fields. The turbulence amplified the field, however, eventually reducing $\beta(\brms)$ to about 0.2 in the simulation by \citet{col11}.
The key observational distinction between weak and strong mean fields is in the field structure:
The field in the weak mean-field case is very tangled, whereas that in the strong mean-field case maintains a correlation in the orientation on large and small scales \citep{ost01}.
Observations by \citet{li09} of the the orientations of the field on large and small scales led them to conclude that the mean field is strong--even an \alfvenic Mach number $\ma=2$ is too large to be consistent with the data.
One of the goals of our work is to revisit the issue of the orientation correlation in simulations that include self gravity.

\renewcommand{\thefootnote}{\alph{footnote}}
\begin{table*}
\caption{Glossary}
\label{tab:1}
\begin{tabular}{lll}
\hline
\hline
Parameter\fnm[1] & ~~~Definition~~~~~ & Reference \\
\hline
$B_0$ & Magnitude of initial magnetic field: $31.6\,\mmug$ (strong) $3.16\,\mmug$ (weak) & Sec. \ref{sec:param}\\
 $\blos$ & 
 Magnitude of mass-weighted line-of-sight field in central Gaussian beam~~~~& Eq. (\ref{eq:blos})\\
 $\absbmidz$&Average magnitude of field in central $\pi\rbeam^2$ of clump midplane &Sec. \ref{sec:obsmfr}\\
 $\bmidz$&Magnitude of average axial field in central $\pi\rbeam^2$ of clump midplane &Sec. \ref{sec:obsmfr}\\
 $\btot$& Total mass-weighted field, $(B_{\los,\,x}^2+B_{\los,\,y}^2+B_{\los,\,z}^2)^{1/2}$& Eq. (\ref{eq:btot1})\\
 $B_V$& Magnitude of the field averaged over clump volume, $\avg{|B|}_V$  &  Sec. \ref{sec:compsw} \\
 $\brmsv$&RMS field in clump volume, $(\avg{B^2}_V)^{1/2}$ & Sec. \ref{sec:dist}\\
 $\cs$& Mass-weighted isothermal sound speed & Eq. (\ref{eq:sid})\\
 $c_A$& $A/\ell^2=1\mbox{ (box)},\; \pi/4$ (sphere) & Eq. (\ref{eq:A})\\
 $c_V$& $V/\ell^3=1\mbox{ (box)},\; \pi/6$ (sphere) & Eq. (\ref{eq:V})\\
 $c_\calm$ & $\sid^2/\snt^2=1+3/\calm^2$; $\rightarrow 1$ for highly supersonic flow & Eq. (\ref{eq:sid})\\
 $c_\Phi$& Numerical coefficient in $M_\Phi$; adopt $c_\Phi=1/(2\pi)$ & Eq. (\ref{eq:mphi})\\
 $M_\Phi$&Magnetic critical mass & Eq. (\ref{eq:mphi})\\
 $M_B$&Alternate form of magnetic critical mass, $M_\Phi/\mu_\Phi^2$&Eq. (\ref{eq:mb})\\
 $\calm$&Sonic Mach number, $\vrms/\cs$ & Sec. \ref{sec:scale}\\
 $\ma$&\alfven Mach number, $\vrms/\va$ & Sec. \ref{sec:scale}\\
 $\nbh$&Mean density of H nuclei & Eq. (\ref{eq:btm})\\
 $\nlos$&Mean density of H nuclei in central Gaussian beam &Eq. (\ref{eq:nlos})\\
 $\Nhbeam$&Column density of H nuclei in central Gaussian beam & Eq (\ref{eq:nlos})\\
 $\va$& Mass-weighted \alfven velocity, $\brmsv/(4\pi\bar\rho)^{1/2}$& Sec. \ref{sec:scale}\\
 $\vrms$& Mass-weighted 3D nonthermal velocity dispersion, $\surd 3\snt$& Sec. \ref{sec:scale}\\
 $\avg{x}$; $\avg{x}_t$& Average value of $x$ in 100-clump sample; also averaged over time & App. \ref{sec:100clump}\\
 $\alpha$&Exponent for density scaling of magnetic field, $B\propto\nbh^\alpha$ & Eq. (\ref{eq:btm})\\
 $\beta$& Plasma $\beta,~ = 8\pi\bar\rho\cs^2/\brmsv^2$ & Eq. (\ref{eq:beta})\\
 $\mu_\Phi$&(Normalized) mass-to-flux ratio, $M/M_\Phi=2\pi G^{1/2}\Sigma/B$ & Eq. (\ref{eq:muphi})\\
 $\muplos$&Central-beam line-of-sight mass-to-flux ratio, $2\pi G^{1/2}\Sigma_\beam/\blos$ &  Eq. (\ref{eq:muplos})\\
 $\avg{\muphi}_h$&$\avg{1/\muphi}^{-1}=$ harmonic mean mass/flux ratio = average flux/mass ratio&Eq. (\ref{eq:har}) \\
 $\sid$&Mass-weighted total 1D velocity dispersion, $(\cs^2+\snt^2)^{1/2}$ &Eq. (\ref{eq:sid})\\
 $\snt$ & Mass-weighted 1D nonthermal velocity dispersion & Eq. (\ref{eq:sid})\\
 $\Sigma_\beam$&Surface density in central Gaussian beam &Sec. \ref{sec:comp}\\
\hline
\hline
\end{tabular}

\begin{flushleft}
\fnt{1} {$^a$ Subscripts: beam = central Gaussian beam, $h$ = harmonic mean, H = hydrogen nuclei, los = line of sight, $M$ = mass, med = median, $t$ = time-averaged, $V$ = volume-averaged, $\mu$ = microgauss}
\end{flushleft}
\end{table*}

In this paper, we describe the results of two large-scale adaptive mesh refinement (AMR) simulations with the ORION2 code \citep{li12b} of supersonically turbulent, magnetized, self-gravitating molecular clouds (Li et al. 2015, in preparation). There are two objectives of our study: First, we wish to compare the results of our simulations with the inferred properties of magnetic fields in molecular clouds by \citet{cru10}. Second, we wish to examine the magnetic properties of molecular-cloud clumps under strong and weak mean-field initial conditions in order to determine which agrees better with observation.  
In Section \ref{sec:obs}, we review the results of the Bayesian statistical analysis of the Zeeman observations of
magnetic field strengths by \citet{cru10}.
Section \ref{sec:analysis} presents new results from an analysis of this data.
We describe the simulations, the scaling to physical variables, and the model parameters in Section \ref{sec:sim}, and the
selection of the samples of simulated clumps for comparison with observations in Section \ref{sec:100clump}.
We compare the line-of-sight properties of clumps from our simulations with those from
the 68 OH and CN sources cataloged in \citet{cru10} in Section \ref{sec:comp}.
In Section \ref{sec:obsmfr}, we study the relation between the observed line-of-sight value of
the mass-to-flux ratio with the intrinsic 3D value and describe its evolution.
The long-standing question of whether the mean magnetic fields in molecular clouds is strong or weak is
discussed in Section \ref{sec:is}.
Theoretical results on the comparison of magnetic and
gravitational forces are summarized in Appendix \ref{app:theory}.
Finally, Section \ref{sec:conclusion} summarizes the results of our analysis of the line-of-sight data and our
detailed comparison between simulations and observations of magnetic fields in clumps in molecular clouds.

\section{The Observational Results of Crutcher et al. (2010)}
\label{sec:obs}

The most comprehensive analysis of magnetic fields in the ISM is that of \citet{cru10}, who
analyzed Zeeman data from HI observations by \citet{hei04}, OH observations by \citet{cru99} and
\citet{tro08}, and CN observations by \citet{fal08}. 
The OH data comprises 54 sources, 20 of which are detected at more than 2 sigma; the CN data is from
14 sources, 8 of which are detected at more than 2 sigma.
\citet{cru10} used a Bayesian
analysis to self-consistently include upper limits on the field strengths with a statistical significance less than 2 sigma.
 
The Zeeman signal is proportional both to the integral of the field along the line of sight and to the
column density of the molecule being observed. Let the line of sight be along the $z$ direction. 
We assume that the molecule has a fixed abundance, so that the column density of the molecule
is proportional to the mass column density. Then the magnitude of the Zeeman field measured by a given telescope beam is
\beq
\blos=\frac{1}{\bar\rho V} \left|\int_\beam \rho B_z dV\right| =\frac{1}{M_\beam}\left|\int_\beam B_z \,dM\right|,
\label{eq:blos}
\eeq
where $M_\beam$ is the mass inside the beam
and $B_z$ is the component of the magnetic field along the line of sight. (Definitions are collected together in Table \ref{tab:1}.)
Note that $B_z$ is a component of a vector and can be positive or negative, whereas values of the field that
are not explicitly components of a vector, such as $\blos$, are field magnitudes.
If the molecule being observed is excited only
above some critical density, then $M_\beam$ is the mass in the beam above that density.
It is important to note that if the field is tangled, then $\blos$ can be less than the average magnitude
of $B_z$. Of course, what is of greater interest is the value of the field in the absence of
projection effects, 
\beq
\vecB_{\tot}=\frac{1}{M_\beam}\int_\beam \vecB \,dM,
\label{eq:btot}
\eeq
and this is the focus of Crutcher et al.'s analysis. 

\citet{cru10} carried out a Bayesian analysis to infer the distribution of $\btot$ as a function of the 
mean density of hydrogen nuclei,
$\nbh$, which extends over the range 10~cm\eee$\la\nbh\la 10^7$~cm\eee\ in their sample. 
They considered several different distributions and found that the best-fitting to their data is a uniform
distribution of $\btot$ from $f\btm$ to $\btm$ with
\beq
\btm=\left\{\begin{array}{l} B_0~~~~~~~~~~~~~~~(\nbh<n_0),\\
B_0\left(\dis\frac{\nbh}{n_0}\right)^\alpha~~~~(\nbh>n_0). \end{array} \right.
\label{eq:btm}
\eeq
They found $B_0=10~\mu$G, $n_0=300$~cm\eee, $\alpha=0.65$, and $f=0.03\simeq 0$.

The density $n_0$ dividing the constant density regime from the power-law one
can be understood as the critical density, $n_{\rm H,\,crit}$, at which the thermal pressure in the clump balances the turbulent pressure of the ambient gas,
\beq
P_{\rm turb}=\bar\rho\snt^2=x_\tot n_{\rm H,\,crit}\kb T,
\label{eq:nhcrit}
\eeq
where $\bar\rho$ is the mean density, $x_\tot\equiv n_{\rm tot}/\nbh$ is the total number density relative
to hydrogen, and $\kb$ is the Boltzmann constant.  
Since the magnetic field is about the same inside and outside the clump, it has no effect on the pressure balance. 
The observed clumps are in a variety of different environments, and the minimum ambient pressure is that of
the ISM.  From \citet{bou90}, the 
turbulent pressure in interstellar gas is $P_{\rm turb}/k \approx 1 \times 10^4$ K cm$^{-3}$.  Using $T \approx 50$ K and $x_\tot = 1.1\simeq 1$ for atomic gas including helium, we find $n_{\rm H,\,crit}\approx 200$~cm\eee, very close to the observed power-law relation starting point of 300 cm$^{-3}$.

All the data below $n_0$ is from HI observations; most of the data above that is from OH and CN observations.
Thus, the molecular data is inferred to have a power-law dependence on the density,
\beq
\btm=98\nbhf^{0.65}~~\mmug~
~~~~\mbox{(molecular gas)},
\label{eq:btm1}
\eeq
where $\nbhf=\nbh/(10^4$~cm\eee).
It should be noted that the power-law dependence of $\btot$ on $\nbh$ is not apparent in the OH data alone
($10^3$~cm\eee$\;\la\nbh\la 2\times 10^4$~cm\eee) or the CN data alone 
($2\times 10^5$~cm\eee$\;\la\nbh\la 4\times 10^6$~cm\eee).
A least squares fit of the OH data gives a slope of $\blos$ versus $\nbh$ of $0.40\pm0.45$ and that for CN data is $0.04\pm0.67$;
that is, both the low density and high density molecular lines are consistent with either no density dependence
or with the density dependence inferred from the combined data set.
As they point out, $\alpha\simeq \frac 23$, the value expected for the contraction of a spherical cloud with a weak, uniform
magnetic field \citep{mes66}. An important implication of this model is that
field strengths extend to very low values; for example, about 10 per cent of
the clumps are inferred to have fields that are less than about 10 per cent of the maximum value expected at that density.
The existence of very weak fields has been advocated as enabling the formation of discs around protostars \citep{kru13}.
It should be noted that $\btm$ is not an absolute upper bound on the field strength;
both W3(OH) observed in CN \citep{fal08} and Mon 16 \citep{tro08} have fields exceeding $\btm$, although
the latter result is uncertain since the field measurement is less than 2 sigma.

\section{New Results from Old Data: Analysis of the Line-of-sight Field}
\label{sec:analysis}

\begin{figure}
\includegraphics[scale=0.37,angle=-90]{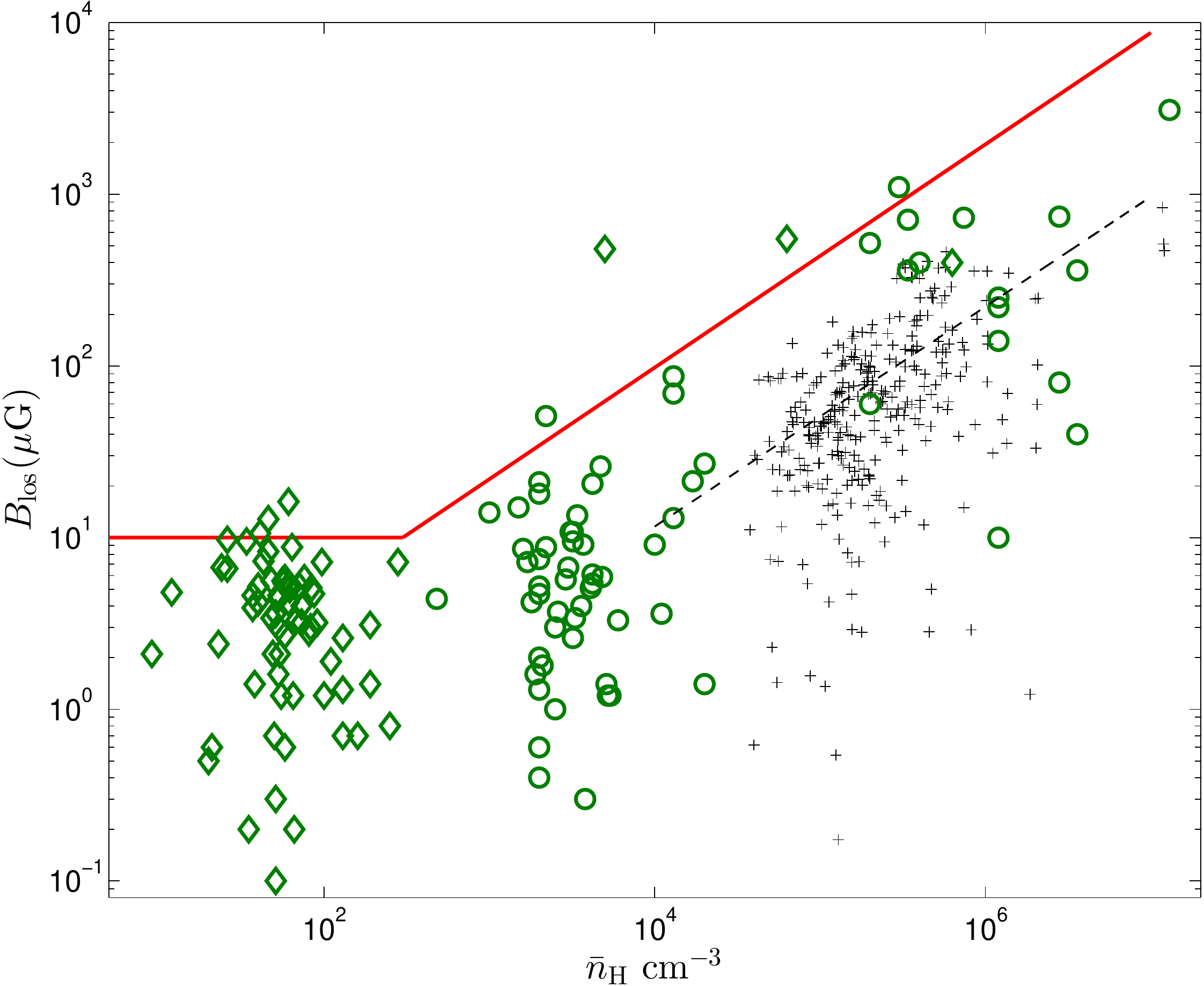}
\caption{Dependence of the observed field strength, $\blos$, on density, $\nbh$. The convolved, mass-weighted, line-of-sight magnetic field strength, $\blosr$, along the three cardinal axes vs volume density in the central-beam regions of clumps from the strong-field simulation (black pluses) are plotted with the observed HI (green diamonds) and OH+CN sources (green circles) from \citet{cru10}.
The dashed line is the best power-law fit to
the simulation data.  The large dispersion is due primarily to projection effects. 
\label{fig3}}
\end{figure}

The results of \citet{cru10} show that the distribution of $\btot/\nbh^\alpha$, with $\alpha\simeq 0.65$,
is independent of density for molecular gas. In other words, while they presented their results as the
scaling of the upper envelope of the total field, their best-fit distribution implies that the scaling applies to
all the fields. We can test this directly:
Since the line-of-sight field is a projection of the total field onto the line of sight, it follows mathematically that the distribution of
$\blos/\nbh^\alpha$, with $\alpha\simeq 0.65$, should also be independent of density in molecular gas.
Indeed, a fit to the data on the line-of-sight fields in molecular gas from \citet{cru10}, shown by the open green
circles in Fig. \ref{fig3}, gives 
\beq
\alpha(\blos)=0.64\pm0.13, 
\eeq
consistent with this expectation. Since this is indistinguishable from the result of \citet{cru10}, we
shall adopt $\alpha=0.65$ as the observed value.
We conclude that on average, fields in molecular gas scale approximately as
$\nbh^{0.65}$.
We emphasize that the uncertainty of all the curve fitting results in this paper is at the 95 per cent confidence level, i.e., about $2\sigma$. We choose to use a 95 per cent confidence level since then any disagreement is likely to
be a true difference instead of a statistical fluctuation.
We estimate the 95 per cent confidence limits on the result of \citet{cru10} for $\alpha$ from their Figure 4: The probability
that $\alpha<0.59$ or $\alpha>0.75$ are each less than 2.5 per cent, so their result is $\alpha=0.65_{-0.06}^{+0.10}$.
\footnote{Values of $\alpha>0.76$ were not tested (Crutcher 2015, private communication), so our estimate of the upper confidence level is based on an extrapolation of the results for
$\alpha\leq 0.76$}

In Appendix \ref{app:sample} we discuss the effects of noise on the inferred value of $\alpha(\blos)$. There we
show that since the noise in the low-density data (OH) is greater than that in the high-density data (CN),
the value of $\alpha(\blos)$ inferred from the data is less than the true value by about 0.02. 
If the field is distributed isotopically with respect to the line of sight, then $\alpha(\blos)=\alpha$. We conclude
that the true value of $\alpha$ corresponding to the observed value $\alpha(\blos)=0.64$ is $\alpha=0.66$. 
This shift is small compared to the uncertainty, so it is not significant, but if the noise in future larger samples
is asymmetric as in the present case this effect should be taken into account.

Since the field scales as a power of the density,  we define
\beq
b\equiv \frac{B_\mu}{\nbhf^{\alpha}},
\label{eq:b}
\eeq
where $B_\mu=B/(1~\mmug)$. The quantity $b$ for the line-of-sight field is denoted $\bblos$ and that for
the total field $\bbtot$. Henceforth, we shall set $\alpha=0.65$ in this expression. For a sample of sources, it
follows that
\beq
\avg{B_\mu}=\avg{b \nbhf^{0.65}}\simeq \avg{b}\avg{\nbhf^{0.65}},
\eeq
where the last step follows if $B$ indeed scales as $\nbh^{0.65}$, so that $b$ is not correlated with $\nbh$.
We find that the average value of normalized line-of-sight field is $\blosnfa=22.4$, and that
$\avg{\blosmu}/\avg{\nbhf^{0.65}}=20.2$.
We conclude that the average line-of-sight field in the observed sample is related to the density by
\beq
\avg{\blos}(\nbh)\approx 21 \nbhf^{0.65}~~~\mmug.
\eeq
As noted after
Equation (\ref{eq:psia}) in Appendix \ref{app:proj}, 
the average total field is just twice the average line-of-sight field.
We therefore conclude that the average magnetic field in the molecular clumps in the study of
\citet{cru10} is
\beq
\avg{\btot(\nbh)}\approx 42 \nbhf^{0.65}~~~\mmug.
\eeq
Insofar as the clumps in this study are broadly representative, this is the average field molecular clumps in the interstellar medium.

\begin{figure}
\includegraphics[scale=0.37,angle=-90]{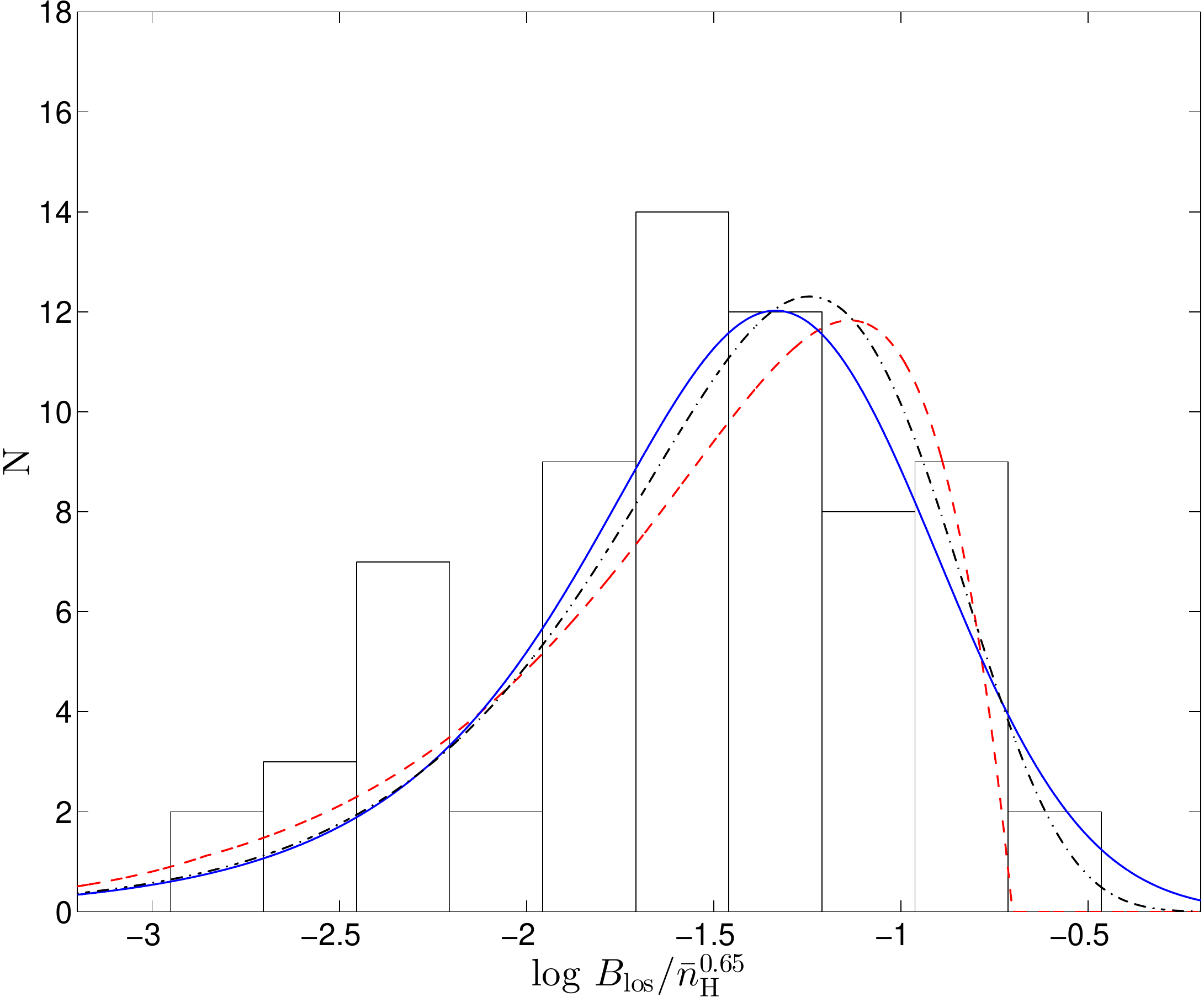}
\caption{
Model fits to the distribution of $\blos/\nbh^{0.65}$ for the observed OH+CN sources: lognormal (blue solid line), exponential (black dot-dash line), and flat distribution (red dashed line). The differences among the three models are small (see Section \ref{sec:analysis} for discussion).
More observations are required to determine the actual distribution.
\label{fig2}}
\end{figure}

The Bayesian analysis of \citet{cru10} enables one to infer the range of possible 3D magnetic fields that are consistent with the observed distribution of line-of-sight fields. However, it is also possible to infer the properties of the fields directly from the line-of-sight data. As shown in Appendix B,
the mean value of the total field can be inferred from the mean value of the line-of-sight
field, $\avg{\btot}=2\avg{\blos}$, and the density dependence of the field can be directly measured as
well. Furthermore, instead of comparing the inferred 3D fields with a model distribution for those fields, we shall
convert the 3D distribution to the line-of-sight distribution and compare that directly with the data,
thereby eliminating any uncertainties introduced
by the statistical inference of the 3D properties of the field.
Since the physical conditions we have chosen for our simulation are appropriate for molecular clouds,
we shall compare our results only with the 68 molecular clumps in the \citet{cru10} sample.
Our sample therefore excludes three HI sources with densities in the power-law region ($\nbh>300$~cm\eee)
found by \citet{cru10}; in fact, these sources have densities $\geq 5000$~cm\eee, overlapping those of molecular
clouds. Only 28 of the 68 molecular clumps in this sample 
have magnetic fields detected at a significance greater than 2 sigma.
For the measurements that are less than 2 sigma, we use the measured value, which is appropriate
provided that the errors are statistical and not systematic. Since this is also a necessary condition for the
validity of the analysis of \citet{cru10}, nothing is lost by this procedure. 
This method may be thought
of as analogous to stacking optical or X-ray data to bring out results that are buried in the noise in individual frames.
However, there is a difference in the present case: whereas photon noise can result in the observed value being less than
the noise, it is always positive; for Zeeman observations, on the other hand, measurement errors could change the sign of the observed magnetic field if the noise is larger than the true field strength.  In Appendix C we discuss a numerical experiment that demonstrates that this effect is small.

We have considered three theoretical models for the distribution of field strengths: (1) a uniform distribution of total
normalized field strengths from $fB_{\max}/\nbhf^{0.65}$ to $B_{\max}/\nbhf^{0.65}$ \citep{cru10}; 
(2) a log-normal distribution of total normalized field strengths;
and (3) an exponential distribution of line-of-sight normalized field strengths.
Explicit expressions for the total fields and line-of-sight fields for each are given in Appendix \ref{app:proj}.
The distribution of the observed values of $\blosmu/\nbhf^{0.65}$ are compared with these models in Fig. \ref{fig2}.
The striking feature of this figure is that the three very different distributions of total field strength, $\btot$, look 
remarkably alike when viewed along the line of sight, providing graphic evidence of the difficulty in accurately inferring
the distribution of $\btot$ from observations of the line-of-sight field.

We conclude that the log normal and exponential models provide slightly better fits to the data than the uniform
distribution. This does not contradict the conclusion of \citet{cru10}:
For the molecular data, they considered three cases: (1) OH data alone with $B$ independent of
density; they found that a uniform distribution was better than one obtained from a numerical simulation,
which was approximately log normal; (2) OH + CN with an assumed density dependence $B\propto n^{1/2}$;
here they found that a uniform distribution was better than a delta function, but they did not consider
a log normal; and finally, (3) all the data, including HI, with a density dependence that was determined
from the data, $B\propto n^{0.65}$; they did not consider a log normal fit to the data in their paper. Thus in
no case did they compare a log normal vs. a uniform distribution for the molecular data with $B\propto n^\alpha$,
as we have done here.
More accurate Zeeman measurements and more sources are required for a better comparison.

\begin{figure}
\includegraphics[scale=0.37,angle=-90]{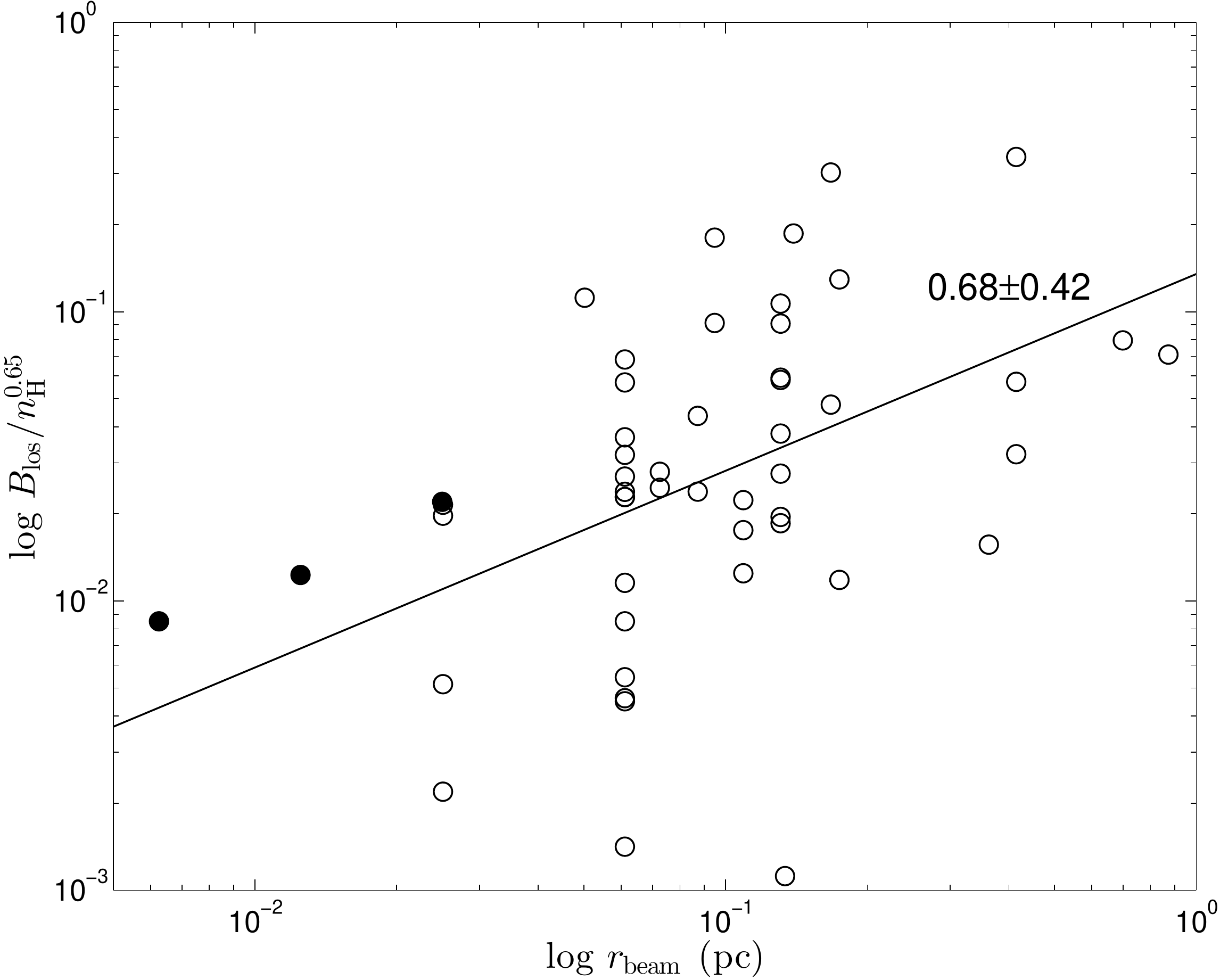}
\caption{The best-fitting power-law relation between $\blos/\nbh^\alpha$ and the telescope-beam size $\rbeam$ of 
34 OH and and 14 CN sources (open circles) from \citet{tro08} and \citet{fal08}.  The three solid circular symbols are from the 100-clump strong-field sample, using 3 different beam sizes
(0.025 pc, 0.0125 pc, and 0.00625 pc) at $t=0.4\tff$.
See the discussion in Sections \ref{sec:obs} and \ref{sec:den}.
\label{fig1}}
\end{figure}

Before concluding our discussion of the observations, we note that
we have found evidence that the measured values of $\blos/\nbh^{0.65}$ for the OH sources studied by \cite{tro08} and the CN sources studied by \citet{fal08} depend on
size of the beam, $\rbeam$, used to study them. (We do not know the beam sizes for the sources listed by \citet{cru99}, which came from a variety of references.)
In Fig. \ref{fig1}, we plot the measured values of $\blos/\nbh^{0.65}$ for these 48 sources as a function of $r_\beam$.  The correlation coefficient is 0.271 and the probability that there is no correlation is $p=0.062$.  The correlation is weak, as the scattered data points show in Fig. \ref{fig1}, but $p$ is small enough that this correlation is marginally significant. 
Further observations are needed to determine if this correlation is real.

\section{Simulations}
\label{sec:sim}

We have carried out two large-scale AMR simulations to study the formation 
and structure of Infrared Dark Clouds (IRDCs) (Li et al. 2015, in preparation).  These simulations 
use the ORION2 adaptive mesh refinement (AMR)
code \citep{li12b} with ideal MHD and self-gravity to follow the evolution
of driven turbulence on a base grid of $512^3$ with two levels of refinement,
with each level having a factor of 2 increase in resolution.
We adopt periodic boundary conditions.
We drive the system for two crossing times with no self-gravity in order to produce 
self-consistent turbulent density and velocity fields. The first crossing time is
at the base level and the second has two levels of refinement.
The refinement thresholds for pressure jumps (thermal pressure + magnetic pressure) and shear flows are chosen so that about $12-15$ per cent of the volume is refined at all times during the second crossing time.
This ensures that before gravity is turned on, the high-mass end of the probability density function (PDF) of volume density is equivalent to that which would be obtained in a $2048^3$ single-grid simulation \citep{li12b}.
As we shall see in Section \ref{sec:param} below, the simulation box has a size of 4.55 pc for fiducial values of the parameters, so the cell size at the finest resolution is $2.2\times 10^{-3}$~pc.
The driving is at the largest scale, with wave numbers $k = 1-2$, and it uses the \citet{mac99} recipe.   After two crossing times, we turn on gravity and set $t = 0$ at this moment.  The driving continues with a constant energy injection rate that maintains the system globally at a constant sonic Mach number, $\calm$.
Dense regions inside the system undergo gravitational collapse, so it is necessary to add the Jeans condition for refinement \citep{tru97}; we adopt a Jeans number of 1/8.
The finest resolution remains unchanged from the first part of the simulation, and no sink particles are introduced.

\subsection{Scaling Relations for a Turbulent Box}
\label{sec:scale}

Simulations of isothermal MHD turbulence are scale-free and are defined by two dimensionless parameters, the three-dimensional (3D) thermal Mach number, $\calm=\vrms/\cs$,
where $\vrms$ is the mass-weighted rms velocity,
and the \alfven Mach number, $\ma=\vrms/\va$, where $\va=\brmsv/(4\pi\bar\rho)^{1/2}$ is the mass-weighted
 \alfven velocity and $\brmsv=(\avg{B^2}_V)^{1/2}$ is the square root of the volume-weighted mean-square magnetic field. Alternatively, the strength of the magnetic field can be characterized by the plasma-$\beta$ parameter, 
\beq
\beta=\frac{8\pi \bar\rho \cs^2}{\brmsv^2}=2\left(\frac{\ma^2}{\calm^2}\right).
\label{eq:beta}
\eeq

\begin{table*}
\caption{System Properties for the Strong- and Weak-Field Simulations}
\label{tab:3}
\begin{tabular}{lccccccc}
\hline
\hline
 & \multicolumn{3}{c}{Strong} & & \multicolumn{3}{c}{Weak}\\
\cline{2-4} \cline{6-8} \\
 & Initial\fnm[1] & $t=0$\fnm[2]  & Final\fnm[3] & & Initial\fnm[1] & $t=0$\fnm[2] & Final\fnm[3]\\
\hline
 \alfven Mach number, $\ma(\brms)$& 1 & 0.85 &0.83 & & 10 & 1.86 & 1.61\\
 Plasma $\beta=8\pi \rho_0 \cs^2/\brms^2$& 0.02 & 0.015 & 0.014 & & 2.0 & 0.069 & 0.055\\
 Mass-to-flux ratio, $\mu_{\Phi,\,z}$& 1.62 &  1.62 & 1.62 & & 16.2 & 16.2 & 16.2\\
 Avg magnetic field ($\mu$G), $\avg{|B_\mu|}_V$ &31.6 & 36.2 & 36.8 && 3.16 & 14.9 & 16.2 \\
 RMS magnetic field ($\mu$G), $\brmsmu$  & 31.6 & 37.2 & 38.3 & & 3.16 & 17.1 & 19.2\\
 $\brmsmu/\nbhf^{2/3}$& 151 & 177.4 & 182.7 & & 15.1 & 81.6 & 93.0\\
\vspace{-0.4 cm}\\
\hline
\hline
\end{tabular}

\begin{flushleft}
\fnt{1} {$^{\rm a}$ The initial field is uniform so that $B_0=\avg{|B|}_V=\brms$ then.}\\
\fnt{2} {$^{\rm b}$ Gravity is turned on at $t=0$, after the system is driven for two crossing times.}\\
\fnt{3} {$^{\rm c}$ The final values are time averages over $(0.4-0.64)\tff$ for the strong-field case and $(0.5-0.94)\tff$ for the weak-field case.}
\end{flushleft}
\end{table*}

In order to set the scale, we need three dimensional relations \citep{mck10}.
Interstellar molecular gas is observed
to have a temperature $T=10T_1$~K with $T_1\simeq 1$,
corresponding to an isothermal sound speed $\cs=0.188T_1^{1/2}$~km~s\e; this gives one relation. 
Most interstellar
gas obeys the turbulent linewidth-size relation \citep{mck07},
\beq
\snt=0.72\spcs R_{\rm pc}^{1/2}=0.51\spcs\ell_{0,\,\pc}^{1/2}~~~\mbox{km s\e},
\label{eq:lws}
\eeq
where $\snt$ is the mass-weighted 1D nonthermal velocity dispersion in a sphere of radius $R$,
which we take to be the same as in a box of size $\ell_0=2R$, and where
$\spcs \sim 1$ allows for deviations from the typical linewidth-size relation.  
When self-gravity is included, a third dimensional parameter enters, the gravitational constant, $G$,
together with a corresponding dimensionless parameter that measures the effect of self-gravity, the
virial parameter, 
$\avir\simeq 2\times$ kinetic energy/(gravitational energy) (see Equation (\ref{eq:avir})). 
We assume that the turbulent box is large enough that the gas is highly supersonic so that
the mass-weighted 1D total velocity dispersion, $\sid$, obeys $\sid^2=\snt^2+\cs^2\simeq \snt^2$. 
For a highly supersonic,
fully molecular gas, the physical parameters of the turbulent system---the size of the turbulent box, $\ell_0$,  
the flow time (or crossing time), $t_f$, the mass of the box, $M_0$, and the column density,
$\NH=\nbh\ell_0$---can
then be expressed as \citep[see the Appendix in][]{mck10}:
\begin{eqnarray}
\ell_0&=& \frac 23\;\frac{\calm^2 \cs^2}{(\spc^2/\mbox{ 1pc})}=0.0455\left(\frac{\calm^2 T_1}{\spcs^2}\right)~~\mbox{pc},
\label{eq:lscale}\\
t_f&=&\frac{\ell_0}{\vrms}=2.36\times 10^5\left(\frac{\calm T_1^{1/2}}{\spcs^2}\right)~~~\mbox{yr},
\label{eq:tscale}\\
\nbh&=&9.6\times 10^4\left(\frac{\spcs^4}{\avir\calm^2T_1}\right)~~~\mbox{cm\eee},
\label{eq:nscale}\\
M_0&=& 0.311\left(\frac{\calm^4T_1^2}{\avir\spcs^2}\right)~~M_\odot,
\label{eq:mscale}\\
\NH&=&1.34\times 10^{22}\left(\frac{\spcs^2}{\avir}\right)~~~\mbox{cm\ee}.
\label{eq:Nscale}
\end{eqnarray} 
It should be noted that the only assumptions entering the derivation of these relations
is that the Mach number is large and that the velocity dispersion scales as $\ell_0^{1/2}$.
Observe that the size scale for a simulation of self-gravitating ($\avir\sim 1$) interstellar molecular gas
($T_1\sim 1$) that obeys the standard linewidth-size relation ($\spcs\sim 1$) is set solely by the
Mach number, $\calm$. Equation (\ref{eq:Nscale}) shows that the column density of such gas
is independent of the Mach number and is therefore approximately constant, one of the three
relations discovered by \citet{lar81}.
With different coefficients that allow for the difference in shape between an interstellar cloud and 
a turbulent box, these relations apply to actual interstellar clouds as well.

The magnetic field is given by
\beq
B=4.56\;\left(\frac{\nbht T_1}{\beta}\right)^{1/2}\mmug
=31.6\left(\frac{\spcs^2}{\avir^{1/2}\ma}\right)\mmug.
\label{eq:bscale}
\eeq
where $\nbht=\nbh/(1000$~cm\eee). Magnetic fields in gas that obeys the standard linewidth-size relation can be signficantly less than 30~\mug\ only if self-gravity is weak ($\avir\gg 1$) and/or if the
motions in the gas are highly super-\alfvenic ($\ma\gg 1$).

The relative importance of magnetic fields and gravity is discussed in Appendix \ref{app:theory}.
Clouds with masses exceeding the magnetic critical mass, $M_\Phi$, can undergo gravitational
collapse; those with smaller masses cannot. We adopt $M_\Phi=$(magnetic flux)/$2\pi G^{1/2}$, which
is exact for a sheet threaded by a perpendicular field \citep{nak78}. The mass-to-flux ratio 
normalized to the critical value is then
\beq
\muphi\equiv \frac{M}{M_\Phi}=2\pi G^{1/2}\frac{\Sigma}{B}=\frac{3.80}{B_\mu}\left(\frac{\NH}{10^{21}\mbox{ cm\ee}}\right).
\eeq

\subsection{Model Parameters}
\label{sec:param}

The models we consider here have
$\calm=10$, $T=10$~K, and $\avir = \spcs=1$, so that
the size of the simulated turbulent region is $\ell_0 = 4.55$ pc, the number density $\nbh = 960$ cm$^{-3}$, the total mass $M_0 = 
3110 \,M_{\sun}$, and the initial column density of the system $\NH = 
1.34\times 10^{22}$ cm$^{-2}$.  
The free-fall time of the system based on the mean density,
\beq
\tff=\left(\frac{3\pi}{32G\bar\rho}\right)^{1/2}=1.37\times 10^6\nbht^{-1/2}~~~\mbox{yr},
\label{eq:tff}
\eeq
is $1.4\times10^6$ yr, which is 
$0.59 t_{\rm f}$.

We present two simulations in this paper, one with a strong initial mean field
($\mao=1$, $\mu_{\Phi,\,0}=1.62$), which is only slightly supercritical, and one with a weak initial mean field ($\mao=10$, $\mu_{\Phi,\,0}=16.2$), which is initially very supercritical.  The parameters of the runs are given in
Table \ref{tab:3}.
The initial magnetic field is along the $z$-axis.
We ran the strong-field simulation to $0.64 \tff$ $\approx 900,000$~yr , which provided enough data to time-average the parameters that do not vary systematically with time for comparison with observation. We ran the weak-field model somewhat longer, to $t\simeq 0.94 \tff$,
since the maximum column density in our simulation is comparable to that in the observed sample at that time.
A detailed discussion of the time evolution of the whole system will be given in a subsequent paper (Li et al. 2015, in preparation); here we focus on the properties of the magnetic fields in dense clumps during the self-gravitating phase of the simulations. 

\section{The 100-Clump Sample for Comparison with Observation}
\label{sec:100clump}

Since the magnetic fields in the Crutcher molecular clump sample are measured in condensations in molecular clouds, we select a sample of dense clumps from our simulation. The most massive dense clumps are the most
readily observable, so we select them; in order to have a good statistical sample, we choose the 100 most massive dense clumps and term
this the `100-clump sample.' It is important to note that during the course of the simulation, clumps accrete matter and can merge, so the clumps in this sample at one time are not the same as those at a different time. We identify
the dense clumps from the simulations using a modified CLUMPFIND algorithm \citep{wil94}.  The clumps are defined as connected regions with a density greater than the mass-weighted median density of the turbulent 
region, $(\nbh)_{M,\,\rm med}$; half the mass in the region has a higher density than this and half a lower one.
No cell with density below this threshold will be included as part of the clumps that are identified.
This density rises from about $3\nbh$ at $t = 0$ to about $6\nbh$ at the end of the simulations as a result of gravitational contraction, although
there is no requirement that the clumps be gravitationally bound.  We note that the clumps for which Zeeman observations are available are not necessarily bound either \citep[e.g.,][]{fal08}.

To assemble a sample of simulated clumps for comparison with observations, we impose requirements on the resolution, density, and mass as described in Appendix \ref{app:100clumps}.  {\it Resolution}:
The minimum radius of the clumps must exceed the minimum radius of the OH and CN sources,
which is 0.03 pc; all clumps closer than 0.06 pc are merged together. {\it Density}: As noted in Section \ref{sec:obs} above, the power-law behavior of the field begins at a density $n_0\simeq 300$~cm\eee\ that can be understood as the critical density at which the thermal pressure of the gas balances the typical turbulent pressure of the interstellar medium. Our simulation is meant to represent the interior of a GMC, so the pressure is significantly higher; for a molecular gas at $T=10$~K, the corresponding critical density is $\sim 3.2 \times 10^4$ cm$^{-3}$. We require the clumps to have densities slightly higher than this, $\nbh>4\times 10^4$~cm\eee. With this choice of density threshold, the simulated clumps have densities closer to those of the CN sources than of the OH sources. {\it Mass}: As noted above, we select the 100 most massive clumps that satisfy the above two requirements.

Although we have selected our clumps from simulations based on the typical conditions observed in molecular clouds (in particular, the simulations have $\mao=\avir=1$), there are important differences between our simulations and reality.
The molecular clumps in the \citet{cru10} sample are from a variety of clouds with varying line width-size relations, virial parameters, and mass-to-flux ratios.  The initial conditions of our two simulation models have only one line width-size relation and one virial parameter. We know exactly the duration of the gravitational collapse phase of the simulations, but the cloud clumps observed can be of any age. Our simulations are isothermal, but there are temperature variations in real clouds.  Observed cloud clumps are obtained from OH and CN emission-line observations, but our clumps are identified using CLUMPFIND and the rules listed above.
Some observed clouds have masses of the same order as the total mass of our whole simulated region, whereas most of the clumps in the simulations are less than $10 M_{\sun}$. On the other hand, the densities in our clump samples are comparable to to those in some OH sources and in all the observed CN sources.  Despite these differences, we shall show in the following sections that there is excellent agreement between the properties of
clumps in the simulations and those of the observed clumps.

To compare with observations or to compare the two models, we use the mean, denoted  $\langle x \rangle$ for a quantity $x$, and median, $(x)_{\rm med}$, from the 100-clump samples. All observed magnetic fields are mass weighted. When ambiguity is possible,
mass-weighted quantities are denoted by $\avg{x}_M$ and volume-weighted ones by $\avg{x}_V$.
When comparing time-dependent quantities, we use data
from $t=0.57\tff$ for the strong mean-field simulation and $0.94\tff$ for the weak mean-field simulation since the maximum
column densities in the 100-clump samples are comparable to the maximum value in the \citet{cru10} sample at those times.
When comparing time-independent quantities, we evaluate the time average, denoted by the subscript $t$. For example, the time-averaged mean is $\langle x \rangle_t$. For the strong mean-field model, we determine the time averaged value from data at $t= 0.4$, 0.5, 0.57, and $0.64 \tff$; for the weak mean-field model, which takes somewhat longer to reach an approximate steady state, we use data at $t=0.5$, 0.6, 0.7, 0.8, 0.9, and $0.94 \tff$ for the time average.  

\begin{figure*}
\includegraphics[scale=0.7]{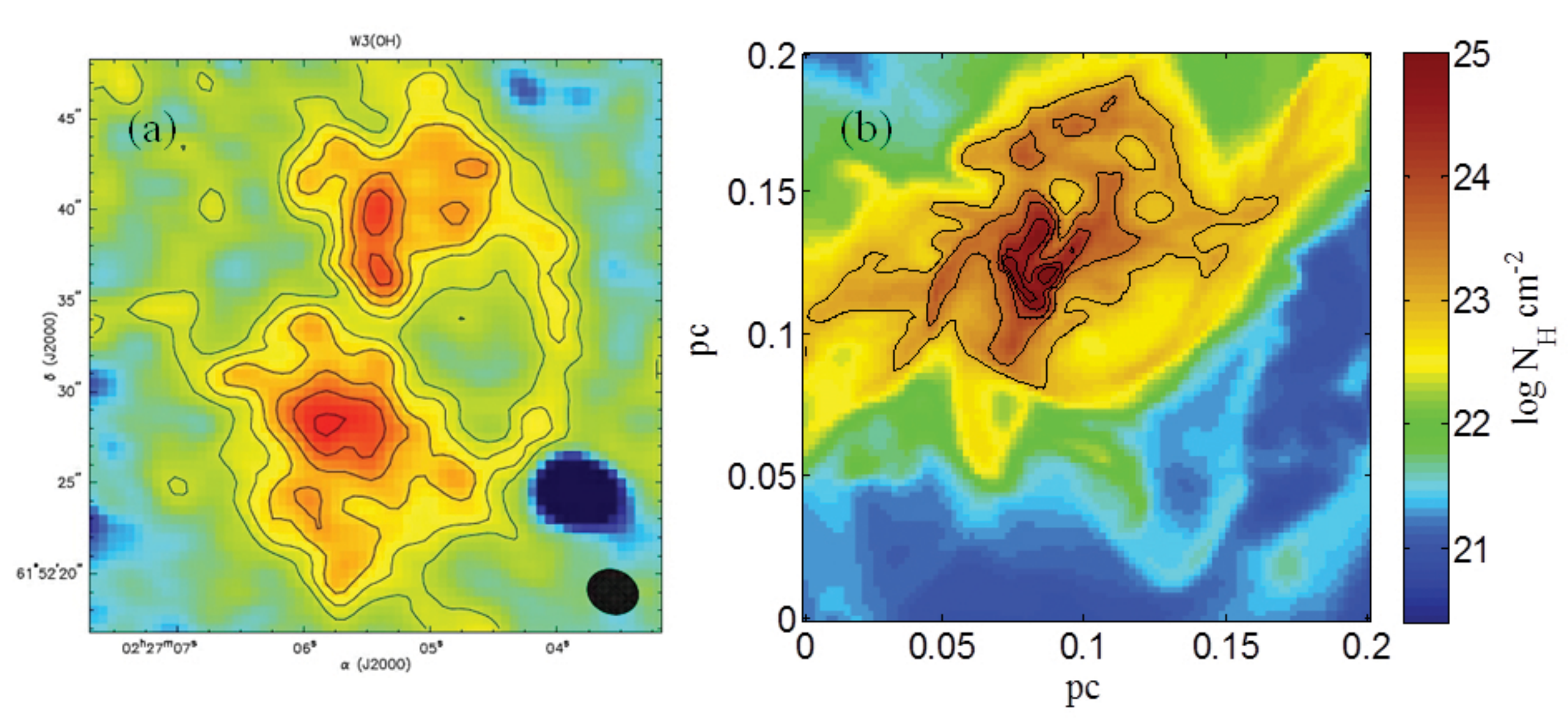}
\caption{(a) The CN image of molecular cloud W3(OH) using CARMA  \citep{hak11}, showing clumpy structure.  (b) The column density map of a massive clump from the strong-field simulation at $0.57 \tff$, showing structure inside the clump.
\label{fig4}}
\end{figure*}

We did not use sink particles in these simulations, so some of the clumps may violate the Jeans condition,
which is that the Jeans length must be resolved by enough cells to prevent artificial fragmentation
\citep{tru97}. In our 100-clump sample, we require the clumps to be at least 0.03 pc in radius, which eliminates the possibility of separating small artificial fragments into separate clumps. To quantify how much clumps that violate the Jeans condition could affect our results, we removed the clumps in which cells with more than 25 per cent of the mass violated the Jeans condition and replaced them with clumps of lower mass so as to maintain a 100-clump sample. We found that the power-law index describing the relation between $B$ and $\nbh$ discussed in the next section changed by only 0.02, well within the uncertainties. It thus appears that our results are not sensitive to artificial fragmentation.

In Appendix \ref{app:convergence} we show that the results of our simulation for the 100-clump strong mean-field sample are converged
by comparing with results of a simulation with 3 levels of refinement. The convergence test was carried out
at $t=0.4\tff$, the latest time that we could afford to run with 3 levels of refinement.
Almost all of the quantities from the 2-level and 3-level 100-Clump samples in Tables A.1 and A.2 are converged within the error of the means.
The one exception is the mass-weighted rms field, $\brmsr$, which has a difference slightly larger than the error of the individual means, although within the rms error.
We repeated the test described above, in which we analyzed the results after eliminating the clumps that violated the Jeans condition at the 25 per cent level, for both the 2-level and 3-level 100-clump samples and found that the results are still converged, showing once again that even though there are cells that violate the Jeans condition, this has no significant effect on our results.

\section{Comparison of the Strong Mean-Field Model with Observations}
\label{sec:comp}

We wish to compare our simulations both with each other and with observation.
In Section \ref{sec:is} we shall see
that although many of the results from the two simulations are similar, the weak-field simulation
does not reproduce the observed correlation of the magnetic field orientation on large and small scales
\citep{li09}.
Since the strong-field model is favored by observation, in this section we shall compare the
results of that simulation with the observations of \citet{cru10}.
Equations (\ref{eq:blos}) and (\ref{eq:btot}) show that we need the mass-weighted fields averaged
over the telescope beam in order to compare with the observations. 
Zeeman observations require long observing times, so that only the single 
region of a given clump with the strongest emission in the Zeeman molecule is observed (Crutcher, private communication).
Fig. \ref{fig4}a shows a
CN map of a clump in the molecular cloud W3(OH) observed by \citet{hak11}.
Fig. \ref{fig4}b is a column-density map of one of our massive clumps. Like the clump in W3(OH), it shows considerable internal structure.  The complexity of the structure in both the observed
clump and the simulated one in Fig. \ref{fig4} shows that, like the observers, 
we are obtaining only the average of one component of the field in the clumps.

As discussed in the selection of the 100-clump sample in Section \ref{sec:100clump}, the smallest known beam size of the CN and OH sources in \citet{cru10} is about 0.025 pc in radius.  We `observe' our clumps also using a Gaussian beam with radius of 0.025 pc. For column density, we create column density maps of a clump along each cardinal axis ($x$, $y$, and $z$-axes) and convolve the maps using a Gaussian beam of radius 0.025 pc. The peak column density of each convolved map is the `observed' column density of the clump along that direction:
$\Nhbeam$ is the number of H nuclei per unit area and $\Sigma_\beam=(2.34\times 10^{-24}\mbox{ g})\Nhbeam$
is the mass per unit area in the beam.  To obtain the observed magnetic field, $\blos$,
we create a density-weighted map of the line-of-sight magnetic field integrated along each cardinal axis first,
convolve the map with the same Gaussian beam, and then take the magnitude.
The `observed' field strength is the value of $\blos$
at the location of the peak `observed' column density in the convolved column density map. We thus obtain 300 measurements of $\Nhbeam$ and the corresponding $\blos$ from our 100-clump sample. The mean volume density of each clump is obtained from the measured column density as 
\beq
\nlos = \Nhbeam / 2 r_{\rm clump}, 
\label{eq:nlos}
\eeq
where $r_{\rm clump}$ is the mean radius obtained from the contour of the half maximum of $\NH$.  

\subsection{Density Scaling of the Magnetic Field}
\label{sec:den}

One of the primary results obtained by
\citet{cru10} is that in molecular gas there is a power-law relation between the value of $\btm$ that they infer (but do not directly measure) and the density, $\btm\propto\nbh^\alpha$, with $\alpha=0.65$.
As we discussed in Section \ref{sec:analysis}, this relation applies to the observed line-of-sight fields
and to the total fields as well.
In many cases the density of the molecular gas was estimated as in Equation (\ref{eq:nlos}), although in some cases the density was inferred from observations of density-sensitive lines.
At low densities, $\nlos \la 300$ cm$^{-3}$, they found that the clumps (which are all HI at these
densities) form a horizontal branch with low field strengths,
$\blos\la 10$~\mug, that do not scale with density. In creating our 100-clump sample, we eliminated the clumps in the horizontal branch since our focus is on clumps in molecular clouds, which are observed to be in the power-law regime. The density range of the central-beam region of the simulated clumps is in the range $(5 \times 10^4-1.1 \times 10^7)$ cm$^{-3}$,
which is somewhat above the density range of the OH observations but covers the range of the CN observations.
As discussed in Section \ref{sec:100clump} above, the transition from the horizontal
$\blos(\nbh)$ relation to the power-law regime
begins at a much higher density in our simulations than in the observations because the transition
occurs when the thermal pressure in the clumps is comparable to the ambient turbulent pressure, and this occurs at a much higher density in the molecular clouds we are simulating than in the diffuse ISM that is included in the observations. 

\begin{figure}
\includegraphics[scale=0.37,angle=-90]{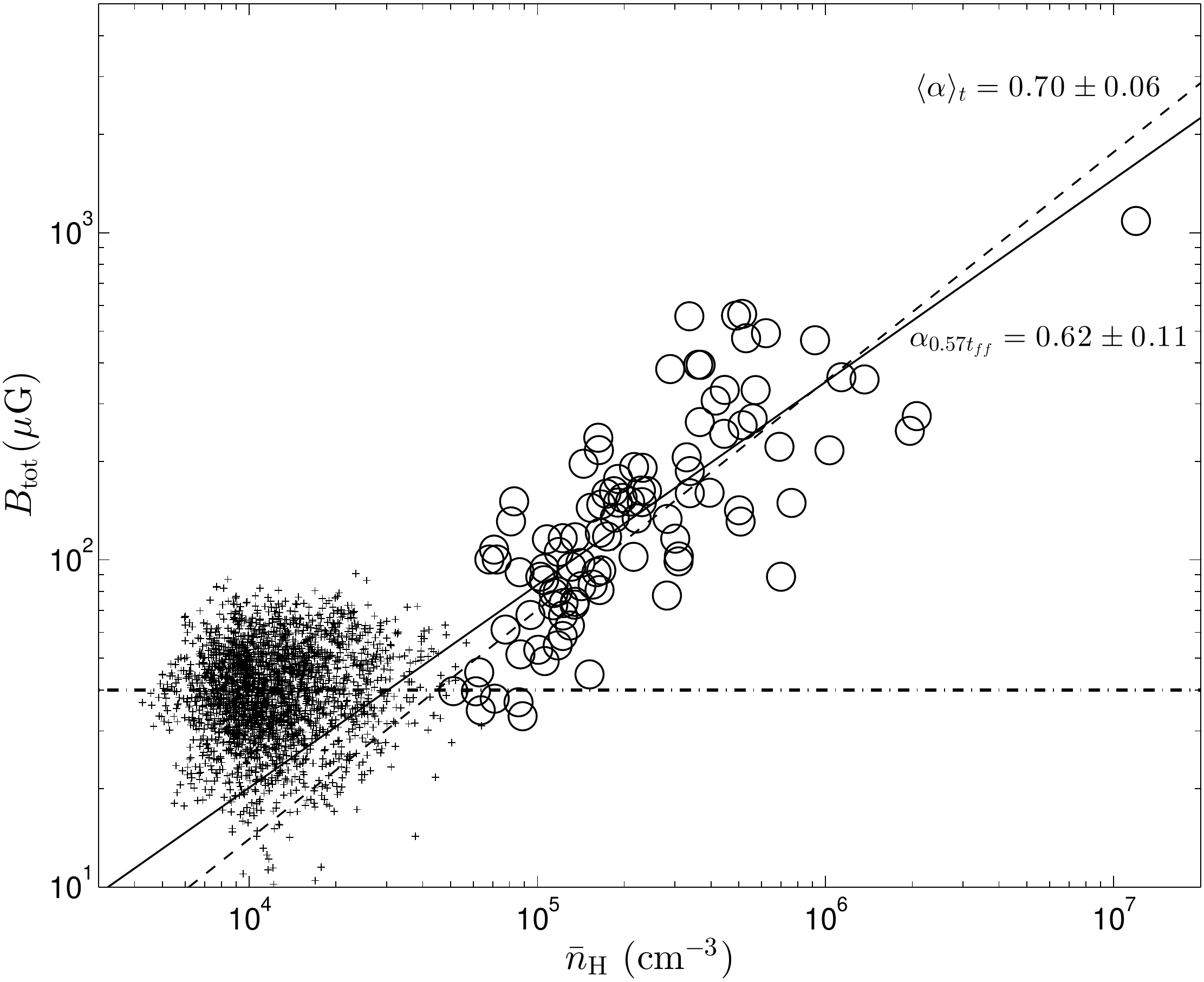}
\caption{Power-law relation between total field strength, $\btotr$, and volume density of the central beam region in the strong-field 100-clump sample at $0.57 \tff$ (open circles and solid line).  The power-law index relating $\btot$ to $\nbh$ at that time is $\alpha_{0.57 \tff} = 0.62\pm0.11$. The time-averaged [$(0.4-0.64)\tff$] value from the strong-field 100-clump samples is shown by the dashed line labeled $\at = 0.70\pm0.06$. Clumps that have column densities of $\Nh \geq 2 \times 10^{21}$ cm$^{-2}$ at $t = 0$,
prior to the time that self-gravity has acted, are shown by pluses.  
At that time, the magnetic field strength of clumps in the simulation is independent of the volume density, similar to the horizontal branch (HI cloud clumps) in the $\btotr$ vs $\nbh$ diagram in Fig. \ref{fig3}.
The dash-dot line indicates the average field strength of these clumps.
The power-law branch of $\btotr$ vs. $\nbh$ is the result of gravitational collapse of clumps to higher density and magnetic field strength.
\label{fig5}}
\end{figure}

\begin{figure*}
\includegraphics[scale=0.75]{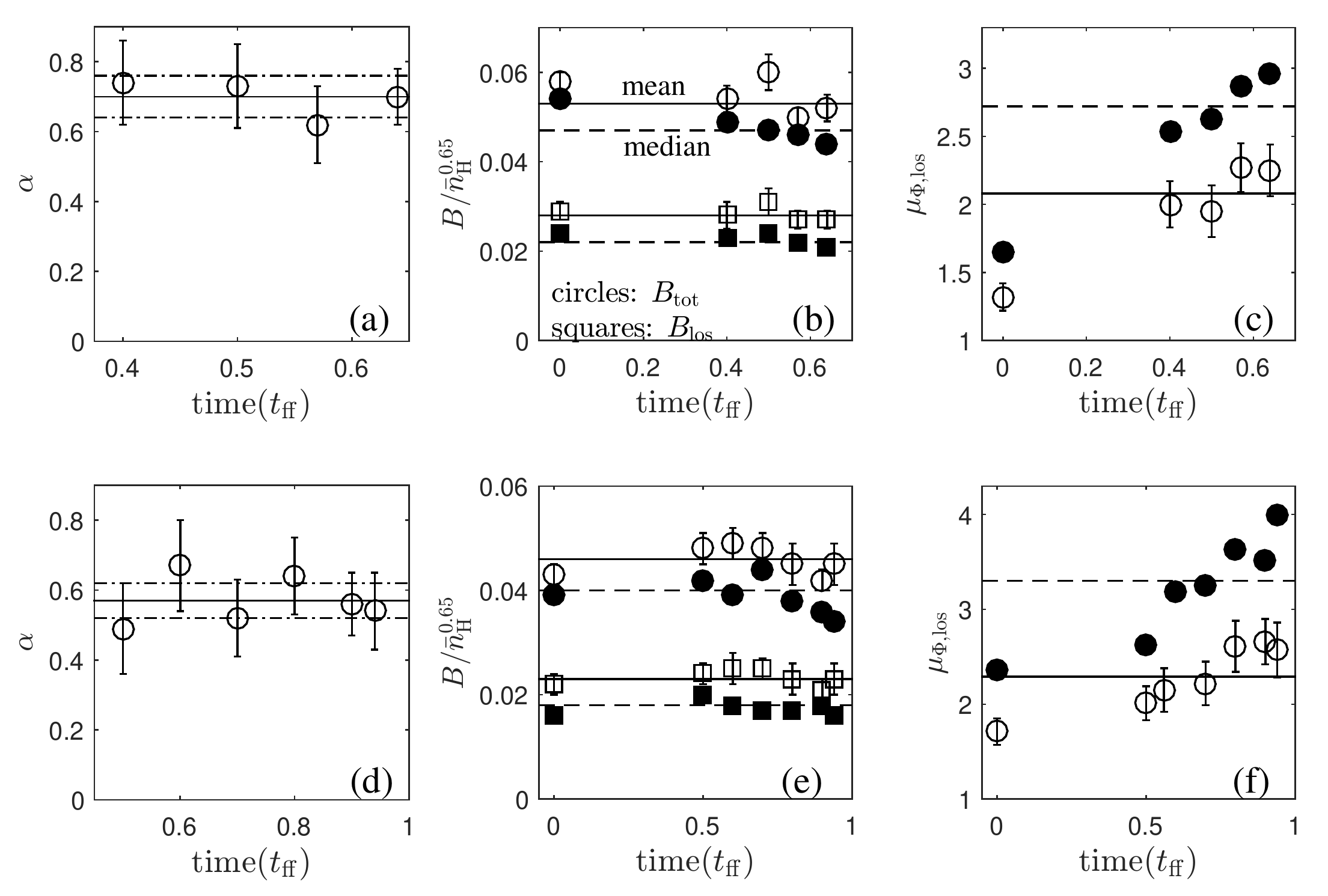}
\caption{Approximate time independence of key quantities: The power-law index $\alpha$, 
normalized magnetic field strengths, $\btot/\nbh^{\alpha}$ and $\blos/\nbh^\alpha$, 
and line-of-sight mass-to-flux ratio, $\muplos$, from the 100-clump samples in the two simulations at different times in the gravitational collapse phase.
The strong-field results are in the top row (a-c) at times 0.4, 0.5, 0.57, and $0.64 \tff$;
the weak-field results are in the bottom row (d-f) at times 0.5, 0.6, 0.7, 0.8, 0.9 and 0.94 $\tff$. Values at
$t=0$ are shown for $\btot/\nbh^\alpha$ and $\muplos$.
(a) The horizontal solid line is the time-averaged mean, $\at$, over this period of time, and the two dot-dashed lines show the uncertainty of $\at$. (b) The means (open circles) and medians (solid circles) of $\btotr/\nbh^{0.65}$ and those (squares) of $\blosr/\nbh^{0.65}$. Comparison with the time-averaged means (solid lines) and medians (dashed lines) of $\btotr/\nbh^{0.65}$ and $\blosr/\nbh^{0.65}$ (including $t = 0$), respectively, show that they do not vary much from the initial value at $t = 0$. The value of $\blosr/\nbh^{0.65}$ is about half that of $\btotr/\nbh^{0.65}$, as expected from a random distribution of field orientations. (c) The harmonic means (open circles) and medians (solid circles) of $\muplos$ show a slow increase with time. The time-averaged harmonic means (solid lines) and medians (dashed lines), excluding $t = 0$, are shown for comparison.
(d), (e), and (f) are the same as (a), (b), and (c), respectively, but for the weak-field model.
\label{fig6}}
\end{figure*}

We now demonstrate that our simulations show that $\btot\propto\nbh^\alpha$, with
a value of $\alpha$ consistent with that found by 
\citet{cru10} and obtained in Section \ref{sec:analysis}.  
We begin with a discussion of how we obtain $\btot$ from our simulations.
Since each core in the simulation gives us three different
line-of-sight fields to analyze,
we infer $\btot$ based on observing
the field in the three cardinal directions,
\beq
\vecB_\tot=(B_{\los,\,x},\,B_{\los,\, y},\, B_{\los,\,z}).
\label{eq:btot1}
\eeq
Thus, $\vecB_{\tot}$  in Equation (\ref{eq:btot}) is the field associated with the mass in the actual telescope beam,
which we have assumed is on the $z$ axis, whereas $\vecB_{\tot}$ in Equation (\ref{eq:btot1})
is the field that would be measured by an Earthbound observer and two hypothetical
observers along the $x$ and $y$ axes, who would observe slightly different volumes of gas unless the
emission were unresolved. We have shown that the distributions of $\btot=|\vecB_{\tot}|$ calculated with
Equations (\ref{eq:btot}) and (\ref{eq:btot1}) are very nearly the same.
Furthermore, the time-averaged 
median value of $\btot/\blos$ using this procedure
is 2.05, very close to the expected value of 2 for the case in which the same region of the clump is used to determine the relation between $\btot$ and $\blos$.

\begin{table*}
\caption{Observations Compared with Time-Averaged Properties of 100-Clump Samples}
\label{tab:4}
\begin{tabular}{lllll}
\vspace{-0.3cm}\\
\hline
\hline
\vspace{-0.2cm}\\
    &             & Strong field\fnm[1]    & Weak field\fnm[1]& Observed\fnm[3]\\[0.5ex]
\hline
\vspace{-0.2cm}\\
Central Beam\fnm[2]& $\avg{\alpha(\blos)}_t$&$0.63\pm0.07$&$0.58\pm0.05$&
   $0.64\pm0.13$\\
	&$\blosnfat$  & $11.1\pm0.4$ & $9.2\pm0.4$ & $22.4\pm3.3$\\
              \vspace{0.7ex}
             & $\blosnfamt$ & 8.9           & 7.1 & 11.9\\
		& $\langle \alpha(\btotr) \rangle_t$   & $0.70\pm0.06$   & $0.57\pm0.05$&
		$0.65_{-0.06}^{+0.10}~$\fnm[4]\\
             & $\btotnfat$  & $21.1\pm0.8$ & $18.2\pm1.6$& $44.8\pm 6.6$ \\
              & $\btotnfamt$ & 18.7         & 15.7 & \\
\vspace{-0.4cm}\\
\hline
\vspace{-0.2cm}\\
Whole Clump  & $\langle \alpha(\blos) \rangle_t$   & $0.67\pm0.10$   & $0.65\pm0.08$\\
             & $\blosnfat$  & $10.7\pm0.4$ & $11.9\pm0.4$&\\
             \vspace{0.7ex}
             & $\blosnfamt$ & 8.8       & 8.6&\\
		      & $\langle \alpha(\btotr) \rangle_t$   & $0.70\pm0.06$   & 
$0.69\pm0.07$\\
             &$\btotnfat$  & $21.9\pm0.4$ & $23.7\pm0.8$&\\    
             & $\btotnfamt$ & 19.1      & 19.7&\\      
         \vspace{-0.4cm}\\
\hline
\hline
\end{tabular}

\begin{flushleft}
\fnt{1} {$^a$ See Table \ref{tab:3}. The time averages cover $(0.4-0.64)\tff$ in the strong-field case and $(0.5-0.94)\tff$ in the weak-field case.}\\
\fnt{2} {$^b$ Both $\blos$ and $\nbh$ are convolved with a Gaussian beam 0.025 pc in radius (see Section \ref{sec:comp}).}\\
\fnt{3} {$^c$ Based on data from \citet{cru10}. The time averages of the quantities in the table do not apply to the observed results.}\\
\fnt{4} {$^d$ These 95 per cent confidence limits are
inferred from Fig. 4 of \citet{cru10} under the assumption that the count above $\alpha=0.76$ is small.}\\
\end{flushleft}
\end{table*}

The values of $\blos$ at $t=0.57\tff$ in the strong mean-field simulation are plotted as a function of $\nbh$ in Fig. \ref{fig3} along with the line-of-sight field measurements of \citet{cru10}. 
The distribution of the line-of-sight fields versus 
the mean density determined along each line of sight
in the 100-clump sample fully covers the distribution of observed CN clumps in terms of $\nbh$ and the top envelope of the distribution is consistent with a power-law relation between $\blos$ and $\nbh$, similarly to the power law obtained from the Bayesian analysis of \citet{cru10}. The distribution is also widely scattered, similar to that of the observed CN and OH clumps. Below, we will see that $\blosnam$ for the 100-clump sample is close to the observed value.

The relation between the field strength and the density is much clearer for $\btot$ since the scatter due
to projection along the line of sight is removed. 
For the density, we take the mean value in the central beam region along the three cardinal axes. 
As shown in
Fig. \ref{fig5}, the strong-field 100-clump sample at $t=0.57\tff$ shows
a clear power-law relation between $\btot$ and $\nbh$ with $\alpha = 0.62\pm0.11$, consistent with 
the value inferred from observation.  This result is explored further in Fig. \ref{fig6}, 
which shows $\alpha$ at several different times
(results of the weak mean-field model are also shown). Note that $\alpha$ is approximately independent of time
once self-gravity has had time to act;
by contrast, $\alpha$ cannot be determined at $t=0$ since $\btot$ and $\nbh$ are uncorrelated then, as shown in Fig. \ref{fig5}.
The time-averaged value of $\alpha$ over the period $(0.4-0.64)\tff$ is $\at=0.70\pm0.06$, consistent
with the value of $\alpha=0.65$ from \citet{cru10}.
The line-of-sight fields in the simulations exhibit a power-law relation with density as they should, since they are the projection of fields that obey such a relation; for the strong-field case, we find
$\avg{\alpha(\blos)}_t=0.61\pm0.07$, consistent with the value $\alpha(\blos)=0.64$ from
a least-squares fit to the data of \citet{cru10}. The results for the power-law dependence
of the magnetic field on the density are summarized in Table \ref{tab:4}. The important conclusion to be drawn from
these results is that the observations and simulations are consistent, and both
show that $\btot$ and $\blos$ scale as $\nbh^{2/3}$ within the errors.

In addition to determining the power law in the relation between the field and the density, it is important
to determine the coefficient. In Appendix \ref{app:theory} we show that $B/\nbh^{2/3}$ is proportional to
a magnetic critical mass, $M_B$, that does not depend explicitly on the size of the clump (Eq. \ref{eq:mb}).
Equivalently, for gravitationally bound ($\avir\sim 1$), supersonically turbulent ($\calm\gg 1$) clumps,
$B/\nbh^{2/3}\propto\calm^{4/3}/\muphi$ is inversely proportional to the mass-to-flux ratio $\muphi$.
These results change very little when the exponent 2/3 is replaced by the observed value of $\alpha=0.65$.
Thus, the coefficient in the power-law relation depends on the mass-to-flux ratio.

We have collected the time-averaged mean and median values of both
$\btotmu/\nbhf^{0.65}$ and $\blosmu/\nbhf^{0.65}$
from our simulations together with the observed values in Table \ref{tab:4}.
Here we briefly compare the strong-field central-beam values with the observations; we defer discussion of the 
the whole-clump values to Section \ref{sec:obsmfr} and the weak-field values to Section \ref{sec:is}.  The time-averaged
median value of $\blosmu/\nbhf^{0.65}$ is about 75 per cent of the observed value. This agreement is extraordinary, both because of the many differences between the simulated clumps and the observed clumps discussed in Section \ref{sec:100clump} and because the simulated clumps and the observed clumps each have a large dispersion in $\blos$ (see Fig. \ref{fig3}).  
On the other hand, the average observed value of $\blos/\nbhf^{0.65}$ is about twice that in the simulation.

What causes the power-law relation between the magnetic field and the density? In our simulation, it is clearly self-gravity. Fig. \ref{fig5} shows the 100-clump sample from the strong mean-field simulation at $t=0.57\tff$
in the $\nbh$--$\btot$ plane, and the figure also shows all the clumps with 
$\Nh = 2 \times 10^{21}$ cm$^{-3}$ before gravity is turned on
($t=0$). The power-law tail is present only after self-gravity has had time to act.
The same is likely to be true in the observations: The HI clumps are not self-gravitating and have
$\btot$ independent of density, whereas many of the molecular clumps, which do show
$\btot\propto\nbh^\alpha$, are self-gravitating
(e.g., \citealp{cru99}). In Equation (\ref{eq:nhcrit}) we showed that the power-law region begins at
about the point at which the turbulent and thermal pressures are the same; \citet{kru05} showed that
this corresponds to the condition for the clump to be self-gravitating, provided it is embedded in
a self-gravitating cloud.
The effect of self-gravity on the 100 clumps can be assessed by determining if they are collapsing and, if so, how
the collapse velocity compares with the free-fall velocity. 
We compare the mean value of the density-weighted infall velocity averaged over a sphere
of radius $r$ to the free-fall velocity there, $v_{\rm ff}(r)=[Gm(r)/r]^{1/2}$,
\beq
\frac{\int \rho v_r dA}{v_{\rm ff}(r) \int \rho dA}.
\label{eq:vinvff}
\eeq
We have evaluated this ratio for the 100 clumps at $0.57 \tff$ 
at a radius of 0.05 pc. The median value is 0.29, showing that these clumps 
are contracting at a significant fraction of the free-fall velocity and confirming that self-gravity is important.

\citet{cru10} pointed out that $\alpha\simeq \frac 23$ is consistent with the suggestion by \citet{mes66} that the spherical contraction of a cloud with a weak magnetic field would lead to $B \propto n^{2/3}$. In Section \ref{sec:actual}, we shall see that the median mass-to-flux ratios of the 100-clump samples from the two simulations are around 2,
with the strong-field model being somewhat less than 2 and the weak-field one somewhat greater. The median magnetic field is
therefore not very weak compared to gravity.
However, the contraction of the clumps is approximately spherical:
The time-averaged median value of the aspect ratio $Z/R$ of the 100-clump samples from the strong and weak mean-field models 
at the end of the simulations are 1.19 and 1.08, respectively.
(Recall that $Z$ is along and $R$ is perpendicular to the mean field direction of the clump.) 
In fact, all clumps in both samples have $0.5 < Z/R < 2$ at all times $t\geq 0$ and the medians remain close to unity throughout the simulations, so that their collapse is approximately spherical.
Note that the range of $Z/R$ implies that there are clumps with prolate, oblate, and triaxial shapes in the 100-clump samples, as seen in other ideal MHD simulations with turbulence and self-gravity \citep[e.g.][]{gam03,li04}. 

Independent of the strength of the field, a necessary condition to obtain $B\propto \nbh^{2/3}$
is that the coefficient in the $B(\nbh)$ relation
must be about the same for all the clouds. The results of our simulation show that
the ratio of the time-averaged dispersion of 
$\bbtot=\btotmu/\nbhf^{2/3}$, 
which we label $\sigma(\bbtot)$,
to the median value is 
\beq
\frac{\avg{\sigma(\bbtot)}_t}{\avg{b_{\tot,\,\med}}_t}= 0.34,
\eeq
which is small enough so that this is approximately true.
This ratio for $b_{\los}$ is 0.60; it is larger because of the additional dispersion added by projection effects.

We conclude that our simulations show that the density dependence of the field found by \citet{cru10} is due to 
gravitational contraction, that this contraction is approximately spherical, and that different clumps have similar
values of the ratio $B/\nbh^{2/3}$. These characteristics are all consistent with the formation of molecular clumps due to
the gravitational contraction of clouds that initially had similar values of $B$ and $\nbh$ (e.g., in the cold neutral
medium of the ISM) under the assumption that the field is so weak that it does not affect the contraction.
However, although the field is weak enough that almost all the clumps are magnetically
supercritical, it is not so weak that it has no dynamical effects. Indeed, our simulations give the counter-intuitive result that the Lorentz
force in the initially weak-field simulation has a larger dynamical effect than that in the initially strong-field simulation.
The density dependence of the magnetic field in both molecular clumps and in our simulations appears to be an
emergent property associated with turbulent, magnetized, self-gravitating systems. A more complete understanding
must await further research.

\subsubsection{Scaling with Velocity Dispersion}

\begin{figure*}
\includegraphics[scale=0.65]{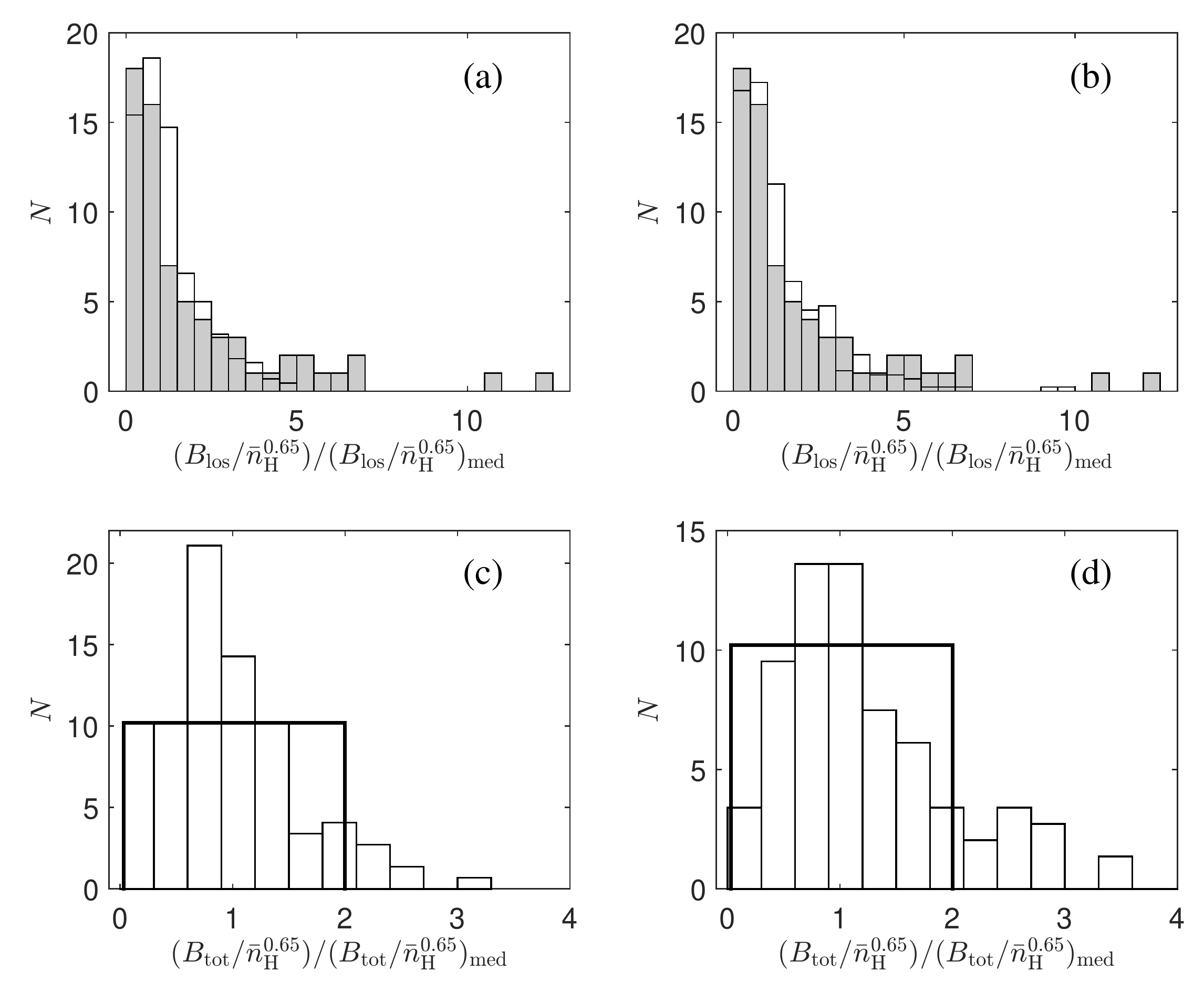}
\caption{(a) The PDFs of $\blosr/\nbh^{0.65}$,
normalized by the median value of the clump sample, from the strong-field model at $0.57 \tff$ (shaded histogram) and from the OH+CN sources (open histogram) in \citet{cru10}.  K-S tests show a good match of the PDFs at this time and at different times as well. (b) Same as (a) for the weak-field model at $0.94 \tff$.  K-S tests show that the weak-field model agrees with observation about as well
as the strong-field model. (c) The PDF of the normalized 
values of $\btotr/\nbh^{0.65}$ of the 100-clump 
sample from the strong-field model at $0.57 \tff$ compared with the uniform distribution proposed in \citet{cru10}.  (d) Same as (c) for the weak-field model at $0.94 \tff$.
\label{fig7}}
\end{figure*}

\citet{mou91} suggested that the magnetic field in molecular clumps depends on the velocity dispersion,  $\sigma_v$, as well
as on the density, $B\propto \bar\rho^{1/2}\sigma_v$.  Based on HI and OH Zeeman data from \citet{cru99}, 
a significantly smaller sample than that in \citet{cru10}, \citet{bas00} showed that this magnetic field scaling is in excellent agreement with observation. He also found that the velocity dispersion is independent of density. \citet{bas00} derived this
relation under the assumption that clouds are planar and concluded that the agreement with observation supported this hypothesis. In Appendix \ref{app:theory} we show that this relation comes from an identity and
applies to 
clouds of a variety of shapes provided that they
are gravitationally bound and are close to being magnetically critical, and provided that $\muphi$ and $\avir$
are independent of density.  Now with all 68 OH+CN sources from \citet{cru10}, we confirm the prediction of \citet{mou91} and \citet{bas00}, 
$\blos/\sigma_v \propto n^{0.50\pm0.12}$, although there is a significant uncertainty in the exponent. However, as we have noted
above, the magnetic field-density relation is in fact $\blos\propto n^{0.64\pm0.14}$, implying that the velocity dispersion scales
weakly with density.
Indeed, a direct fit of $\sigma_v$ vs. $n$ gives $\sigma_v \propto n^{0.14\pm0.05}$, as expected from these two relations.
We also find that the observations imply that the values of $\muphi$ and $\avir$ inferred from line-of-sight data 
are independent of density to within the errors.

The simulations give somewhat different results.
For the strong-field model, we find that the time-averaged $\sigma_v \propto n^{0.35\pm0.03}$ and $\muphi \propto n^{0.13\pm0.04}$; $\avir$ is independent of density to within the errors.
Despite the fact that the two power-law indexes increase slowly with time as the clumps undergo gravitational collapse,
the value of $\alpha$ does not exhibit a secular variation with time for $t>0.4\tff$, although it does fluctuate somewhat (Fig. \ref{fig6}).
The 100-clump sample in the weak field model behaves similarly, with small differences in the power-law indexes.  
The results of both models are consistent with Equation (\ref{eq:bsid}). 
Observationally, we do not have information on the time evolution of these relations. Fitting these power-law relations to the OH and CN sources separately results in very large uncertainties because of the small number of sources and 
the limited range of densities in each group of sources.
Therefore, although it is plausible that the CN sources, being denser, represent a later stage of evolution than the OH sources, current observations
do not enable us to determine if the power-law indexes in actual molecular clumps evolve in time.

\subsection{Distribution of Magnetic Field Strengths}
\label{sec:dist}

\begin{figure*}
\includegraphics[scale=0.7]{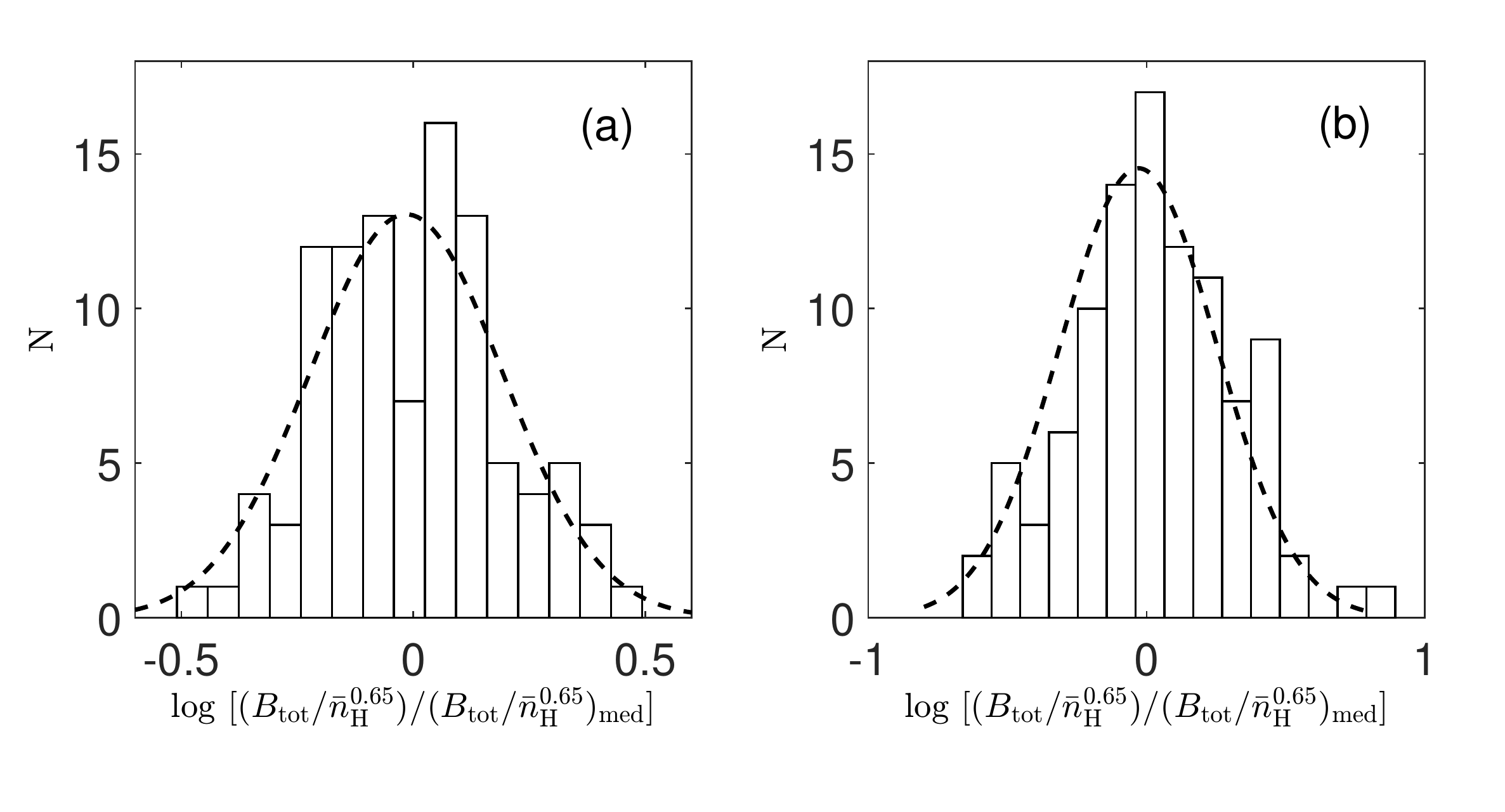}
\caption{Lognormal fit of the normalized values of $\btot/\nbh^{0.65}$ of the 100-clump sample from (a) the strong-field model at $0.57 \tff$ and (b) the weak-field model at $0.94 \tff$.  K-S tests show a good match of the PDFs with $p-$values of 0.89 and 0.98, respectively.
\label{fig8}}
\end{figure*}

We first show that distribution of line-of-sight fields in our simulations agrees remarkably well with observation.
In Fig. \ref{fig7}a, we plot the probability density function (PDF) of $\blosr/\nbh^{0.65}$, normalized by the median $\blosnam$, of our 100-clump sample from the strong-field model at $0.57 \tff$ 
together with that of the total 68 OH+CN sources from \citet{cru10}.  The PDF of the 100-clump sample is normalized to a total of 68 sources for direct comparison.  We performed a K-S test on these two PDFs, with the null hypothesis being that the two samples are from the same continuous distribution.  The result is that the probability that this could be true (the $p$-value of the test) is 0.89.  We performed K-S tests on the simulation data at different times with different binning sizes.  The tests gave $p$-values between 0.50 - 0.99 in the strong-field model.
We have performed K-S tests on the weak-field model as well, and the results are similar to those for the strong field model.  The comparison of the strong- and weak-field model results will be discussed in Section \ref{sec:is}.
We conclude that the PDFs of $\blos/\nbh^{0.65}/\blosnam$ from the simulations
and from the observed OH+CN sources are consistent with being drawn from the same continuous function.

An important conclusion from the Bayesian study in \citet{cru10} is that the inferred distribution of the total field strength,
$\btot$, is best described by a uniform distribution
over the range $f\btm<\btot<\btm$ with $f=0.03$ and $\btm\propto \nbh^\alpha$.  Here we shall look at the shape of
the distribution based on the values of $\blosr$, $\btotr$, and $f$ from our simulations.  
We begin by considering how the observed and simulated line-of-sight fields agree with either a uniform or
log-normal distribution.
As we found in Section \ref{sec:analysis}, a $\chi^2$ test suggests that a uniform distribution 
for $\btot$ does not provide a good fit to the observed line-of-sight data, since $\chi^2=1.85$. On the other hand,
a K-S test, which is weaker, indicates that a uniform distribution is quite consistent with the observed data, since the $p$-value is 0.97.
A log-normal distribution provides a slightly better fit to the observed line-of-sight 
data, with $\chi^2=1.45$; the $p$-value for the K-S test is 0.98. 
These results are inconclusive: the observed line-of-sight data does not allow us to infer whether the 
underlying distribution of total field strengths has a uniform, log normal, or some other distribution.
Analysis of the line-of-sight results from our two simulations yields a similar result,
although both a uniform distribution and a log-normal distribution provide better fits than they do to the observed data.

However, the 3D data available from the simulation provides a much more powerful discriminant than the line-of-sight data:
The distribution of the total field strength, $\btot/\nbh^{0.65}/\btotnam$, from our simulations
is {\it not} consistent with a uniform distribution--the null hypothesis is rejected with
a $p$-value of 0.004 in the strong-field case and 0.001 in the weak-field case. This is shown graphically in Fig. \ref{fig7}c and
\ref{fig7}d.
On the other hand, the distribution of normalized total field strengths from the simulation is well fit by a log-normal distribution
(Figs. \ref{fig8}a,b):  A K-S test gives a $p$-value of 0.89 for the strong-field case and 0.98 for the weak-field case,
with $\sigma=0.21,\, 0.28$ for the two cases, respectively.
The peak of the lognormal distribution is very close to the median value of the observed distribution in both cases.

We conclude that both the observed and simulated {\it line-of-sight} data are consistent with each other and with a
uniform distribution of $\btot$. On the other hand, the simulated data for the {\it total} normalized field strength is not consistent with a uniform distribution, but it is consistent with a log-normal distribution. 
If we knew that at least one of our simulations was a reasonable representation of reality, we could conclude that
observed fields do not have a uniform distribution of normalized total field strength.
However, the many differences between the simulation and the observations discussed in Section
\ref{sec:100clump} prevent us from drawing such a conclusion.

An important implication of the the work of \citet{cru10} is that magnetic field strengths in molecular
clumps extend to very low values, $\btmin\simeq 0.03\btm$. From Equation (\ref{eq:btm}), this
implies min($\btotmu/\nbhf^{0.65})=2.9$. 
Now, the median value of $\blos/\nbhf^{0.65}=11.9$
from the data in \citet{cru10}, so their results imply
\beq
\frac{\mbox{min }(\btotmu/\nbhf^{0.65})}{(\blos/\nbhf^{0.65})_\med}=0.24~~~~~~~~\mbox{\citep{cru10}}.
\eeq
In other words, they infer that the minimum normalized total field is about 1/4 of the observed
median line-of-sight field. We can actually measure this ratio in our simulation, and in our strong-field case we find that this ratio is 0.69, almost 3 times as large.

In addition, the total field is subject to cancellation, since it is an average over a Gaussian beam.
The total field strength is better characterized by the rms field. Averaging over the whole
clump (and not just the central beam as in the observations), we find
\beq
\frac{\mbox{min }(\brmsmu/\nbhf^{0.65})_{\rm clump}}{(\blos/\nbhf^{0.65})_\med}=1.07,\;0.93
\eeq
for the strong- and weak-field cases at 0.57 $\tff$ and 0.95 $\tff$, respectively.
(Note that the numerator is for the whole clump, whereas the denominator is the observed line-of-sight value for 
the central beam.) Thus we find no evidence for very weak
fields in our simulation; instead, the minimum rms field is comparable to the median line-of-sight field.  We also find that
\beq
(\brmsv/\blos)_{\rm med} = 2.03,\;1.94
\eeq
for the strong- and weak-field cases, respectively, where $\brmsv$ is the volume-weighted $\brms$ averaged over the volume of the whole clump.  
Since $\blos$ is mass-weighted and on average is half of $\btot$ (also mass-weighted), it follows that
$\brmsv$ is very nearly the same as the mass-weighted $\btot$ averaged over a 0.025 pc radius central Gaussian beam.

\subsection{Possible Dependence of $\blos/\nbh^\alpha$ on Beam Size}

In Section \ref{sec:analysis}, we pointed out that the 48 OH and CN sources for which we could determine the size of the telescope beam show a weak but significant dependence of $\blos/\nbh^{0.65}$ on $\rbeam$.  
Fig. \ref{fig1} shows that we see a similar effect in our strong-field simulation.
The three solid circle symbols show $(\blos/\nbh^{0.65})_{\rm med}$ for beam sizes of 0.025 pc (the one used in the analysis above), 0.0125 pc and 0.00625 pc. For our 2-level simulation, the results with a beam size of 0.00625 pc are not converged, so the value of $(\blos/\nbh^{0.65})_{\rm med}$ for this case is obtained from our 3-level run in the convergence study.  The three data points from the simulations match almost exactly the best-fitting slope of the 48 observed cloud clumps.  Since $\nbh$ almost certainly decreases with $\rbeam$, the increase of $\blos/\nbh^{0.65}$ with $\rbeam$ implies that the value of $\alpha$ is less than 0.65 within individual clumps,
possibly due to a diffusion mechanism operating within the clumps.  Our simulations are based on ideal MHD, so
the agreement between the simulation results and observation suggests that, at least down to about 0.025 pc, non-ideal MHD diffusion mechanism such as ambipolar diffusion are not responsible for the reduction in $\blos/\nbh^{0.65}$ on small scales.  A possible candidate for this is reconnection in a turbulent
medium (sometimes termed `reconnection diffusion'---see \citet{laz15} and references therein), which will be discussed in Section \ref{sec:obsmfr}.  However, because of the large scatter of the observed data in Fig. \ref{fig1}, we cannot draw a firm conclusion on this possible beam-size effect.
Future high resolution Zeeman observations will provide more information on this issue.

\section{Observed vs. Actual Mass-to-flux Ratio}
\label{sec:obsmfr}

As discussed in Appendix \ref{app:theory}, the mass-to-flux ratio, $\muphi$, is of fundamental importance
in star formation since it determines whether gravity can overcome magnetic stresses and
initiate gravitational collapse. The {\it observed} mass-to-flux ratio is based on the mass-weighted line-of-sight field, $\blos$, and the column density, $\Sigma_\beam$,
in the central beam:
\beq
\muplos=2\pi G^{1/2}\left(\frac{\Sigma_\beam}{\blos}\right).
\label{eq:muplos}
\eeq
For the moment, let us assume that the clump is static. There are then two main corrections that
are needed in order to obtain the actual mass-to-flux ratio: First, the field and column density
must be measured along the direction of the mean field, which we label $z'$. This de-projected value, 
termed the `intrinsic value' by \citet{hei05a}, is $\muphi(\btot)$ in
our notation. Second, the magnetic field must be changed from $\btot$, the
mass-weighted field in a central Gaussian beam, to $\bmidz$, the field in, and normal to, the
midplane of the clump with no mass weighting.
The method for determining the midplane in
a non-uniform clump will be described in Section \ref{sec:actual} below.
For a static, equilibrium cloud embedded
in a uniform magnetic field, $\bmidz$ is the total field.
We define $\bmidz$ so that it is
averaged over a disc of radius $\rbeam$ at the center of the clump; this is preferable to
averaging over the whole clump since the size of the clump depends on the somewhat
arbitrary criterion adopted
to define the clump. In addition to correcting the field, the column density must also be
corrected so that it refers to the mass inside a cylinder of radius $\rbeam$ rather than that in
a Gaussian beam of that characteristic radius;
we label this $\Sigma_{\midp,\,z'}$.
The result, $\muphi(\bmidz)$, is the {\it actual} mass-to-flux ratio for a static
clump.

In the real world, clumps are not static, and the field can become tangled. The critical value of
the mass-to-flux ratio is not known in this case; 
we shall continue to use the value for a static cloud here.

\subsection{Projection Effects}
\label{sec:projection}

\begin{table*}
\caption{Line-of-Sight and Actual Mass-to-Flux Ratios of the 100-Clump Samples}
\label{tab:5}
\begin{tabular}{llcccccccc}
\hline
\hline
Model\fnm[1] & $t/\tff$ & \multicolumn{2}{c}{$\muplos$\fnm[2]} & & \multicolumn{2}{c}{$\muphi(\bmidz)$\fnm[3]} & & \multicolumn{2}{c}{$\muphi(\absbmidz)$\fnm[4]}\\
\cline{3-4} \cline{6-7} \cline{9-10} \vspace{-0.2cm}\\
& & H. Mean & Med & & H. Mean & Med & & H. Mean & Med\\
\hline
Strong  & 0    & $1.32\pm0.10$  & 1.65 && $0.95\pm.05$ & 0.86 && $0.91\pm.05$ & 0.81\\
field   & 0.4  & $2.00\pm0.17$  & 2.53 && $1.57\pm0.09$ & 1.69 && $1.52\pm0.08$ & 1.67\\
        & 0.5  & $1.95\pm0.19$  & 2.63 && $1.56\pm0.10$ & 1.83 && $1.51\pm0.10$ & 1.70\\
        & 0.57 & $2.27\pm0.18$  & 2.87 && $1.80\pm0.12$ & 2.05 && $1.68\pm0.10$ & 1.94\\
        & 0.64 & $2.25\pm0.19$  & 2.96 && $1.75\pm0.13$ & 1.98 && $1.63\pm0.11$ & 1.86\\
        &\mbox{Time avg.\fnm[5]} & $2.08\pm0.10$ & 2.72 && $1.63\pm0.06$ & 1.86 && $1.56\pm0.05$ & 1.77\\
\hline
Weak   & 0    & $1.71\pm0.14$ & 2.36 && $0.99\pm0.05$ & 1.02 && $0.92\pm0.04$ & 0.97\\
field  & 0.5  & $2.01\pm0.18$ & 2.62 && $1.32\pm0.12$ & 1.48 && $1.25\pm0.11$ & 1.43\\
       & 0.6  & $2.15\pm0.23$ & 3.19 && $1.55\pm0.11$ & 1.82 && $1.45\pm0.10$ & 1.72\\
       & 0.7  & $2.22\pm0.23$ & 3.26 && $1.52\pm0.14$ & 1.92 && $1.42\pm0.12$ & 1.70\\
       & 0.8  & $2.61\pm0.27$ & 3.63 && $1.99\pm0.23$ & 3.09 && $1.82\pm0.19$ & 2.70\\
       & 0.9  & $2.66\pm0.24$ & 3.51 && $1.83\pm0.20$ & 2.76 && $1.66\pm0.19$ & 2.37\\
       & 0.94 & $2.57\pm0.29$ & 4.00 && $2.04\pm0.20$ & 3.02 && $1.88\pm0.18$ & 2.51\\
       &\mbox{Time avg.} & $2.29\pm0.10$ & 3.30 && $1.56\pm0.06$ & 2.27 && $1.45\pm0.06$ & 2.02\\
\hline
Observed\fnm[6]  & & $2.40\pm0.29$ & 4.10 && - & - && - & -\\
\vspace{-0.4 cm}\\
\hline
\hline
\end{tabular}

\begin{flushleft}
\fnt{1} {$^{\rm a}$ See Table \ref{tab:3}.}\\
\fnt{2} {$^{\rm b}$ Observed $\muplos$, which is derived from the line-of-sight column density and $\blos$ by convolving the clump with a Gaussian beam of 0.025 pc in radius (see Section \ref{sec:comp}).}\\
\fnt{3} {$^{\rm c}$ Actual $\muphi$, which is based on the field normal to the clump midplane (see Section \ref{sec:obsmfr}).}\\
\fnt{4} {$^{\rm d}$ Value of $\muphi$ based on the magnitude of the field normal to the clump midplane (see Section \ref{sec:obsmfr}).}\\
\fnt{5} {$^{\rm e}$ The time average excludes $t=0$.}\\
\fnt{6} {$^{\rm f}$ From the 68 OH+CN clumps in \citet{cru10}}
\end{flushleft}
\end{table*}

The distribution of the observed field, which is the line-of-sight field, to the distribution of
the intrinsic field is given by Equation (\ref{eq:psia}).
One can immediately show that this implies that $\avg{1/\blos}$ is infinite. As a result,
it is preferable to work with the flux-to-mass ratio, $\muplos^{-1}\propto\blos$, 
in evaluating the average properties of the observed mass-to-flux ratio.
Even when dealing with the intrinsic mass-to-flux ratio, which does not
suffer from this divergence, it is better to average $\muphi^{-1}$ if
the total field can assume very small values, as inferred by \citet{cru10}.
The mean of the flux-to-mass ratio is the inverse of the harmonic mean of the mass-to-flux ratio,
\beq
\avg{\mbox{flux-to-mass ratio}}=\left\langle\frac{1}{\muphi}\right\rangle\equiv \frac{1}{\avg{\muphi}_h}.
\label{eq:har}
\eeq

For spherical clouds, the column density is not affected by projection, 
so that $\avg{\muplos}_h=2\avg{\muphi}_h$ since $\avg{\blos}=\frac 12\avg{\btot}$. 
The median values of $\blos$ and
$\muplos$ depend on the distribution of field strengths, $\phi(\btot)$; for a delta-function
distribution, $B_{\los,\,\med}=\frac 12 \btot$ and $\mu_{\Phi,\,\los,\,\med}=2\muphi(\btot)$.
At the other extreme, if clouds are flattened
into sheets, then the observed column density is larger than that normal to the sheet by
a factor $1/\cos\theta$, so that the average of the line-of-sight values is related to that of the total-field values by
$\avg{\muplos}_h=
3\avg{\muphi}_h$
(this is a corrected version of the result given in \citealp{hei05a}); for a delta-function distribution of 
field strengths, the median is given by
$\mu_{\Phi,\,\los,\,\med}=
4\muphi(\btot)$.
\citet{li11} have pointed out that gravitational contraction along field lines can increase the
apparent mass-to-flux ratio, but the maximum possible value in their model is that for a sheet, so
their discussion fits within the existing framework.
The medians and harmonic means of $\muplos$ of the 100-clump samples at different times from the strong and weak field models are given in Table \ref{tab:5}.

\subsection{The Actual Mass-to-flux Ratio}
\label{sec:actual}

The clumps in a turbulent medium do not have the axisymmetric shape assumed in theoretical treatments of the mass-to-flux ratio. Lacking an axis of symmetry, along what direction do we evaluate $\muphi$? 
Imagine evaluating $\muphi$ along different directions in the axisymmetric case. Along the axis of symmetry, the flux will be a maximum and the column density a minimum, so that $\muphi$ will be a minimum. Requiring
the minimum $\muphi$ to exceed unity provides the most stringent test for whether the clump is supercritical.
We therefore adopt this criterion in the turbulent case. To implement it, we divide the polar angle into 24 divisions ($7.5\degr$ apart) 
and the azimuthal angle into 48 divisions and evaluate $\muphi$ in the central
$\pi\rbeam^2$ at each orientation. The orientation
of the midplane is the one that gives the minimum value of $\muphi$; the 
center of the midplane is defined as the cell with the maximum density, and the midplane itself
is defined by the condition that it
pass through this cell.

To obtain the actual mass-to-flux ratio, we take the steps described above.
The ratio of the actual mass-to-flux ratio to the line-of-sight value is
\beq
\frac{\mu_{\Phi,\,\act}}{\mu_{\Phi,\,\los}}=\frac{\Sigma_{\midp,\,z'}}{\Sigma_\beam}\cdot
\frac{\blos}{\bmidz}.
\label{eq:act0}
\eeq
Recall that $\bmidz$ is the average value of the component of the field normal to the midplane over an
area $\pi\rbeam^2$ at the center of the clump; it is not density weighted. The line-of-sight field, $\blos$, differs
from this in several respects: It is density weighted, it is averaged along the line of sight, and it is also averaged over
a Gaussian beam. One typically assumes that the line-of-sight field is about half the actual field due to projection
effects. In fact, the density weighting of the line-of-sight field increases the ratio; we find that the median
value of $\blos/\bmidz$ is 0.80. As noted above,
spherical clouds have isotropic column densities, whereas sheets lead to an increase in the ratio of the
observed column density to the value that enters $\muphi$. For the clumps in our simulation, the median value of the ratio of
the actual column density to the line-of-sight value (which is averaged over a Gaussian beam) is 1.09, close
to the spherical value. Equation (\ref{eq:act0}) then implies
\beq
\mu_{\Phi,\,\act,\,\med}\simeq 0.7\mu_{\Phi,\,\los,\,\med}.
\label{eq:act}
\eeq
The key result of this analysis is that the actual mass-to-flux ratio is significantly greater than (0.25-0.5) times the observable line-of-sight value, which is what is expected from a naive application of the projection effects discussed above.

Detailed results for the mass-to-flux ratio are given in Table \ref{tab:5}.
The ratio of the median of the actual $\muphi$ to the median of the line-of-sight value is 0.68, very close to the estimate
above. The harmonic means are closer: the actual value is 0.78 times the line-of-sight value.
The weak mean-field results are similar
to the strong mean-field ones. 

It should be noted that clump mergers and the growth of clumps due to flows along field lines imply that the identity of the
100 clumps in the samples can evolve with time. Mergers of clumps in the samples allow new, smaller clumps
with lower mass-to-flux ratios to enter
the samples; field-aligned accretion leads to an increase in the mass-to-flux ratio. The latter effect dominates, so that
there is a slow increase in the mean mass-to-flux ratio with time.

Next, we assess the effects of tangling of the field by evaluating $\muphi$ with the magnitude of the field in the midplane, $\absbmidz$, rather than the component normal to the midplane, $\bmidz$; the results are only slightly smaller, showing that field-line reversals are insignificant in the midplane
(the time-averaged median value of $|\bmidz|/\bmidz$ is about 1.01).
Field reversals are somewhat more important for the line-of-sight field, which averages over a volume along
the line of sight: The time-averaged median value of $(|\blos| / \blos)$ for the 100 clumps is 1.29.
As shown in Table \ref{tab:5}, comparison of the time-averaged $\muphi(\bmidz)$ with $\muphi(\absbmidz)$ shows that the effect of field-line tangling in the weak-field case is a little larger than that in the strong-field case, which is a natural outcome for the weak field model.  

The harmonic mean of $\muplos$ from the observed 68 OH+CN clumps in \citet{cru10} is about
2.4 as shown in Table \ref{tab:5},
close to the values for our 100-clump samples.  (We do not place great weight on this agreement, however, 
since we have found that the
values of $\muphi$ in our samples depend weakly on mass, an effect we shall explore in a subsequent paper.)
The results of our strong-field simulation suggest that the harmonic mean of the actual mass to flux ratio is
about 78 per cent of the line-of-sight value, or $\mu_{\Phi,\,\act,\, h}\simeq 1.5$. However, because the harmonic mean is
disproportionately affected by small values, the median value is more characteristic of the observed
mass-to-flux ratio. Since the observed median is $\mu_{\Phi,\,\los,\,\med}\simeq 4$, we infer that the
actual median mass-to-flux ratio is $\mu_{\phi,\,\act,\,\med}\simeq 2.8$ from Equation (\ref{eq:act}). 
Typical clumps in molecular clouds are
therefore magnetically supercritical by a significant margin ($\sim$ factor 3). 

\begin{figure*}
\includegraphics[scale=0.7,angle=-90]{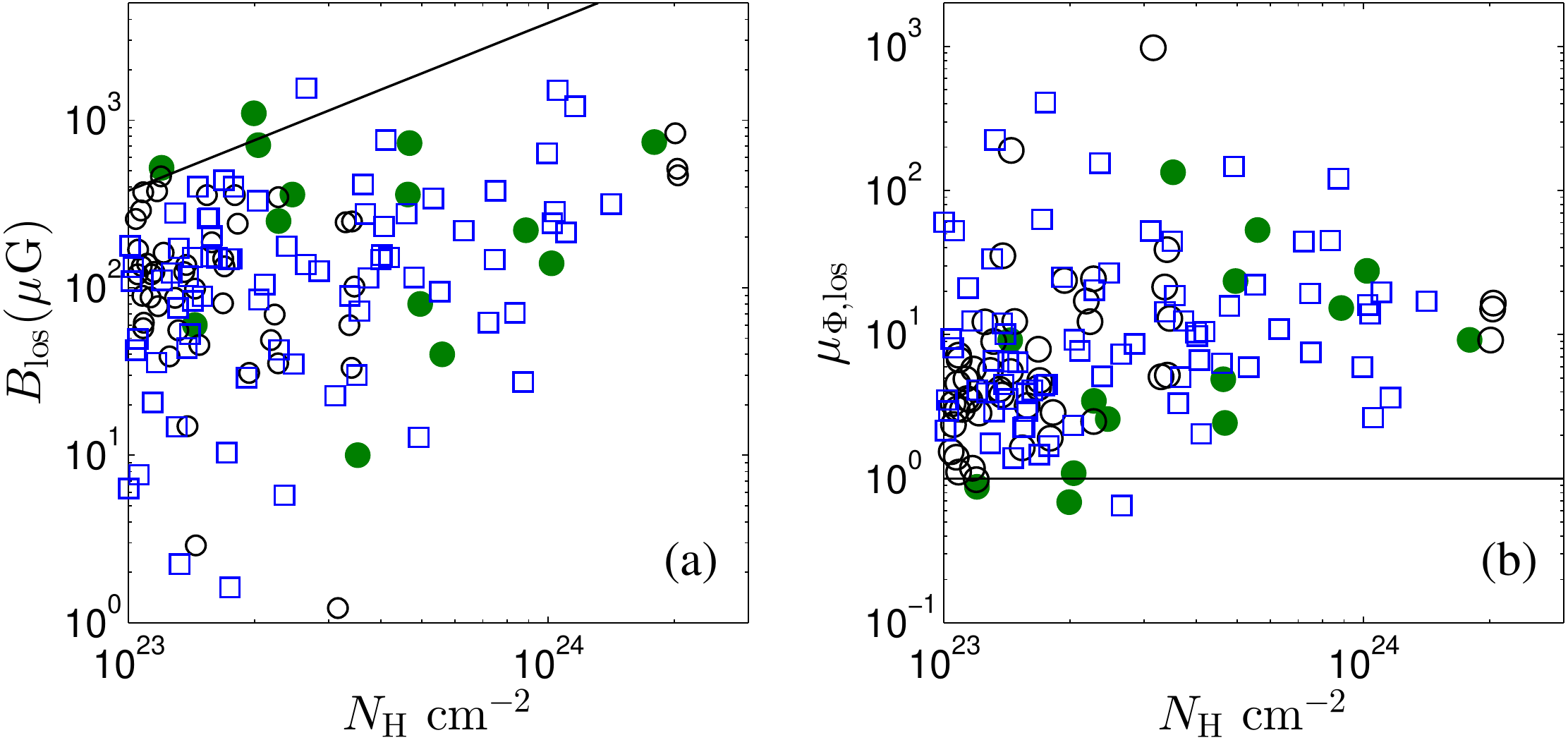}
\caption{Comparison of results from the strong-field simulation
with observations of CN clumps \citep{fal08}, which have $\NH > 1 \times 10^{23}$ cm$^{-2}$ and $\nbh > 2 \times 10^5$ cm$^{-3}$. The 14 observed CN clumps are labeled with green solid circles; 1/3 of the lines of sight from the strong-field 100-clump sample at $0.57 \tff$ are labeled with black circles; and 1/3 of the lines of sight from the weak-field 100-clump sample at $0.94 \tff$ are labeled with blue squares.
Since the beam size for the CN observations is comparable to the clump size, we have calculated the field strength
and column density by averaging over the entire clump for this figure only.  The distribution of (a) $\blosr$ and (b) $\muplos$ of clumps from both models matches well with data for the CN sources.  The straight lines in the panels indicate the magnetic critical condition.
See Section \ref{sec:cncomp} for discussion.
\label{fig9}}
\end{figure*}

A striking result of the strong-field simulation is that
the time-averaged median of the mass-to-flux ratio in the 100 most massive clumps, $\muphi(\bmidz)=1.86$, 
is somewhat greater than the initial value for the entire simulation box, $\muphio = 1.62$.
In ideal MHD, the only way in which the mass-to-flux ratio can increase beyond its initial value is if there
are field-line reversals in the plane used to calculate the flux. However, although
the time-averaged  median of $\muphi(\absbmidz)$, which excludes the field-line reversal effect inside the center of the clumps, is smaller than time-averaged median of $\muphi(\bmidz)$, it is still larger than $\muphio$.
Fully 25 per cent of the clumps have mass-to-flux ratios based on $\absbmidz$ that are at least twice the initial value.
This can occur only due to numerical effects.
These are of two types: One is purely numerical and can be mitigated by increasing the resolution.
The convergence study in the Appendix indicates that the strong-field simulation is reasonably well converged, which implies that this type of numerical effect is not important. The second type is the numerical representation of microphysical processes such
as viscosity and reconnection. For example, simulations 
of turbulence often rely on numerical viscosity to dissipate energy on small scales;
as the resolution is increased, the scale at which the energy is dissipated shrinks, but (provided there is a well-established
inertial range) the rate of dissipation does not change.
Since the increase in the mass-to-flux ratio above the initial value is converged, the numerical effect that allows this is of
the second type. A process that is known to violate flux-freezing is turbulent reconnection (see the review by
\citealp{laz15}), and our results on the increase
in $\muphi(\absbmidz)$ over its initial value are consistent with a numerically enabled version of this process.

\begin{figure*}
\includegraphics[scale=0.48]{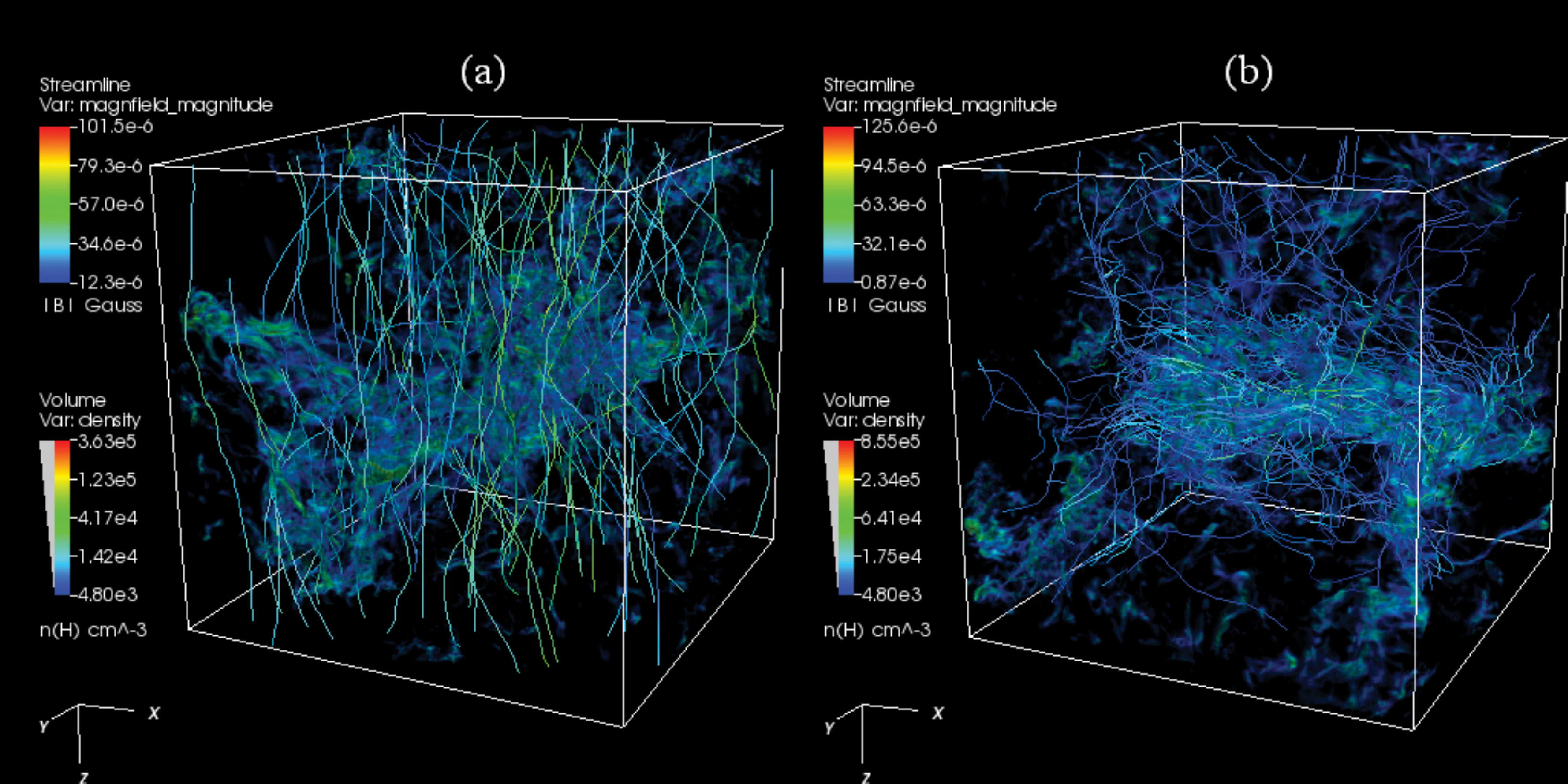}
\caption{Streamline plots of the magnetic field in the two simulations at $t=0$.  The difference is clear between the relatively ordered
magnetic field in the strong-field model (a) and the highly perturbed and stretched field in the weak-field model (b).  The highly perturbed field in the weak-field model makes it more difficult for material to flow along along the field lines.
\label{fig10}}
\end{figure*}

Finally, we note that \citet{ber12} have analyzed the spatial distribution of $\calr\equiv\mu_{\Phi,\,\rm core}/
\mu_{\Phi,\,\rm envelope}$
in simulations of turbulent cores. They found that this ratio is generally $\ga 1$,
although there is a large dispersion with a non-negligible fraction of cases with $\calr<1$. 
A value $\calr>1$ is expected in theories of
star formation that rely on ambipolar diffusion \citet{cru09}, but the simulations of \citet{ber12} were based on ideal MHD.
\citep{cru09} found $\calr\la 1$ in each of the four sources they observed; \citet{ber12} attribute the difference between
their simulations and the observed results to the small size of the observed sample. Our simulations have a higher
resolution than those of \citet{ber12} ($512^3$ base grid plus two levels of refinement vs. $256^3$) and had a longer
driving period (2 crossing times vs. 1 crossing time), but we find a similar result: the value of $\muphi$ for
entire clumps (cores plus envelopes)
is, on average, greater than that for the central beam. Since our results are consistent with those
of \citet{ber12}, we have not carried out a detailed analysis of the spatial dependence of the mass-to-flux ratio.

\subsection{Comparison with CN Zeeman Observations}
\label{sec:cncomp}

Whereas the OH observations come from gas with densities less than that of the clumps in our 100-clump sample, the CN observations \citep{fal08} are from gas with densities that lie within the simulated range, so we consider these sources separately. In fact, only 16
of our 100-clump sample have both a density and a column density that is in the observed range for CN clumps 
($\nbh>2\times 10^5$~cm\eee, $\Nh>10^{23}$~cm\ee), so we create subsamples with these properties.
In Fig. \ref{fig9}, we plot the line-of-sight quantities $\blos$ and $\muplos$ versus $\NH$ for these subsamples together with the results for the 14 observed CN sources.  Since the median beam size for the CN sources is comparable to the size of the sources, we compare the observed values of $\blos$ and $\muplos$ with simulated values from the whole clump instead of the smaller central-beam region.  We can see that the simulated clumps are basically in the same region occupied by the CN sources, with a few more clumps from the weak field model lying on the smaller $\blos$ part of Fig. \ref{fig9}a and larger $\muplos$ part of Fig. \ref{fig9}b compared to the strong field model.  Again, both models reproduce the location of the observed CN sources in Fig. \ref{fig9}.
There is one CN source, out of the 14 observed, that is sub-critical based on the observed value of $\blos$, and two others that are potentially sub-critical since $\blos$ is on or near the critical lines in Fig. \ref{fig9}.  One of these two CN sources, S106OH, would be subcritical if the line-of-sight field is less than 1 sigma below the measured value.  The other one, G10.6, would be subcritical if the actual mean field direction makes an angle larger than $24\degr$  from the line of sight. Interestingly, it is the weak-field model that has
a sub-critical clump and more clumps close to the critical line.  In general,
the distribution of $\blosr$ versus $\NH$ of the high density clumps from our simulation matches well with that of the CN sources.  Future observation using ALMA will be able to increase the number of CN sources for a better statistical comparison with our simulation.

\section{Is the Mean Field Strong or Weak?}
\label{sec:is}

\subsection{Comparison of the Strong- and Weak-Field Models}
\label{sec:compsw}

The global properties of the strong- and weak-field simulations are shown in Table \ref{tab:3}.
Initially the uniform magnetic fields differ in strength by an order of magnitude. After 
turbulent driving for two crossing times, 
the magnetic field in the weak-field model is enhanced enough to
increase $\brmsv$ by more than a factor 5, so that the \alfven Mach number drops from $\ma=10$ to $\ma=1.86$.
By contrast, in the strong-field simulation, $\ma$ drops by only about 15 per cent. These results are portrayed graphically in
Fig. \ref{fig10}, which shows the field lines of the magnetic field inside the turbulent boxes.  The magnetic field is roughly along the mean field direction in the strong-field model (Fig. \ref{fig10}a), but it is highly tangled and stretched in the weak-field model (Fig. \ref{fig10}b).   Initially,
the values of $\brmsmu/\nbhf^{2/3}$ of the two systems  differ by a factor of 10, but
after two crossing times,
the difference is less than a factor of 2.  At the end of the simulations, the rms magnetic field has been enhanced by a few  per cent (strong field) or slightly more than 10 per cent (weak field)
due to gravitational contraction.

\begin{figure}
\includegraphics[scale=0.37,angle=-90]{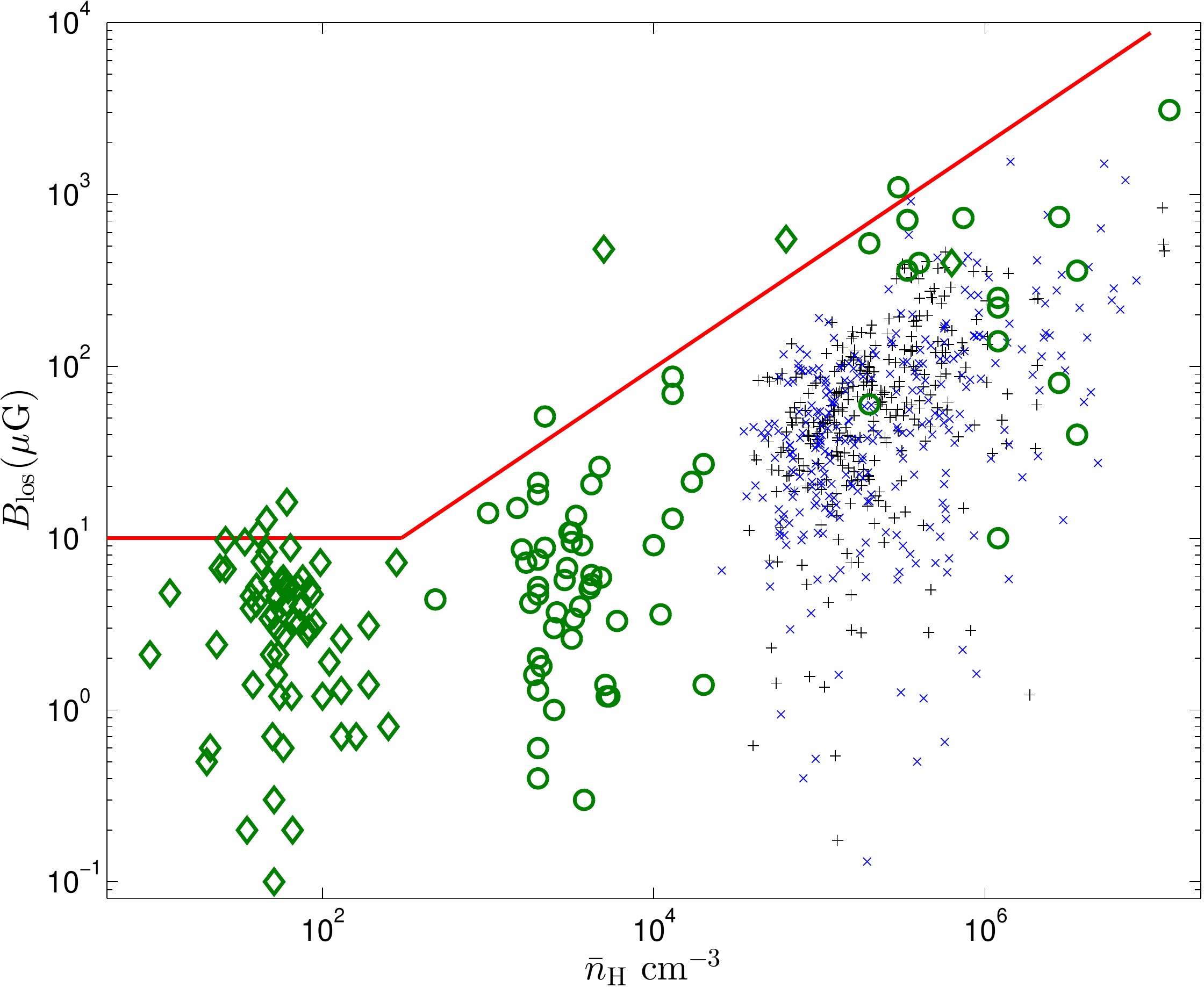}
\caption{Results of the weak-field model at 0.94 $t_{\rm ff}$  (blue crosses) are overplotted on the data in Fig. \ref{fig3}.  Both models produce clumps in the same region of the $\blos$ vs $\nbh$ diagram,
but the median value of $\blos$ in the 
weak-field model is 80 per cent of that in the strong-field model.
\label{fig11}}
\end{figure}

Next, consider the results for the 100-clump samples. As shown in Table \ref{tab:4}
the magnetic-field properties of the clumps in the strong-field model provide a good match with observation.  
One might therefore expect the weak-field model to have results that clearly differ from observation, since it has an initial field strength that is a tenth of that in the strong-field model and, correspondingly, a plasma $\beta$ that is a hundred times that of the strong-field model (Table \ref{tab:3}).  We identify the 100-clump sample in the weak-field model in the same manner as in the strong-field model, 
with the same density threshold of $\nbh = 4 \times 10^4$ cm$^{-3}$.
In Fig. \ref{fig11}, we repeat Fig. \ref{fig3} and then include the
data points from the 100-clump weak-field sample at $0.94 \tff$.  The data points
from the two models overlap substantially and are not readily distinguished from each other.

The time-averaged value of the power-law index $\alpha(\btotr)$
between $\btotr$ and $\nbh$ for the weak-field model is $0.57\pm0.05$, 
significantly less than the
value inferred from observation, $0.65_{-0.06}^{+0.10}$. 
The strong-field model has a significantly larger value, $\alpha(\btotr) =0.70\pm 0.06$, but it agrees with the observed
value within the errors.  
One would have expected the weak-field model to have a density dependence closer to the value of 2/3 predicted
by \citet{mes66} for weak fields. However, as we have noted above, it is only the mean field that is weak in 
the weak-field model; turbulence has amplified the small scale field so that it is comparable to that in the strong-field
model. Furthermore, the tangling of the field can be amplified as the clumps contract under the influence of gravity.
We have confirmed this picture by calculating $\avg{\alpha(|\blos|)}_t$ for the {\it magnitude} of the line-of-sight field. For the
strong-field case, this is $0.67\pm0.04$, which is consistent with the value for the observable line-of-sight field
given in Table \ref{tab:4}, $\avg{\alpha(\blos)}_t=0.63\pm0.07$. By contrast, for the weak-field case, we find
that $\avg{\alpha(|\blos|)}_t=0.64\pm0.03$ for the magnitude of the line-of-sight field, whereas Table \ref{tab:4}
shows that $\avg{\alpha(\blos)}_t=0.58\pm0.05$. We conclude that the magnitude of the field increases with density
at about the same rate for both the strong and weak-field cases (consistent with the observed value,
$\alpha(\blos)=0.64\pm0.13$), but that the rate of increase for the observable line-of-sight field, $\blos$, is less, just
as it is for $\btot$. The increased tangling of the field with compression in the weak-field case prevents the
observable field from increasing at the expected rate of $\alpha=2/3$.

\begin{figure}
\includegraphics[scale=0.37,angle=-90]{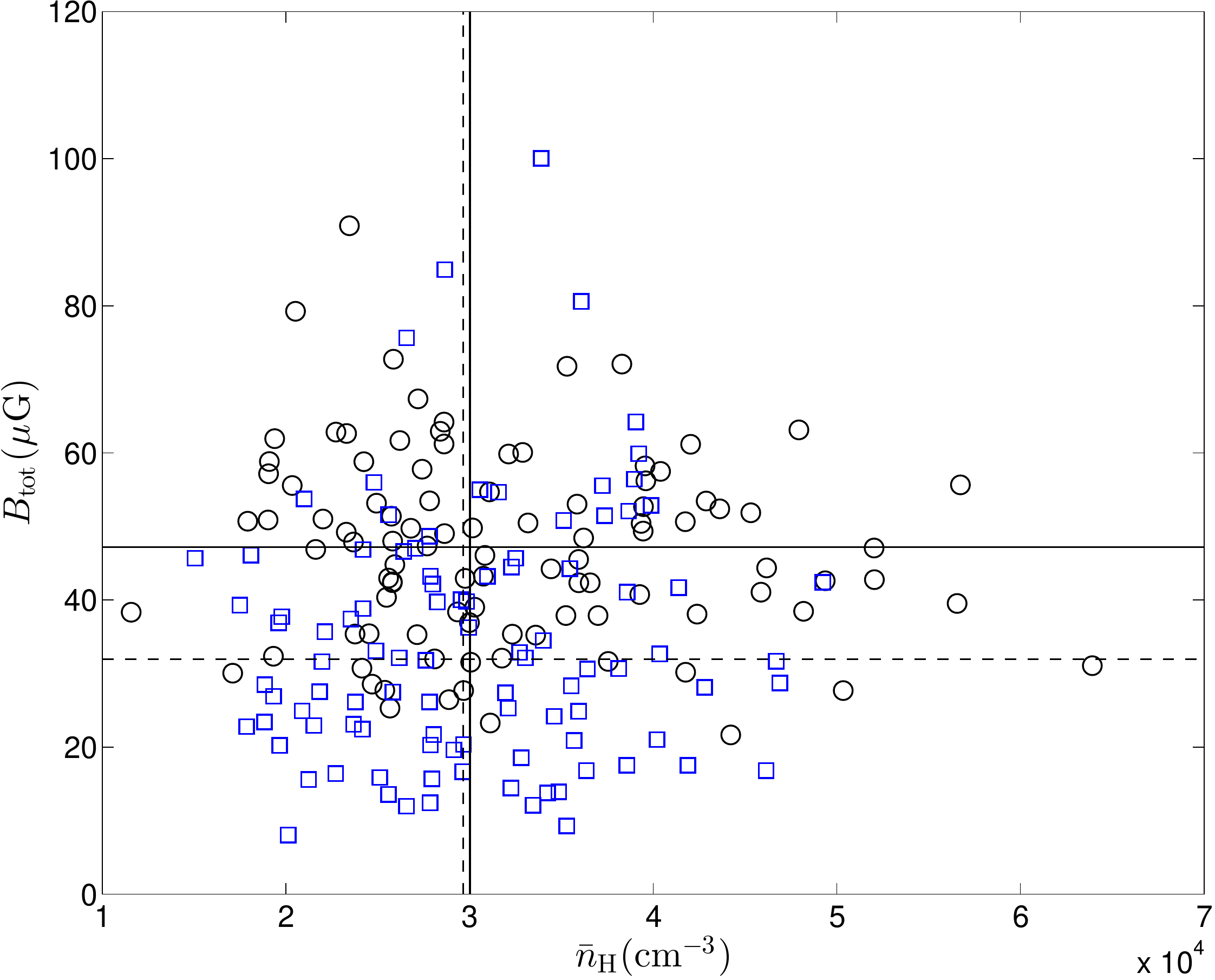}
\caption{Distribution of $\btotr$ vs $\nbh$ in a 0.025 pc radius Gaussian beam for the 100-clump samples from the strong-field (black circles) and the weak-field (blue squares) models at $t = 0$.  The median values of $\btotr$ and $\nbh$ of the strong-field sample (solid lines) and the weak-field sample (dashed lines) are also shown.  
The clumps from the weak-field model occupy the low-$\btotr$ region of the plot.  
\label{fig12}}
\end{figure}

Table \ref{tab:4} also gives
the means and medians of $\btotmu/\nbhf^{0.65}$ and $\blosmu/\nbhf^{0.65}$ for the weak-field sample.
These quantities are plotted
in Fig. \ref{fig6}d-e
as functions of time during the gravitational collapse phase; they are relatively constant
for $t>0.5\tff$. 
The time-averaged means and medians of these quantities agree with the strong-field values
to within about 25 per cent, even though the initial fields differ by a factor 10.
The mean and median values of $\muplos$ for the weak-field model are plotted as a function of time in Fig. \ref{fig6}f; they
are larger than those in the strong-field model by about a factor of 1.4, and 
after $t=0.5\tff$ they increase only slightly over the remainder of the simulation.
In Fig. \ref{fig12},
we plot $\nbh$ and $\btotr$ of the 100-clump samples from the two models at $t = 0$.
The mean magnetic field in the weak-field clumps, $\btotrm$, is about 70 per cent of $\btotrm$ of the strong-field model clumps, and the median densities are even closer.

Finally, we compare the results for the 100-clump samples with the results for the entire system;
in this case the field 
in the clumps is volume-averaged rather than mass-weighted, 
since it is the volume-averaged field that
is relevant for clump structure.
As we discussed above, the rms field changes in the turbulent driving phase much more for the weak-field case than
the strong-field one. Including the subsequent phase in which self-gravity is operative,
we see from Table \ref{tab:3} that from the initial uniform field to the end of the simulations,
$\brmsv$ of the whole system increases by a factor of 1.2 in the strong-field case and about a factor of 6 in 
the weak-field model, mainly due to enhancement of the field by supersonic turbulence.  
On the other hand, the median values of
$\brmsv$ in clumps in the 100-clump sample are about 2.7 and 4.6 times the values of $\brmsv$ for the entire turbulent box for the strong- and weak-field models, respectively, at the end of the simulations.  The higher values of $\brmsv$ in the clumps compared to that in the entire simulation box is simply the result of further enhancement of the field by the gravitational collapse of dense clumps.

\begin{figure}
\includegraphics[scale=0.34]{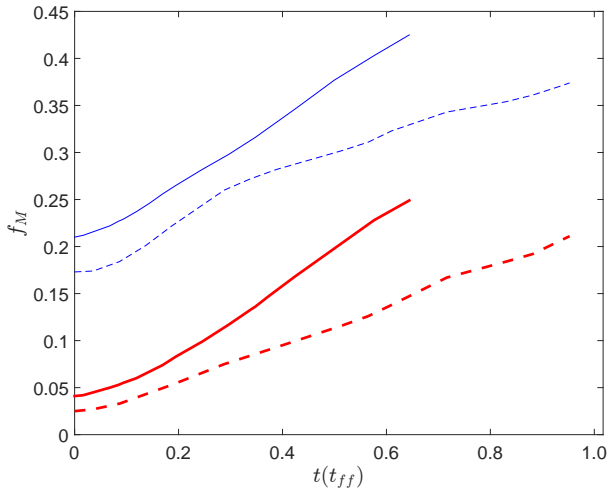}
\caption{The mass fraction, $f_M$, of gas with density $\nh \ge 1 \times 10^4$ cm$^{-3}$ (thin blue lines) and  $\nh \ge 4 \times 10^4$ cm$^{-3}$ (thick red lines) of the strong (solid line) and weak (dashed line) mean field simulations as a function of time.  Material can be accreted onto dense clumps more easily with the ordered magnetic field in the strong-field model than with the highly perturbed field in the weak-field model (see Fig. \ref{fig10}).  The difference in accretion rates onto the dense regions of the two models is about a factor of 1.5.
\label{fig13}}
\end{figure}

The remarkable similarity in these results of the strong-and weak-field simulations 
is due to the greater enhancement of the magnetic field in the weak-field model as the result of both easier compression and greater stretching of the field lines.  
Correspondingly, the turbulent \alfven Mach number in the weak-field model has dropped from 10 to 1.61, not much larger than the the final value in the strong-field case, 0.83, at the end of the simulations. 
Similar results have been found previously \citep[e.g.][]{lun08,pad10}.
However, there are two important differences between the two models. Prior to the time that
gravity is turned on, the median mass-to-flux ratio does not change significantly in the strong-field case,
but it drops by more than a factor 15 in the weak-field case. 
This is consistent with turbulent fragmentation along flux tubes: the median field in the weak-field simulation
has increased by a factor 6.7 at $t=0$, and the median surface density in the 100-clump sample is $0.58\Sigma_0$. 
As a result, $\muphi\propto \Sigma/B$ should typically drop by a factor
of about 11.6.  This is consistent with the results in Table \ref{tab:5}, which gives
$\muphio/\muphi(\bmidz)_{\rm med} = 15.9$, about 1/3 larger than our simple estimate. Note that by
using the mean field in the box rather than in the clumps in our estimate, we implicitly assumed that the
clumps formed primarily by flow along field lines rather than by compression. The agreement with the
simulation results shows that this is approximately correct.

We can develop a simple picture of the turbulent fragmentation model as follows.
In the strong-field case, 
the fact that the mass-to-flux ratio of the most massive
clumps (i.e., the ones in our 100-clump sample) is typically about the same as the initial value implies
that these clumps form by accreting material along a flux tube over
a distance comparable to the size of the box. 
What about the weak-field case?
In this case, the field is stretched and
compressed. So long as flux-freezing holds, we can define flux tubes with dimensions that evolve smoothly
in time. Since the volume in the simulation is fixed, 
the average volume of a flux tube must be constant. Consider a flux tube in the initial uniform field, with a length
$\ell_0$ equal to the box size, an area normal to the field $A_{\ft,0}$, and a volume $V_{\ft,0}=A_{\ft,0} \ell_{\ft,0}$.
After turbulence begins, the average area of a flux tube
shrinks by the same factor as the length increases, $A_{\ft}=V_{\ft,0}/\ell_{\ft}$.
Conservation of magnetic flux then implies $B=B_0(A_{\ft,0}/A_\ft)=B_0(\ell_\ft/\ell_0)$: the field
increases in direct proportion to the length of the flux tube. Mass conservation gives a similar
result, $\Sigma=\Sigma_0(\ell_\ft/\ell_0)$, so that the mass-to-flux ratio $\Sigma/B$ is constant in time.
Fragmentation along the flux tube reduces the mass and surface density by a factor $\ell_\frag/\ell_\ft$,
where $\ell_\frag$ is the length from which the fragment mass is drawn, but fragmentation does not affect the field.
As a result, the mass-to-flux ratio in the fragment is $(\ell_\frag/\ell_\ft)\mu_{\Phi,0}$. For example,
the results of the strong-field simulation discussed above show that the typical flux tube was stretched by
a factor 6.7, increasing $B$ and $\Sigma$ by the same factor, while leaving the mass-to-flux unchanged.
Typical fragments were drawn from a length $\ell_\frag=0.58\ell_0$. The typical mass-to-flux ratio was
then reduced by a factor $\ell_\frag/\ell_\ft=0.58/6.7=1/11.6$.

Once gravity is turned on, a second important difference between the strong- and weak-field cases emerges:
Material accreting onto the clumps
is slower in the weak-field case than in the strong-field one because of strong magnetic field tangling.
Fig. \ref{fig13} shows that the rate of gravitational collapse, as measured by the rate of increase of the mass fractions above densities $\nh \ge 10^4$ and $\ge 4 \times 10^4$ cm$^{-3}$, is about 1.5 times greater in the strong-field case than in the weak-field one. 
(Note that the factor $\sim 1.5$ difference between the rate of increase of dense gas in the strong-field model compared to the weak-field one  does not mean that the star formation efficiencies must differ by a factor 1.5 since the star formation rate 
is affected by feedback processes that we have not considered in this work.)
We interpret this counter-intuitive result as being due to the much greater twisting and stretching of the field in the weak-field case (Fig. \ref{fig10}), which makes it more difficult for material to flow along the field and condense onto dense regions. At $t=0$, before gravity has any effect, the mass fraction of dense gas, $f_M$, in the strong-field model is already higher than that in the weak-field model. Since the turbulent driving is the same in the two models, the only differences in the models are the strength and tangling of the field. In the strong-field model,
compression along relatively straight field lines can create more dense regions than in the weak-field model, in which the more isotropic magnetic forces from the highly tangled field better resist the formation of dense regions. During gravitational collapse, the rate of increase of $f_M$ increases slightly in the strong-field model but remains about the same or even decreases slightly in the weak-field model. 
The median values of the ratio of the infall velocity to the free-fall velocity (Equation \ref{eq:vinvff}) of the 100 clumps in the weak-field model at 0.6 and $0.94 \tff$ are 0.17 and 0.15, respectively, about half the value in the strong-field model (0.29), a direct indication of slower material accretion onto the clumps. We compute the 
degree to which the Lorentz force, $\vecF_\Lor$, in each clump resists self gravity by summing up the component of the force opposite to the direction of self-gravitational acceleration, $-\vecF_\Lor\cdot\hat\vecg'$, cell-by-cell, where $\hat\vecg'$ is the unit vector associated
with the gravitational acceleration due to the clump. Normalizing this to the total gravitational force, we evaluate
\beq
\frac{\int_{\rm clump}\vecF_\Lor\cdot\hat\vecg' dV}{\int_{\rm clump} \rho\vecg'\cdot\hat\vecg' dV}
\eeq
for each clump.
We find that the mean of this quantity for the 100-clump sample of the weak-field model is 
1.35 times that of the strong-field model. This is a quantitative manifestation of how the greater tangling of the field in clumps can inhibit gravitational collapse in the weak-field model.

Table \ref{tab:5} lists the harmonic means and medians of $\muphi$ for the two models.  At $t = 0$, 
after the initial turbulent driving, the median $\muphi$ of the weak-field model is about 1.2 times the value of the strong-field model, instead of the factor of 10 from the initial conditions.  At that time, more than half the clumps in the strong-field sample are sub-critical, whereas the majority of the clumps in the weak-field model are marginally supercritical. These results demonstrate that turbulent fragmentation is effective in reducing $\muphi$ in clumps, particularly in the weak-field case.

\begin{figure*}
\includegraphics[scale=0.7,angle=-90]{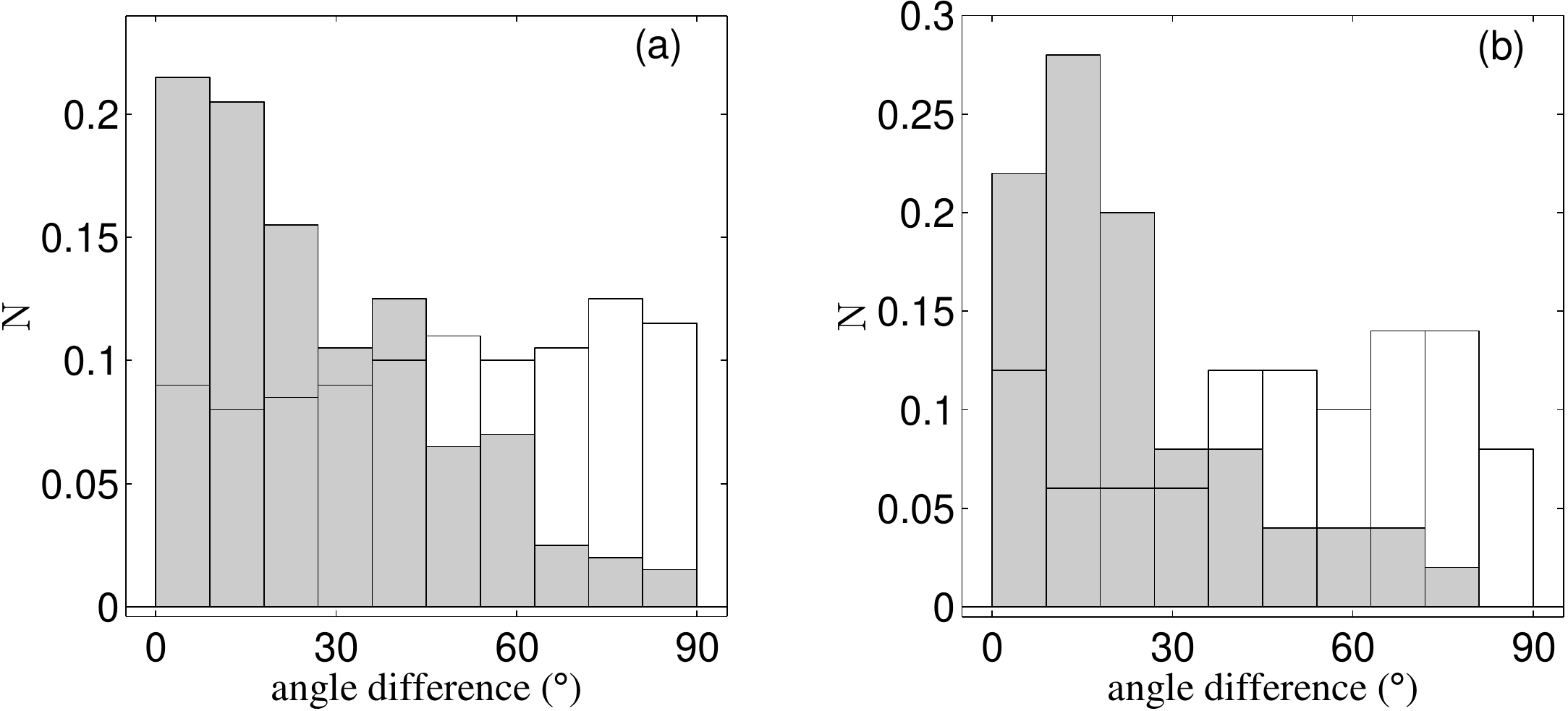}
\caption{Orientation of the projected 3D, mass-weighted magnetic field, $\btot$, relative to the global mean field in the simulations.  (a) The PDF of field orientations from the 100-clump sample in the strong-field (shaded histogram) and weak-field (open histogram) simulations.  For the strong-field simulation, 54 per cent of the clumps have orientations within $26\degr$ of the mean and 81 per cent have orientations within $45\degr$.  The PDF for the weak-field simulation is consistent with a uniform distribution.
(b) Same as (a) but only for the largest 25 clumps with $\nbh > n_{\rm crit}$.  For the strong-field simulation, 70 per cent of the clumps have fields oriented within $26\degr$ of the mean, and 86 per cent are oriented within $45\degr$, close to the observational results of \citet{li09}.  See discussion in Section \ref{sec:bdirection}.
\label{fig14}}
\end{figure*}

\subsection{Magnetic Field Orientation}
\label{sec:bdirection}

\begin{table}
\caption{Large-Scale Orientation of Magnetic Field
\label{tab:6}}
\begin{tabular}{lcc}
\vspace{-0.4cm}\\
\hline
\hline
\vspace{-0.4cm}\\
                & ~~\% $< 26\degr$~~    & ~~\% $< 45\degr$~~\\[0.5ex]
\hline
\vspace{-0.4cm}\\
Observed~~~~~~~~~~&68&84\\
             \vspace{-0.4cm}\\
             \hline
             \vspace{-0.4cm}\\      
Strong-field simulation&&\\
~~~100 clumps &   54 (56)\fnm[1] &  81 (80) \\
~~~25 clumps  &   70 (67) &  86 (91) \\
             \vspace{-0.4cm}\\
             \hline
             \vspace{-0.4cm}\\
Weak-field simulation&&\\
~~~100 clumps &  24 (24) &  45 (42)  \\
~~~25 clumps  &  22 (35) &  42 (51)\\
  \vspace{-0.4cm}\\
             \hline
             \vspace{-0.4cm}\\
Random&29&50\\
         \vspace{-0.4cm}\\
\hline
\hline
\end{tabular}

\fnt{1} {$^a$ Numbers inside brackets are the proportions of magnetic field orientation obtained by integrating along the line of sight across the whole turbulent box assuming that it is optically thin.  See Section \ref{sec:bdirection} for discussion.}
\end{table}

Recent polarimetric investigations of the orientation of magnetic fields in cloud clumps on scales
$\ga 0.04$~pc \citep{li09,li11} show that it is not random, but instead is aligned with the global magnetic field of the surrounding intercloud medium (ICM, the medium between molecular clouds)
on scales of 100-200 pc. \citet{li09} found that almost 
70 per cent of the 25 clumps they observed have magnetic fields oriented within $26\degr$ of the ICM field, and that
84 per cent have fields oriented within $45\degr$ of the ICM field. 
As shown in Table \ref{tab:6}, these percentages are much larger than expected by chance, and \citet{li09}
estimate that the probability that the $45\degr$ alignment occurs by chance is less than $7\times 10^{-5}$.  Comparing with several ideal MHD simulations, they concluded that only sub-Alfvenic models--i.e., models in which the magnetic field is dynamically significant in molecular clouds--can have such a significant correlation between the field orientation in molecular clumps and that in the ambient ICM. 

In Fig. \ref{fig14}, we plot the distributions of the magnetic field orientation with respect to the global mean field of the turbulent box for the two simulations. The global mean field is equivalent to the ICM field since it represents the field orientation of the material out of which the cloud was formed. Because the direction
of the field in the plane of the sky is unknown, the magnitude of the relative orientation is in the range $0\degr -90\degr$. We plot the 200 relative orientations
based on observations of the 100 clumps along the $x$- and $y$-axes, normal to the global mean field.
In Fig. \ref{fig14}a, we plot the angle differences of the 100-clump samples of the strong- and weak-field models.  The histogram is normalized to unity for direct comparison.
As shown in Table \ref{tab:6}, the strong-field simulation shows alignments much greater than expected from
chance, whereas the weak-field simulation shows no alignment.
In fact, a KS test shows that the weak-field results are consistent with a random distribution of orientations with 
$p=0.31$.
We can change the viewing direction so that it is not along the axes normal to the mean field, but this changes the percentage of clumps with alignments less than $26\degr$ and $45\degr$ by less than 3 per cent.  

Most of the clumps in the 25-clump sample in \citet{li09} have sizes $\ge 0.3$ pc, much larger than the clumps in our simulations. If we focus on the largest 25 clumps in the strong-field simulation (Fig. \ref{fig14}b), 
which have sizes of  0.1-0.13 pc, we find much better agreement with the results of \citet{li09}. 
Higher resolution observations in the future will be able to probe the magnetic field orientation closer to the center of the cloud clumps and determine whether fields on those scales are less aligned with the ICM fields.
An initial step in this direction has been taken by \citet{hul13}, who find alignment within $45\degr$
between angular scales of
$2.5\arcsec$ and $20\arcsec$ of 70 per cent. This is less than that found by \citet{li09}, but the errors in the
measurement of the orientations are large enough that this difference is not statistically significant.

The above analysis includes only the material that we identified as being within a clump. Observers do not
have this luxury; they must integrate along the entire line of sight. As an approximation to what observers do,
we have obtained the mass-weighted field orientation by
integrating through the turbulent box over the projected area of each clump in both the $x$ and $y$ directions.
Clumps with mean column densities less than half the total are not included. This eliminates 76 of the 200 clump
observations in the strong-field model.
There are two other differences between our method and that of observers: We did not attempt to use velocity information to 
separate out the gas that is associated with an individual clump; and we integrated only through the turbulent box, not along
the entire line of sight through the clump.
The results for the proportions of clumps with magnetic field orientation within $26\degr$ and $45\degr$ of the mean-field direction are listed in Table \ref{tab:6} inside the parentheses.  For the strong-field model, the new proportions are
similar to those when the orientation of the magnetic field is computed only over the clumps;
for the 25-clump sample, the proportion of sources varies by 3-5 per cent.  
For the weak-field simulation, the alignment increases somewhat in the 25-clump sample, but in all cases remains consistent
with a random orientation. 
Based on these results, one would expect that the line-of-sight field along the mean-field direction is larger than that normal to that direction for the strong-field case but not for the weak-field case, and this is indeed the case:
The time-averaged 
median value of $B_{\rm los,\parallel}/B_{\rm los,\perp}=(1.86,\,0.98)$ for the two cases, respectively; the weak-field case is
consistent with isotropy.

We conclude that the weak-field simulation is inconsistent with the observations of the alignment of
fields on large and small scales by \citet{li09}.
The strong-field simulation has $\ma=1$ and exhibits slightly less 
alignment than observed, consistent with
the motions in observed molecular clouds being slightly sub-\alfvenic, as suggested by \citet{li09}.
It is also possible that in the strong-field case, ambipolar diffusion could bring the small-scale field into closer alignment with the large-scale one, allowing a moderate increase in $\ma$.  However, 
ambipolar diffusion would be ineffective in aligning the field in the
weak-field case since ambipolar diffusion is ineffective in weak fields (e.g., \citealp{mck10}).

\section{Conclusions}
\label{sec:conclusion}

The measurement of interstellar magnetic fields is very challenging: In the data compiled by \citet{cru10},
there were fewer than 140 interstellar clouds with line-of-sight fields determined by Zeeman splitting,
and among the molecular sources (about half the sample), only about 40 per cent had fields detected with a significance greater than 2 sigma. \citet{cru10} carried
out a sophisticated Bayesian analysis of this data, including the sources from which only upper limits are available, in order to infer the properties of the 3D fields. Their principal conclusions were (1) the maximum total field strength at a density $\nbh>300$~cm\eee\ (almost entirely molecular in their sample) 
satisfies 
$\btm\propto\nbh^{\alpha}$ with $\alpha=0.65_{-0.06}^{+0.10}$, where we have estimated the errors from their Figure 4;
(2) the distribution of field strengths is
approximately uniform from very low values to $\btm$; and (3) virtually all of the measured cloud clumps are 
magnetically supercritical, consistent with previous results (see Section \ref{sec:obs}).  Thus nature appears to favor the formation of magnetically supercritical clumps, which is a necessary condition for star formation to be possible.  

\subsection{Data Analysis}

We began by analyzing the data in a manner complementary to that of \citet{cru10}: rather than attempting to infer
the 3D properties of the fields, we analyzed the line-of-sight fields directly. As noted above, a majority of the molecular
sources have only upper limits on their field strengths; by using the measured values, we are effectively stacking low signal-to-noise data in order to draw conclusions. This is valid only if the errors in the data are random and not systematic, but this is a prerequisite for the Bayesian analysis as well. It should be noted that the fields determined by Zeeman
observation are mass-weighted (Eq. \ref{eq:blos}), whereas those used in theoretical models are volume-weighted.

\begin{itemize}

\item[1.] {\it Density dependence.} \citet{cru10} found that the total field in molecular gas has a 
density-dependent maximum value, $\btm=98\nbhf^{0.65}~\mmug$ (Eq. \ref{eq:btm1}). In analyzing
their entire data set, including the HI data, they concluded that the distribution of field strengths was approximately
uniform over the range $f\btm-\btm$, with $f\simeq 0.03$. 
This implies that the typical field strength also scales as
$\nbh^{0.65}$. We tested this directly by showing that $\blos\propto\nbh^{0.64\pm0.13}$
(uncertainty at the 95 per cent confidence level),
which implies that the total field scales this way as well.
Since projection of the field onto the line of sight adds considerable dispersion, the dispersion in $\btot(\nbh)$ from the simulations
is considerably less than that in $\blos(\nbh)$.
In Appendix \ref{app:sample} we show that the density dependence of the magnetic field is most accurately
inferred from the line-of-sight data by including all the data, unweighted by its accuracy.

\item[2.] {\it The normalized field strength}. The normalization of the relation between the field and the density is also
significant. Since the average total field is twice the average line-of-sight field, $\avg{\btot}=2\avg{\blos}$ 
(Sec. \ref{sec:analysis}), we were able to infer the average value of the magnetic field in the observed molecular clumps
as a function of the density in the clumps directly from observation,
\beq
\avg{\btot}\approx 42~\nbhf^{0.65}~~~\mmug.
\eeq
Insofar as the molecular clumps in the sample of \citet{cru10} are representative, this result describes the average field 
in the molecular interstellar medium.

\item[3.] {\it Distribution of normalized field strengths}: 
By comparing the observations of the line-of-sight fields directly with the projected PDFs
for uniform and log-normal field distributions
(Appendix \ref{app:proj})
we concluded that the log-normal distribution is slightly better, but the difference is not statistically significant.  The primary result of \citet{cru10} is that the normalized total field strength $\btot/\nbh^{0.65}$ must have an intrinsically broad distribution;
their data is consistent with a uniform distribution, but other distributions are possible (Crutcher, private communication).
The principal difference between the uniform and log normal distributions 
is in the fraction of very low and very high field strengths: Whereas
a uniform distribution has about 10 per cent of the clumps with fields less than 20 per cent of the median, a log normal
has only 2.6 per cent in that range, and whereas a uniform distribution has no clumps with fields greater than twice the median,
a log normal has 20 per cent in that range.

\end{itemize}

\subsection{Results of the Simulations}

To confront theory with observation, 
we have performed two large scale MHD simulations to study the formation of massive, dense infrared dark clouds (IRDCs) (Li et al. 2015, in preparation).  
We drove Mach 10 turbulence for two-crossing 
times with no gravity; we then turned gravity on, and continued driving the turbulence. For the standard linewidth-size relation in molecular gas, the size of the region simulated is 4.55 pc (Section \ref{sec:sim}). Here we have used the data from the gravitationally collapsing phase of these simulations to study the  properties of the magnetic fields in the 100 most massive clumps identified by CLUMPFIND in each simulation and have compared the results with those of \citet{cru10}.  Our two simulations have the same initial conditions, except that the magnetic field in the weak-field model is 10 times weaker  than that in the strong-field model; the initial \alfven Mach numbers in the two simulations were $\mao=10$ (weak field) and $\mao=1$ (strong field).  The \alfven Mach number observed in molecular clouds is typically of order unity \citep{cru99}; by the end of our simulations, $\ma$ was about 0.8 and 1.6 in the strong and weak-field simulations, respectively. 

The differences between the observations and our simulations must be borne in mind. The observed data are very heterogeneous, coming from different molecular clouds with different initial conditions, size scales, and different ages, whereas we have carried out only two simulations that differ only in the strength of the initial magnetic field. Various telescopes were used to obtain the observed data, whereas we used a single beam size of 0.025 pc to `observe' our 100-clump sample, corresponding to the minimum beam size used to observe the 68 OH+CN sources in the sample of \citet{cru10}.  The observed clouds have masses up to about $1700 M_\odot$, whereas the most massive cloud in the simulations has a mass of about $78 M_\odot$.
This difference in masses is due both to the limited size of the simulated region and to the fact that the resolution of our simulated beam is greater than that used for the observations of the most massive clouds.   However, our 100-clump samples have clumps with densities comparable to some OH sources and to all the CN sources. Despite these many differences, our simulations are in remarkably good agreement with the observations.

\begin{itemize}
\item[1.] {\it Critical density}.
In \citet{cru10}, the HI sources appear to be in a horizontal branch with $\blosr$ independent of $\nbh$; the power-law relation between $\btm$ and $\nbh$ sets in above a density $n_0\simeq 300$~cm\eee, which is close to the critical density from a balance between turbulent and thermal pressure, as discussed in Section \ref{sec:100clump}.  The clumps from our two simulations also have a relation between $\blosr$ and $\nH$ 
with horizontal branch at low density and a power-law branch at high density; as for the observed fields,
the power-law branch begins at about the density at which the thermal pressure in the clumps balances the 
average turbulent pressure.
When gravity is first turned on, there is only a horizontal branch; the 
power-law branch is created solely as the result of gravitational collapse of clumps.

\item[2.] {\it Density dependence,} $\btot\propto \nbh^\alpha$: The time-averaged values of $\alpha$ from our two simulations are $0.70\pm0.06$ and $0.57\pm0.05$ 
(uncertainties at the 95 per cent confidence level)
for the strong and weak-field models, respectively.
By comparison, the value inferred from observation by \citet{cru10} is
$0.65^{+0.10}_{-0.06}$ (these uncertainties are our estimates based on the results presented in \citet{cru10}).
The density scaling of the magnetic field is due to self-gravity since the fields were independent of density
prior to turning gravity on. As pointed out by \citet{cru10}, the observed value of $\alpha$ is very close to the value of $\frac 23$ predicted by \citet{mes66} for spherical contraction of a cloud with a weak field. 
We have confirmed that the contraction of simulated clumps is approximately spherical.
However, although the field in almost all the clumps is magnetically supercritical, the field is not very weak: In
the strong-field case it preserves the memory of the initial field orientation, and in the weak-field case it is stronger relative
to gravity in the clumps than in the strong-field case. The $B-\nbh$ relation appears to be an emergent property of turbulent, magnetized,
self-gravitating media, and a full understanding must await further research. We have also shown
that tangling of the field increases with density in the weak-field case, causing
$\alpha$ to be smaller for the weak-field simulation than for the strong-field one.
As discussed in Appendix \ref{app:sample}, a systematic variation of noise with density can affect the inferred value of
$\alpha$. For the case at hand, in which the low-density, OH data is noisier than the high-density, CN data, the
inferred value of $\alpha$ is smaller than the actual value by about 0.02. This effect is much stronger if data less than 2 sigma is omitted from the sample; in that case, the inferred value of $\alpha$ is smaller than the actual value by 0.06.

\item[3.] {\it Mean and median values of}
$\blos/\nbh^{0.65}$ in the simulations are about half and 75 per cent, respectively, of
the observed values (Table \ref{tab:4}). Given the many differences between the observed and simulated clumps, this is good agreement.  The mean and median values remain approximately constant throughout the self-gravity phase of the simulations. 

\item[4.] {\it Scaling of the magnetic field with velocity dispersion, $\sigma$}:
As noted in Section \ref{sec:den},
\citet{mou91} suggested that $B$ scales with both the density and the velocity dispersion, $B\propto \bar\rho^{1/2}\sigma$,
and \citet{bas00} showed that this is in excellent agreement with observation. 
We have confirmed that the data of \citet{cru10} are consistent with this scaling, although
the density power law has a significant uncertainty. \citet{bas00} derived this
relation under the assumption that clouds are planar and concluded that the agreement with 
observation supported this hypothesis. In Appendix \ref{app:theory} we show that this relation comes from an identity and
applies to clouds of a wide range of shapes (including spherical) that are gravitationally bound and are close to being magnetically
critical.

\item[5.] {\it Distribution of normalized line-of-sight field strengths.} 
A K-S test shows that the median-normalized distributions of $\blosr/\nbh^{0.65}$ of the strong- and weak-field simulation
samples and of the observed OH+CN sources are all consistent with being drawn from the same distribution.

\item[6.] {\it Distribution of normalized total field strengths in the simulations.} The distribution of $\btotr/\nbh^{0.65}/\btotnam$ 
from either simulation is not a uniform distribution (as concluded by \citet{cru10} for the observed clumps) or a delta function,
but it is consistent with a log-normal distribution.
The uniform distribution of normalized total field strengths inferred by
\citet{cru10} from observation implies that min$(\btotmu/\nbhf^{0.65})=0.24 (\blosmu/\nbhf^{0.65})_\med$. Our strong-field simulation has a minimum value almost
3 times larger. Field-line tangling increases the rms field above the total field; averaged over the whole
clump, the minimum value of $\brmsmu/\nbhf^{0.65}$ is about 4 times larger than
the minimum value $\btotmu/\nbhf^{0.65}$ inferred by 
\citet{cru10}; the average value is about twice the minimum.

\item[7.] {\it The typical molecular clump in the ISM is significantly supercritical.} The {\it observed} 
mass-to-flux ratio, $\muplos$, is based on the line-of-sight field,
$\blos$, which is mass-weighted, and the column density in the central beam. 
For an axisymmetric cloud, the {\it actual} mass-to-flux ratio is based on the ratio of the surface density, $\Sigma$, in a central cylinder to the area-weighted field threading that cylinder in the midplane. For a turbulent cloud, we adopt $\mu_{\Phi,\,\act}=\min (\Sigma/\bmidz)$, where the minimum is evaluated with respect to different orientations and where $\bmidz$ is the component of the field in that orientation averaged over the area of the central disc. We find
\beq
\mu_{\Phi,\,\act,\,\med}\simeq 0.7\mu_{\Phi,\,\los,\,\med},
\eeq
substantially larger than the value $(0.25-0.5)\mu_{\Phi,\,\los,\,\med}$ expected from projection effects,
primarily due to the fact that the observed $\muphi$ is based on a density-weighted field, whereas the actual one  has no density weighting. 
Furthermore, the  column densities are approximately isotropic (there is no evidence for
significant flattening along the field lines), possibly
due to the turbulent nature of the field.
Since the observed median value of $\muplos$ is about 4, we infer that the
actual median value is about 3: The typical molecular clump in the ISM is significantly supercritical.
The harmonic
mean of the actual mass-to-flux ratio is predicted to be about 1.5 based on our results from the strong-field model.

\item[8.] {\it Harmonic mean mass-to-flux ratio.} The average value of the observed mass-to-flux ratio is based on the line-of-sight field, and it diverges. It is therefore better to average the
flux-to-mass ratio, which is equivalent to using the harmonic mean of the mass-to-flux ratio. The same holds true for the actual mass-to-flux ratio if the distribution of field strengths extends to very low values, as suggested by \citet{cru10}.

\item[9.] {\it Strong vs. weak mean fields in molecular gas.} The strong-field (initial \alfven Mach number
$\mao=1$) and weak-field (initial \alfven mach number $\mao=10$) give substantially similar results
in almost all the measurements discussed above because turbulence is very effective at amplifying the initially
weak field so that it is comparable to the field in the strong-field case, consistent with previous results (e.g., \citet{lun08, pad10}). 
The mass-to-flux ratio based on the mean field in the turbulent box is 10 times higher in the weak-field case than in the strong-field case, but the mass-to-flux ratio in individual clumps is about the same in the two cases
due to turbulent fragmentation in the weak-field case.
There is a significant difference
in the accretion rates onto the clumps in the two simulations, with the rate in the weak-field simulation
being slower due to the more complex field geometry.

\item[10.] {\it Non-ideal MHD effects.} 
The agreement of our results with observations of magnetic fields in molecular cloud clumps
is consistent with a relatively minor role for ambipolar diffusion in the observed clumps.
We did find some evidence
for non-ideal effects in our simulations, however:
The time-averaged value of $\muphi(\bmidz)$ in the strong-field case
is larger than the initial value of $\muphi$ for the entire box. This can occur only due to numerical effects.
We presented evidence that our simulations are converged, so that our results are consistent with the increase
being due to reconnection in a turbulent medium 
\citep{laz15}.

\item[11.] {\it The mean field in the ISM is strong.} Based on observations of the alignment of polarization
on large and small scales, \citet{li09} concluded that the mean field in the vicinity of molecular clouds 
is strong, corresponding to an \alfven Mach number $\ma(\avg{B})\la 1.$ Our simulations support this conclusion: 
There is negligible correlation between the orientation of the clump fields with the mean field in the weak-field simulation, whereas the correlation in the strong-field simulation is comparable to the observed values.
\end{itemize}

\noindent
\section*{Acknowledgments}

We would like to thank R. Crutcher for his many helpful comments on this work. 
We also thank E. Falgarone, C. Heiles, and T. Troland for discussions on Zeeman measurement observations, C. Hull for comments on field orientations, and both R. Klessen and the referee for a number of helpful comments on our work.
We thank Philip Stark for discussing statistical methods for fitting data with large uncertainties.
Support for this research was provided by NASA through NASA ATP grant NNX13AB84G (RIK, CFM, and PSL), the US Department of Energy at the Lawrence Livermore National Laboratory under contract DE-AC52-07NA 27344 (RIK), and the NSF through grant AST-1211729 (CFM and RIK).  This research was also supported by grants of high performance computing resources from the National Center of Supercomputing Application through grant TG-MCA00N020, under the Extreme Science and Engineering Discovery Environment (XSEDE), which is supported by National Science Foundation grant number OCI-1053575, the computing resources provided by the NASA High-End Computing (HEC) Program through the NASA Advanced Supercomputing (NAS) Division at Ames Research Center, and the computing resources of the National Energy Research Scientific Computing Center, which is supported by the Office of Science of the U.S. Department of Energy under Contract No. DE-AC02-05CH11231.

\appendix

\section{Theory: The Magnetic Field vs. Gravity}
\label{app:theory}

The relative importance of magnetic fields and gravity is usually parametrized by
the ratio of the mass to the magnetic critical mass \citep{mou76}, 
\beq
M_\Phi=c_\Phi\left(\frac{\Phi}{ G^{1/2}}\right), 
\label{eq:mphi}
\eeq
where $\Phi$ is the magnetic flux and $c_\Phi$ is a numerical constant. 
The value of $c_\Phi$ depends on whether $M_\Phi$ is for the entire cloud or for the central
flux tube; in the former case, it also depends on the flux distribution. The flux distribution can
be described in terms of the equilibrium shape of the cloud in the absence of gravity, so that the
field is uniform.
\citet{mou76} considered the flux distribution corresponding to a cloud that is spherical in the absence of gravity;
\citet{tom88} considered several cases, including both a spherical and a cylindrical initial cloud.
For the spherical case and taking $\Phi$ to be the flux threading the entire cloud,
\citet{mou76} found $c_\Phi=0.126$.
Subsequent calculations by \citet{tom88} gave $c_\Phi\simeq 0.12$ for this case. On the other hand, if
one focuses on the central flux tube, \citet{tom88} showed that $c_\Phi=0.17-0.18$ for the three
cases they considered, so that it is relatively independent of the flux distribution. 
This value is very close to 
the value appropriate for 
a slab threaded by a uniform, perpendicular field,
$c_\Phi=1/2\pi$ \citep{nak78}. Since this is very close to the value for the central flux tube and
since \citet{cru10} used this value, we shall adopt it also,
both for the clumps that form in our simulation and for the simulation box (Section \ref{sec:sim}) as a whole.
The normalized mass-to-flux ratio (henceforth, we shall refer to this as just the `mass-to-flux ratio') is then
\beq
\mu_\Phi\equiv\frac{ M}{M_\Phi}=2\pi G^{1/2}\;\frac{\Sigma}{B}=3.80\;\frac{N_{\rm H,\,21}}{ B_\mu},
\label{eq:muphi}
\eeq
where $\Sigma$ is the mass surface density,
$N_{\rm H,\,21}$ is the column density of H nuclei in units of $10^{21}$~cm\ee,
$B$ is the magnitude of the magnetic field, 
and $B_\mu=B/(1\;\mu$G). 

It is convenient to relate $\mu_\Phi$ to other dimensionless parameters that characterize the system.
Since we wish to do this for the simulation box as well as in clumps, both simulated
and observed, we first define two geometrical factors. For the
simulation box, let $\ell=\ell_0$ be the box size, whereas for a spherical cloud, let $\ell$ be the diameter.
We then define
\beqa
A&\equiv& c_A\ell^2,\label{eq:A}\\
V&\equiv& c_V \ell^3,
\label{eq:V}
\eeqa
where $c_A=c_V=1$ for a box; for a sphere, $c_A=\pi/4$ and $c_V=\pi/6$;
and for an ellipsoid that is axisymmetric about the $z$ axis with semi-major axes
$R=\ell/2$ and $Z$, $c_A=\pi/4$ for the surface normal to $z$ and $c_V=(\pi/6)Z/R$.
For example, the surface density, $\Sigma$, 
of a cloud is related to its mean density, $\bar\rho$, by $\Sigma=M/A=(c_V/c_A)\bar\rho\ell$.

Next, we compare the kinetic and gravitational energies of a cloud.
Let $\snt$ be the mass-weighted, non-thermal
1D velocity dispersion, $\vrms=\surd 3 \snt$ the 3D non-thermal velocity dispersion,
$\cs$ the isothermal sound speed, and
$\calm=\vrms/\cs$ the 3D Mach number.
The mass-weighted, total
1D velocity dispersion, $\sid$, is then given by
\beq
\sid^2=\cs^2+\snt^2=\left(1+\frac{3}{\calm^2}\right)\snt^2\equiv c_\calm\snt^2.
\label{eq:sid}
\eeq
Since the motions in molecular clouds are often highly supersonic, $c_\calm$ is generally of order unity.
To measure the relative importance of 
kinetic energy (including thermal motions) and self-gravity, we use the virial parameter \citep{ber92}
\beq
\avir\equiv \frac{5\sid^2 \ell}{2GM}= \frac{5c_\calm\snt^2 \ell}{2GM},
\label{eq:avir}
\eeq
where we have replaced the radius in the original definition by $\ell/2$. For
centrally concentrated clouds, $\avir$ is somewhat greater than twice the ratio of the kinetic
energy to the gravitational energy \citep{ber92}.
The mass-to-flux ratio can then be expressed as
\beq
\mu_\Phi=\left(\frac{5\pi c_Vc_\calm}{6c_A^2 \avir}\right)^{1/2}\ma,
\label{eq:muphi1}
\eeq
where  $\ma=\vrms/\va$ is the 3D \alfven Mach number
and $\va=B/(4\pi\bar\rho)^{1/2}$ is the \alfven velocity. 
For a sphere, the numerical factor is $(5\pi c_V/6c_A^2)^{1/2}= (20/9)^{1/2}=1.49$, only slightly less than
the value for a box, $(5\pi/6)^{1/2}=1.62$. Furthermore, both these factors are of order unity: It follows that
highly supersonic clouds with $\avir\sim \ma\sim 1$ are close to magnetically critical, $\muphi\sim 1$.
In Section \ref{sec:obsmfr} 
we evaluated $\muphi$ with several different values of $B$, and it must be kept in mind
that the same value of $B$ must be used in both $\muphi$ and $\ma$ when applying Equation (\ref{eq:muphi1}).

Note that solving Equation (\ref{eq:muphi1})
for $B$ gives a relation of the form derived by \citet{mou91} and \citet{bas00} for planar clouds,
\beq
B=\left(\frac{10\pi^2 c_V}{c_A^2\avir}\right)^{1/2}\frac{\bar\rho^{1/2}\sid}{\muphi}.
\label{eq:bsid}
\eeq
Note that this equation is actually an identity; it follows directly from the definitions of the quantities entering it.
Physics enters when it is assumed that the cloud is self-gravitating ($\avir\simeq 1$) and
magnetically critical ($\muphi\simeq 1$), as
\citet{mou91} did.
\citet{bas00} argued that the fact that this relation provided a good fit to the data suggested
that clouds are indeed flattened; however, we see that a relation of the same form
applies to spherical clouds as well.

\subsection{Relation between $B/\nbh^{2/3}$ and the Magnetic Critical Mass $M_B$}

An alternative form for the magnetic critical mass is 
\beq
M_B\equiv \frac{M}{\mu_\Phi^3}\equiv\frac{M_\Phi}{\mu_\Phi^2},
\label{eq:mbo}
\eeq
which was found by \citet{mes56} and related to $M_\Phi$ by \citet{mou76}.
Note that since $M\propto\ell^3$ and $\mu_\Phi\propto M/\Phi\propto \ell$, the critical mass
$M_B$ is independent of $\ell$:
\beq
M_B=2.15\times 10^{-4}\left(\frac{c_A^3}{c_V^2}\right)\frac{B_\mu^3}{\nbhf^2}~~M_\odot.
\label{eq:mb}
\eeq
For an ellipsoid we have $c_A^3/c_V^2=(9\pi/16)(R/Z)^2$.
Equations (\ref{eq:mbo}) and (\ref{eq:mb}) imply
\beq
\frac{B_\mu}{\nbhf^{2/3}}=16.7\left(\frac{c_V^{2/3}}{c_A\mu_\Phi}\right)\left(\frac{M}{M_\odot}\right)^{1/3}.
\eeq
Generalizing the scaling relations (\ref{eq:nscale}) and (\ref{eq:mscale}) given in Section \ref{sec:scale}
to include thermal motions ($c_\calm>1$), this becomes
\beq
\frac{B_\mu}{\nbhf^{2/3}}=11.3\left(\frac{c_V^{2/3}c_\calm^{1/3}\calm^{4/3}T_1^{2/3}}{c_A\spcs^{2/3}\avir^{1/3}\mu_\Phi}\right).
\label{eq:Bn}
\eeq
Thus, the ratio of the magnetic field to a power of the density, which is dimensional, is determined by the
temperature, the coefficient in the linewidth-size relation and the dimensionless parameters that
characterize the problem. 

It is important to note that the observed mass-to-flux ratio is $\mu_\Phi(\blos)$, where $\blos$ is the
{\it mass-weighted} field along the line of sight, whereas the value used in theoretical calculations, 
which assume axisymmetry, 
is $\mu_\Phi(B_p)$, where $B_p$ is the poloidal field threading the midplane of the entire cloud 
\citep{mou76} or the central region of the cloud \citep{tom88} and is
not density-weighted.

\section{Projected Field Distribution}
\label{app:proj}

\begin{table*}
\caption{Models of the Observed Fields}
\label{tab:2}
\begin{tabular}{lcccc}
\hline
\hline
 & Observation~~~~~ &\multicolumn{3}{c}{Model}\\
\cline{3-5}  \vspace{-0.1 cm}\\
 & & Uniform & Log normal & Exponential\\
 & & $f= 0.007$ & $\sigma= 0.825$ & \\
\hline
 $\blosnfa$& 22.45 & 19.91 & 24.68 & 22.69 \\ 
 $\blosnfam\equiv \bblosm$& 11.94 & 14.93 & 15.01 & 15.73 \\
 $\blosnfa/\bblosm$& 1.88 & 1.33 & 1.64 & 1.44 \\
 $\btotnfa=2\blosnfa$ & 44.90 & 39.81 & 49.37 & 45.38 \\  
 $\btotnfam$ &-- & 39.81 & 35.15 & 38.10 \\
 $\btotnfam/\bblosm$&-- & 2.67 & 2.34 & 2.42 \\
 fraction of $\btotnfa < \bblosm$ & -- & 18\% & 15\% & 15\% \\
 fraction of $\btotnfa > 3\bblosm$ & -- & 44\% & 38\% & 39\% \\
Reduced $\chi^2$ of fit & -- & 1.85 & 1.45 & 1.35 \\
\vspace{-0.4 cm}\\
\hline
\hline
\end{tabular}
\end{table*}

The distribution of the observed values of $\blosmu/\nbhf^{0.65}$ is shown in Fig. \ref{fig2}.
In order to relate a given model for the total fields to the observed line-of-sight fields, we must determine
the probability distribution of the line-of-sight field, $\psi(\blos)$, 
from that of the total field,
$\phi(\btot)$. This has been done by \citet{hei05a},
\beq
\psi(\blos)=\int_{\blos}^\infty \frac{\phi(\btot)}{\btot}\; d\btot.
\label{eq:psia}
\eeq
One can readily show that this implies that 
\beq
\avg{\blos^n}=\frac{1}{n+1}\avg{\btot^n};
\eeq
in particular, $\avg{\blos}=\frac 12\avg{\btot}$.
Equation (\ref{eq:psia}) can be differentiated to give
\beq
\phi(\btot)= -\frac{d\psi(\blos)}{d\ln\blos}\Bigg\vert_{\blos=\btot},
\label{eq:phi}
\eeq
but this is not very useful in analyzing data because the derivative amplifies the noise.
In terms of the normalized field, $b$, defined in Equation (\ref{eq:b}), Equation (\ref{eq:psia}) becomes
\beq
\psi(\bblos)=\int_{\blos}^\infty \frac{\phi(\bbtot)}{\bbtot}\; d\bbtot.
\label{eq:psib}
\eeq

In order to relate a given model for the total fields to the observed line-of-sight fields, we must determine
the probability distribution of the line-of-sight field, $\psi(\blos)$, 
from that of the total field,
$\phi(\btot)$. This has been done by \citet{hei05a},
\beq
\psi(\blos)=\int_{\blos}^\infty \frac{\phi(\btot)}{\btot}\; d\btot.
\label{eq:psia}
\eeq
One can readily show that this implies that 
\beq
\avg{\blos^n}=\frac{1}{n+1}\avg{\btot^n};
\eeq
in particular, $\avg{\blos}=\frac 12\avg{\btot}$.
Equation (\ref{eq:psia}) can be differentiated to give
\beq
\phi(\btot)= -\frac{d\psi(\blos)}{d\ln\blos}\Bigg\vert_{\blos=\btot},
\label{eq:phi}
\eeq
but this is not very useful in analyzing data because the derivative amplifies the noise.
In terms of the normalized field, $b$, defined in Equation (\ref{eq:b}), Equation (\ref{eq:psia}) becomes
\beq
\psi(\bblos)=\int_{\blos}^\infty \frac{\phi(\bbtot)}{\bbtot}\; d\bbtot.
\label{eq:psib}
\eeq

We shall consider three possible distributions of field strengths. The first is the uniform distribution of \citet{cru10},
\beqa
\phi(\bbtot) &\simeq& \frac{1}{(1-f)\bbtm} \nonumber\\
&&\left[\Hea(\bbtot-f\bbtm)-\Hea(\bbtot-\bbtm)\right],
\eeqa
where $\Hea(x)$ is the Heaviside step function. For such a distribution,
the distribution of the line-of-sight field (Eq. \ref{eq:psia}) is 
\beq
\psi(\bblos)=\frac{1+f}{(1-f)4\avg{\bblos}}\ln\min\left[\frac{4\avg{\bblos}}{(1+f)\bblos},\,\frac{1}{f}\right]
\eeq
for $\bblos\leq\bbtm$ and $\psi(\bblos)=0$ otherwise. In expressing this result,
we have taken advantage of the general result $\avg{\blos}=\frac 12 \avg{\btot}$, which
implies
\beq
\bbtm=\frac{4\avg{\bblos}}{1+f}.
\eeq
The mean and median of the total density-normalized field are 
\beq
\avg{\bbtot}=(\bbtot)_\med=\frac 12 (1+f)\bbtm.
\eeq
By contrast, the median of the line-of-sight field is less than the mean; for $f=0$, it is
$(\bblos)_\med=0.748\avg{\bblos}$.
Since $\avg{\bbtot}=2\avg{\bblos}$, and since $\avg{\bbtot}=(\bbtot)_\med$, it follows that 
for $f=0$ the median line-of-sight field is related to the median total field by $(\bblos)_\med=0.374(\bbtot)_\med$.

A second possible form for the distribution of total field strengths is a log normal, for which the distribution of field strengths is given by
\beq
\phi(\bbtot)d\bbtot=\frac{1}{(2\pi)^{1/2}\sigma}\exp\left[-\frac{\left(x+\frac 12\sigma^2\right)^2}{2\sigma^2}\right]
dx,
\eeq
where 
\beq
x\equiv\ln\frac{\bbtot}{\avg{\bbtot}}.
\eeq
Note that $\phi$ is not a log normal itself, but rather a log normal times $dx/d\bbtot=1/\bbtot$.
The distribution of the line-of-sight field (Eq. \ref{eq:psia}) is then
\beqa
\psi(\bblos)&=&\frac{ e^{\sigma^2}}{(8\pi)^{1/2}\sigma\,\avg{\bblos}}\nonumber\\
 &&\int_\xlos^\infty \exp\left[-\frac{\left(x +\frac 32 \sigma^2\right)^2}{ 2\sigma^2}\right] dx, \\
&=& \frac{e^{\sigma^2}}{4\avg{\bblos}}{\rm erfc}\left(\frac{\xlos+\frac 32 \sigma^2}{\surd 2\, \sigma}\right),
\eeqa
where erfc$(x)$ is the complementary error function and
\beq
\xlos\equiv\ln\frac{\bblos}{\avg{\bbtot}}=\ln\frac{\bblos}{2\avg{\bblos}}.
\eeq
The median value of $\bbtot$ is related to the average value by
$\bbtotm=\avg{\bbtot}\exp(-\sigma^2/2)$. There does not appear to be a simple
expression for the median value of the line-of-sight field.
However, we can obtain an approximation as follows. For $\sigma\rightarrow 0$, the distribution of $\bbtot$ is
a delta function and $\psi(\bblos)$ is uniform between 0 and $\avg{\bbtot}=\bbtot$; as a result, $(\bblos)_\med=\frac 12 
\bbtot$. On the other hand, in the limit $\sigma\gg 1$, the dispersion in $\bbtot$ dominates the
dispersion due to the projection onto the line of sight, and $(\bblos)_\med=\avg{\bblos}e^{-\sigma^2/2}$, just
as in the case of $\bbtotm$ and $\avg{\bbtot}$ (this can also be shown by direct calculation). 
An approximation that reduces to these two limits is $(\bblos)_\med\simeq
\avg{\bblos}e^{-\sigma^2/2}$. It follows that the dispersion of the total field can be inferred directly from
observations of the line-of-sight field from
\beq
\sigma^2\simeq 2\ln\left[\frac{\avg{\bblos}}{(\bblos)_\med}\right].
\eeq

 The third distribution we shall consider is an exponential distribution of the line-of-sight field,
 \beq
 \psi(\bblos)=\frac{1}{\avg{\bblos}}\exp\left(-\frac{\bblos}{\avg{\bblos}}\right).
 \eeq
 Equation (\ref{eq:phi}) then implies that this corresponds to a distribution of the total field,
 \beq
 \phi(\bbtot)=\frac{\bbtot}{\avg{\bblos}^2}\exp\left(-\frac{\bbtot}{\avg{\bblos}}\right).
 \eeq
This distribution differs from the first two in that it is not symmetric.

Model fits to the observed line-of-sight data from \citet{cru10} are summarized in Table \ref{tab:2}.
We evaluated the goodness of fit with a $\chi^2$ test in $\log(\blosmu/\nbhf^{0.65})$.
For the uniform distribution, $\chi^2=1.85$. The best-fit value of $f=\mbox{min}(\btot/\btm)$ is 0.007, smaller than the value of 0.03 found by \citet{cru10}, which included HI sources as well.
The reduced $\chi^2$ values of the lognormal and exponential distribution (at the bottom of the table) are slightly smaller than that of the uniform distribution, implying that they provide slightly better fits.
Table \ref{tab:2} also gives several characteristic values of $\blos$ for the three theoretical models,
as well as for the observations summarized in \citet{cru10}.
Both the uniform and lognormal distributions have means that differ by about 10 per cent from the observed mean value, whereas the exponential function has a mean that is significantly closer.  The median values of all three models are larger than the observed median value by about 25 per cent.  The ratio of the mean to median,
which is a measure of the shape of the distribution, 
for the lognormal distribution is closer to the observed value than for the other two distributions.

\section{Selection Requirements of the 100-Clump Sample}
\label{app:100clumps}
To select clumps from simulations for comparison with observations, we impose several requirements:

\begin{itemize}
\item[1.] Resolution: The 68 OH and CN Zeeman observations in \citet{cru10} employed several different radio telescopes.  There is information on the beam sizes for the 34 OH Zeeman observations using the Arecibo telescope \citep{tro08} and the 14 CN Zeeman observations using the IRAM-30 m telescope \citep{fal08}.  
It is important to note that $\blos$ is measured in a single telescope beam
(see Section \ref{sec:obs}), and for resolved clumps this is smaller than the whole clump, which is usually defined as the region inside the half-maximum contour in the intensity map. Higher resolution observations will thus probe the dense inner region of a clump better than observations with a larger beam.
The observed clumps we compare with have a minimum 
radius of 0.03 pc.  The minimum beam size of the 48 OH and CN clumps is 0.025pc. The remaining OH sources are most likely observed with a larger beam size.  Taking into account the projection effect from observations, the 3D distance between two clumps of minimum projected distance will be $\approx 0.058$ pc.  We therefore merge all clumps that are separated by 0.06 pc or less, since they would be seen as single clumps in the observations. A possible correlation between the clump properties and the beam size is discussed in Section \ref{sec:den}.

\item[2.] Density: In Fig. 1 of \citet{cru10}, the observed OH and CN clumps appear to have a power-law relation between the maximum
$\blos$ and the density, whereas the HI clumps are on a horizontal branch, with the maximum
$\blos$ independent of density.  The power-law relation starts roughly at $\nbh \approx 300$ cm$^{-3}$.    
As discussed in Section \ref{sec:obs}, this density corresponds approximately to the critical density,
$n_{\rm H,\,crit}$, at which the thermal pressure
of the gas balances the turbulent pressure of the ambient interstellar medium.
Our models have a much higher turbulent pressure, $P_{\rm turb}/k = 1.92 \times 10^5$ K cm$^{-3}$,
because we do not have sufficient dynamic range to include the diffuse ISM as well as dense CN clumps.
At $T = 10$ K and $x_\tot =0.6$ for molecular gas, this implies $n_{\rm H,\,crit} = 3.2 \times 10^4$ cm$^{-3}$.  
Since the focus of our research is on magnetic fields in molecular gas, all of which is above the critical density,
we therefore require our clumps to have minimum density larger than $4 \times 10^4$ cm$^{-3}$.
Below, we shall see that this density corresponds approximately to the beginning of a power-law relation
between $\blos$ and density, just as in the observations. Furthermore, these high densities
are a result of self-gravity; only a few clumps have densities above this threshold prior to the time at which
self-gravity is turned on.

\item[3.] Mass: There are still a few hundred clumps after filtering out the clumps that are too close to adjacent clumps or are too low in mean density.  On the other hand, the observed sample of molecular clumps with fields measured by the Zeeman effect is less than 100. In order to have a sample that is comparable to the observed one, but has enough clumps to obtain reasonably accurate measured mean values, we choose the 100 most massive of the simulated clumps that survive the filtering process. All these clumps have at least 10000 grid cells, so they are well-resolved; fortuitously, they also have sizes $R\ga 0.03$ pc, the minimum resolution of the observations that we are comparing with. Note that our clumps are less massive than the observed ones: The total mass in the simulated region is only $3110 M_{\sun}$ and not many of the simulated clumps have masses larger than $10 M_{\sun}$, whereas the most massive clumps in the observed sample exceed $10^3 M_\odot$.
We also looked at the 100 clumps with the largest volume, 40 per cent of which were not in the most massive sample, and found that 
$\avg{\bbtot}$, $\avg{\bblos}$, and $\avg{\muphi(\blos)}$ in the two samples agree to within approximately 1 sigma.

\end{itemize}

In order to study the properties of the clumps in the absence of self-gravity, we also consider non-self-gravitating 100-clump samples from the strong and weak field simulations, 
in which there is no density threshold. The clumps in these samples have densities
$\nbh\ga 2\times 10^4$~cm\eee, 
about half the minimum in the self-gravitating case.

\section{100-Clump Sample Convergence Study}
\label{app:convergence}

We compare magnetic field properties of clumps obtained from CLUMPFIND in this paper with observations.  Therefore, our convergence study focuses on the magnetic field properties 
of clumps for the whole clump, the central region of the clumps, and the central region of the clump convolved with a Gaussian beam.  We compare the clumps from the 2-level refinement, strong-field model with a 3-level refinement run with exactly the same initial conditions.  With our computing resources, we can afford to run the 3-level model only up to $0.4 \tff$, when the power-law branch of the clumps is well developed; we therefore compare the two runs at that time.  See Section \ref{sec:100clump} on how the 100-clump samples are obtained using CLUMPFIND and Section \ref{sec:comp} on how the size of the convolved Gaussian beam region is chosen.

\setcounter{table}{0} \renewcommand{\thetable}{A.\arabic{table}} 
\begin{table}
\caption{Convergence Study of 100-Clump Sample Convolved With a 0.025 pc Gaussian Beam at 0.4 $\tff$\label{tab:con-1}}
\begin{center}
\begin{tabular}{lll}
\hline
\hline
Model\fnm[1]                        & 2-level        & 3-level\\
\hline
$\langle \blosr \rangle$\fnm[2]     & $76.0\pm7.4$   & $70.4\pm7.4$    \\
$\blosrmed$                         & 43.6           & 42.4            \\
$\langle \btotr \rangle$            & $151.6\pm21.1$ & $148.2\pm20.6$  \\
$\btotrmed$                         & 95.7           & 90.9            \\
$\blosna$                           & $0.028\pm0.003$   & $0.028\pm0.003$   \\
$\blosnam$                          & 0.023          & 0.021            \\
$\btotna$                           & $0.054\pm0.003$& $0.055\pm0.004$ \\
$\btotnam$                          & 0.049          & 0.045           \\
$\langle \muplos(\blosr) \rangle_h$\fnm[3] & $2.0\pm0.1$ & $2.0\pm0.2$       \\
$(\muplos(\blosr))_{\rm med}$       & 2.6            & 2.7             \\
$\langle \NH \rangle^d$             & $4.04\pm0.29$  & $4.05\pm0.23$   \\
$(\NH)_{{\rm med}}$\fnm[4]          & 2.61           & 2.74            \\
$\langle n_{\rm H,los} \rangle$\fnm[5] & $1.91\pm0.13$  & $1.89\pm0.13$   \\ 
$(n_{\rm H,los})_{\rm med}$\fnm[5]  & 1.10           & 1.11            \\
\hline
\end{tabular}
\end{center}
\footnotesize
\fnt{1}{$^a$ Two models labeled by the number of refinement levels.}\\
\fnt{2}{$^b$ All magnetic field strengths are in units of $\mu$G.}\\
\fnt{3}{$^c$ Harmonic mean of the line-of-sight mass-to-flux ratio is computed using $\blosr$.}\\
\fnt{4}{$^d$ Column density in unit of $10^{22}$ cm$^{-2}$.}\\
\fnt{5}{$^e$ Volumn density computed from $\NH / (2 \times r_{\rm clump})$ in unit of $10^5$ cm$^{-3}$.}\\
\end{table}

Table \ref{tab:con-1} shows the convergence results for the 100-clump sample convolved with a Gaussian beam of 0.025 pc radius.  Only line-of-sight quantities
and quantities, such as $\btot$ that are computed based on line-of-sight quantities,
are included in this table.
The uncertainties given in each case are the standard error of the mean; all quantities are converged to within this value.
Table \ref{atbl-2} shows the means and medians of the volume- and mass-weighted magnetic field strength, the mass-to-flux ratio, the column density, and the \alfvenic Mach number, both for whole clumps and for small central cylinders of radius 0.025 pc.  We use a cylinder rather than a Gaussian beam here since we are testing for convergence
of the actual mass-to-flux ratio, not the observed one.
Almost all quantities from the two runs are converged within 1 sigma, 
except the mass-weighted $\brmsr$ is slightly larger than 1 sigma but well within the sum of the two errors of the mean.  With the same sample size for both the 2-level and 3-level 100-Clump samples, the difference is not statistically significant.  Therefore, the 2-level refinement run is converged at 0.025 pc radius size scale.

\setcounter{table}{1} \renewcommand{\thetable}{A.\arabic{table}} 
\begin{table}
\caption{Convergence Study of 100-Clump Sample of the Whole Clump and Inside the Central 0.025 pc region at 0.4 $\tff$\label{atbl-2}}
\begin{center}
\begin{tabular}{llll}
\hline
\hline
Model\fnm[1]   &              &    2-level     &    3-level\\
\hline

Clump Center   & $\langle \blosr \rangle$\fnm[2] & $70.4\pm6.2$   & $67.7\pm7.0$   \\
     & $(\blosr)_{\rm med}$          & 41.8           & 43.3           \\
     & $\langle \muplos(\blosr) \rangle$\fnm[3]  & $2.1\pm0.1$    & $2.0\pm0.1$    \\
     & $(\muplos(\blosr))_{\rm med}$ & 2.8            & 2.7            \\
     & $\langle \NH \rangle$\fnm[4]  & $4.64\pm0.29$  & $4.46\pm0.29$  \\
     & $(\NH)_{\rm med}$             & 2.84           & 2.78           \\
     & $\langle \btotv \rangle$  & $83.7\pm3.1$   & $83.5\pm3.1$   \\
     & $(\btotv)_{\rm med}$          & 79.2           & 77.7           \\
     & $\langle \brmsv \rangle$  & $117.6\pm4.5$  & $121.3\pm4.8$  \\
     & $(\brmsv)_{\rm med}$          & 113.3          & 114.3          \\
     & $\langle \btotr \rangle$  & $143.0\pm17.0$ & $141.3\pm19.6$ \\
     & $(\btotr)_{\rm med}$          & 93.7          & 92.7          \\
     & $\langle \brmsr \rangle$\fnm[5] & $223.5\pm31.5$ & $270.4\pm41.5$ \\
     & $(\brmsr)_{\rm med}$          & 129.9          & 127.6          \\
\hline
Whole Clump & $\langle \btotv \rangle$ & $68.0\pm2.4$   & $67.7\pm2.3$   \\
     & $(\btotv)_{\rm med}$          & 65.0           & 66.6           \\
     & $\langle \brmsv \rangle$  & $79.8\pm2.5$   & $78.4\pm2.4$   \\
     & $(\brmsv)_{\rm med}$          & 78.2           & 77.5           \\
     & $\langle \btotr \rangle$  & $108.7\pm11.7$ & $98.6\pm8.6$   \\
     & $(\btotr)_{\rm med}$          & 78.5           & 76.2           \\
     & $\langle \brmsr \rangle$  & $154.4\pm20.7$ & $165.2\pm23.8$ \\
     & $(\brmsr)_{\rm med}$          & 96.5           & 91.9           \\
     & $\langle \muphi(B_{{\rm tot},\,V}) \rangle_h$\fnm[6] & $1.70\pm0.09$  & $1.62\pm0.09$  \\
     & $(\muphi(B_{{\rm tot},\,V}))_{\rm med}$ & 1.83   & 1.77          \\
     & $(\muphi(|B|_{{\rm tot},\,V}))_{\rm med}$ & 1.75 & 1.68          \\
     & $\langle \ma \rangle$     & $0.99\pm0.05$      & $1.01\pm0.05$     \\
     & $(\ma)_{\rm med}$         & 0.86               & 0.86      \\
\hline
\end{tabular}
\end{center}
\footnotesize
\fnt{1}{$^a$ The two models are labeled by the number of refinement levels.}\\
\fnt{2}{$^b$ All magnetic field strengths are in units of $\mu $G.}\\
\fnt{3}{$^c$ The line-of-sight mass-to-flux ratio is computed using $\blosr$.}\\
\fnt{4}{$^d$ Column density in units of $10^{22}$ cm$^{-2}$.}\\
\fnt{5}{$^e$ We have added the subscript `$M$' to the mass-weighted rms field to emphasize the distinction from $\brmsv$, which appears in the main text.}\\
\fnt{6}{$^f$ The harmonic mean mass-to-flux ratio for whole clumps. For each clump, the mass-to-flux ratio is determined along 1152 different orientations using the volume weighted field, $B_{{\rm tot},\,V}$, and the minimum value is adopted. (Section \ref{sec:actual})}\\
\end{table}

\section{Sampling the distribution of the line-of-sight magnetic field}
\label{app:sample}

The OH and CN sources discussed in this paper are from three separate papers by \citet{cru99}, \citet{tro08}, and \citet{fal08}, using different telescopes at different wavelengths.  From Fig. 1 in \citet{cru10}, the measured line-of-sight magnetic field strength $\blos$ ranges from a few to a few thousand \mug.  Forty of the 68 OH+CN sources have measured values of $\blos < 2 \sigma$ uncertainty.  Because the noise level is so large, the {\it true} values of $\blos$ could be significantly different than the values listed in Table 1 in \citet{cru10}, which could have a significant effect on the inferred the power-law index $\alpha$ that relates $\btot$ and $\nh$.  Here we use a numerical experiment to determine how to infer the value of $\alpha$ from noisy data.

In this experiment, we assume that the normalized total magnetic fields, $\btotr/\nbh^{\alpha}$,
have a uniform distribution between 0 
and $\btm/\nbh^{0.65}$, close to the distribution inferred by \citet{cru10}.
We then generate an 8000-point sample of $\btot/\nbh^{0.65}$ from this distribution. Since the observed OH sources have densities in the range $\nbh = 1\times10^3 - 2\times10^4$ cm$^{-3}$, and the CN sources are in the range $\nbh = 2\times10^5 - 4\times10^6$ cm$^{-3}$, these 8000 points are generated in two equal groups in these two density ranges.   
We then randomly project $\btot$ to the line-of-sight to obtain 8000 points in the $\blos-\nbh$ plane.
A least squares fit of $\log(\blos)$ vs. $\log(\nbh)$ gives $\alpha=0.65 \pm 0.01$, 
where it should be recalled that 0.01 is 2$\sigma$. This confirms that we have correctly implemented
the model of \citet{cru10}.

Next, we select 40 random samples of $\blos$, each with 
the same number of low-density points (54) and high-density points (14) as in the observed data.
These points represent the `true' values of this quantity.
We then add noise to these samples. Since
the CN data has less noise than the OH data---8 of the 14 CN sources are greater than $2 \sigma$, whereas only
20 of the 54 OH sources are---we determine the values of the standard deviation 
$\sigma$ for the high- and low-density groups separately so 
that each sample has the same number of values of $\blos$ above $2\sigma$ as in the data.  
We then determine the `observed' $\blos$ for each data point by randomly drawing from a Gaussian distribution, using the `true' $\blos$ and $\sigma$ as the mean and standard deviation, respectively.  With the noise, the `observed' value of $\blos$ can be negative; when this happens we retain only the magnitude of $\blos$ since we are interested in the distribution of field magnitudes.

Once we have the 68 data points including noise for each of the 40 samples, we  determine $\alpha$ in each sample from a least squares fit of $\log(\blos)$ vs. $\log(\nbh)$ in which we treat all the data points equally. The result is $\alpha=0.63\pm0.13$; the standard error of the mean is 0.01. This value is close to, but slightly less than, the
value $\alpha=0.65$ used in generating the initial 8000-point data set. The decrease in $\alpha$ is to be expected:
since the low density points are noisier than the high density ones, the negative values are larger, as are the corresponding
magnitudes. As a result, noise increases the average magnitude of the low density points more than 
that of the high density ones and the slope is reduced. Although we have been careful to reproduce the sampling of
the data, that does not affect the value of $\alpha$: Adding noise to the 8000 points in the same manner as we did
for the samples (i.e., so that 34/54 of the 4000 low density points are less than 2$\sigma$ and 6/14 of the 4000
high density ones are), a least squares fit gives
$\alpha=0.63\pm 0.01$, just as before. We conclude that the
greater noise in the low-density data reduces the inferred value of $\alpha$ by about 0.02 below the true value.

The effect of the different accuracies of the low and high density data can also be seen
if we exclude data points that are $< 2 \sigma$; we then find $\alpha$ is $0.59\pm0.05$, 
more than $2\sigma$ less than the correct value of 0.65.
Elimination of the data below 2$\sigma$ increases the average value of the retained OH data more than
of the retained CN data, thereby reducing the slope.
Furthermore, this value of $\alpha$ is
more than 1$\sigma$ less than the value obtained
by treating all points equally. We conclude that it is more accurate to treat all the points equally in the least squares
fitting than to omit the points that are less than $2\sigma$.

\label{lastpage}

\end{document}